\documentclass[aps,prd,twocolumn,superscriptaddress,nofootinbib]{revtex4-1}

\usepackage{latexsym}
\usepackage{amsmath}
\usepackage{amssymb}
\usepackage{amsfonts}
\usepackage{bm}

\usepackage{color}
\definecolor{purple}{rgb}{0.5,0,0.5}
\definecolor{blue}{rgb}{0.0,0,0.9}
\definecolor{prdblue}{rgb}{0.133,0.118,0.498}
\usepackage[colorlinks=true, pdfstartview=FitV, linkcolor=prdblue, citecolor= prdblue, urlcolor=prdblue]{hyperref}

\usepackage{supertabular} 
\usepackage{placeins}
\usepackage{epsfig}
\usepackage{graphicx}

\begin{document}


\title{Charmoniumlike tetraquarks in a chiral quark model}


\author{Gang Yang}
\email[]{yanggang@zjnu.edu.cn}
\affiliation{Department of Physics, Zhejiang Normal University, Jinhua 321004, China}

\author{Jialun Ping}
\email[]{jlping@njnu.edu.cn}
\affiliation{Department of Physics and Jiangsu Key Laboratory for Numerical Simulation of Large Scale Complex Systems, Nanjing Normal University, Nanjing 210023, P. R. China}

\author{Jorge Segovia}
\email[]{jsegovia@upo.es}
\affiliation{Departamento de Sistemas F\'isicos, Qu\'imicos y Naturales, Universidad Pablo de Olavide, E-41013 Sevilla, Spain}



\begin{abstract}
The lowest-lying charmonium-like tetraquarks $c\bar{c}q\bar{q}$ $(q=u,\,d)$ and $c\bar{c}s\bar{s}$, with spin-parity $J^P=0^+$, $1^+$ and $2^+$, and isospin $I=0$ and $1$, are systematically investigated within the theoretical framework of complex-scaling range for a chiral quark model that has already been successfully applied in former studies of various tetra- and penta-quark systems.
A four-body $S$-wave configuration which includes meson-meson, diquark-antidiquark and K-type arrangements of quarks, along with all possible color wave functions, is comprehensively considered. Several narrow resonances are obtained in each tetraquark channel when a fully coupled-channel computation is performed.
We tentatively assign theoretical states to experimentally reported charmonium-like signals such as $X(3872)$, $Z_c(3900)$, $X(3960)$, $X(4350)$, $X(4685)$ and $X(4700)$. They can be well identified as hadronic molecules; however, other exotic components which involve, for instance, hidden-color channels or diquark-antidiquark structures play a considerable role. 
Meanwhile, two resonances are obtained at $4.04$ GeV and $4.14$ GeV which may be compatible with experimental data in the energy interval $4.0-4.2$ GeV. Furthermore, the $X(3940)$ and $X(4630)$ may be identified as color compact tetraquark resonances.
Finally, we also find few resonance states in the energy interval from $4.5$ GeV to $5.0$ GeV, which would be awaiting for discovery in future experiments.
\end{abstract}


\maketitle


\section{Introduction}

The past two decades have been witnessed a golden age of meson spectroscopy, many new exotic charmonium- and bottomonium-like hadrons, which are collectively called $XYZ$ states, have been reported by worldwide experimental collaborations.\footnote{Pentaquark candidates with heavy quark content, named $P_{c(s)}$ hadrons, have also been identified; we shall not refer to them herein because they are usually considered as part of the baryon spectrum puzzle.} Particularly, it can be dated back to $2003$, when a hidden-charm tetraquark candidate, $X(3872)$, was first announced by the Belle Collaboration~\cite{skc:2003prl}. After then, along the successive years, more charmonium-like states such as $Y(4260)$~\cite{ba:2005prl}, $Y(4140)$~\cite{ta:2009prl}, $Z_c(3900)$~\cite{ta:2013prl}, etc., have been experimentally reported. Furthermore, in the last five years, a plethora of exotic structures, that are good candidates of $Q\bar{Q}q\bar{q}$ $(q=u,\,d,\,s;\,Q=c,\,b)$ tetraquarks, have been found. The masses of these exotic states are generally located near particular thresholds of two conventional open-heavy-flavor mesons. Examples of this fact are the charged charmonium-like states with strangeness $X_{0,1}(2900)$~\cite{LHCb:2020pxc, LHCb:2020bls}, $X(2600)$~\cite{BESIIICollaboration:2022kwh}, $Z_{cs}(3985)$~\cite{BESIII:2020qkh}, $Z_{cs}(4000)$, $Z_{cs}(4220)$, $X(4630)$ and $X(4685)$~\cite{LHCb:2021uow}; the hidden charm structures $\psi_2(3823)$~\cite{LHCb:2022oqs, BESIII:2022yga} and $X(3960)$~\cite{LHCb:2022vsv}; the charmonium-like vector meson states $Y(4230)$, $Y(4500)$, $Y(4660)$ and $Y(4710)$~\cite{BESIII:2023unv, Ablikim:2022yav, BESIII:2022joj}; and the $J/\psi$-$J/\psi$ bound-state candidates~\cite{LHCb:2020bwg}.

From the theoretical side, an enormous effort devoted to reveal the nature of the unexpected exotic hadrons has been deployed using a wide variety of approaches. In fact, one can already find in the literature many extensive reviews~\cite{Dong:2020hxe, Chen:2016qju, Chen:2016spr, Guo:2017jvc, Liu:2019zoy, Yang:2020atz, Dong:2021bvy, Chen:2021erj, Cao:2023rhu, Mai:2022eur, Meng:2022ozq, Chen:2022asf, Guo:2022kdi, Ortega:2020tng}, which explain in detail a particular theoretical method and thus capture a certain interpretation of the $XYZ$ states. One can find either compact or molecular tetraquark interpretaions, or even relate the signals with simple kinematical effects such as cusps.

The charmonium-like states are the most investigated ones. The $X(3872)$ has been identified as a $D\bar{D}^*$ molecular state when using effective field theories~\cite{Ding:2020dio, Xu:2021vsi} and QCD sum rules~\cite{Wang:2019tlw}. Its interpretation as a combination of $\omega J/\psi$ and $\rho J/\psi$ hadro-charmonia has recently been proposed by a Monte Carlo analysis~\cite{Gordillo:2021bra, Gordillo:2022nnj}. The $X(3872)$ can be explained within phenomenological quark models as either a mixture of $D\bar{D}^*$ molecule and diquark-antidiquark component~\cite{Lebed:2022vks} or a coupling between charmonium and di-meson states~\cite{Ortega:2009hj, Tan:2019qwe, Kanwal:2022ani, Chen:2023wfp}. The description of the $X(3872)$ as a compact tetraquark can be found in, for instance, Refs.~\cite{Maiani:2004vq, Esposito:2016noz}. Another possibility is that the $X(3872)$, as most of the exotic signals, can be just a triangle singularity effect~\cite{Sakai:2020ucu, Molina:2020kyu, Nakamura:2019nch, Guo:2020oqk}. Other relevant theoretical investigations on the nature of the $X(3872)$ state can be found in Refs.~\cite{Ortega:2020lhr, Ortega:2019fme, Ortega:2021yis, Giacosa:2019zxw, Giron:2020fvd}. Beyond its nature, other properties of the $X(3872)$ meson have been investigated as its production via various high energy processes, such as $e^+ e^-$ and $p \bar{p}$ annihilation~\cite{Braaten:2019gwc, Sakai:2020crh}, $B$ and $B_s$ decays~\cite{Maiani:2020zhr, Wang:2022xga}, lepton-proton scattering~\cite{Yang:2021jof, Xie:2021sik} and heavy-ion collisions~\cite{Zhang:2020dwn, Chen:2021akx}. The $X(3872)$'s strong decays have also been studied in Refs.~\cite{Yu:2019ing, Dai:2019hrf, Chen:2019eeq, Bruschini:2022bsh}.

Theoretical analysis using effective field theories~\cite{Ding:2020dio, Wang:2020dko, Wang:2022ztm}, QCD sum rules~\cite{Chen:2021erj, Wang:2022fdu} and neural networks~\cite{Chen:2022ddj} have identified the $Z_c(3900)$ and $Z_c(4020)$ states as $D\bar{D}^*$ and $D^* \bar{D}^*$ molecular resonances, respectively. Furthermore, to explain the mentioned states, charmonium-like compact tetraquarks and threshold bumps have been proposed in constituent quark model investigations~\cite{Zhao:2020jvl, Ortega:2019uuk}. The strong and weak decays of the $Z_c(3900)$ and $Z_c(4020)$ have been studied~\cite{Chen:2019wjd, Xiao:2019spy, Huang:2022zsy}, their properties in cold dense matter have been analyzed~\cite{Azizi:2020itk} and a lattice QCD simulation has been performed too~\cite{Liu:2019gmh}. 

Other exotic hadrons within the hidden-charm sector and without strangeness have triggered many theoretical studies. Particularly, the $Z_c(4000)$ state, which is probably a $D^* \bar{D}^*$ molecule, has been studied through a $B$ decay process~\cite{Zhang:2020rqr}. The $X(3940)$, $X(4020)$ and $X(4050)$ states have been identified as hidden-charm tetraquarks with dominant diquark-antidiquark components in a quark model framework~\cite{Shi:2021jyr}. The $X(4014)$ has been assigned as a $D^* \bar{D}^*$ molecule with $J^{PC}=0^{++}$~\cite{Duan:2022upr, Yue:2022gym}. Three more hidden-charm states, $\psi(4230)$, $\psi(4360)$ and $\psi(4415)$, have been predicted to be $J^{PC}=0^{--}$ $D^{(*)} \bar{D}^{(*)}$ molecular bound-states within an effective field theory approach~\cite{Ji:2022blw}.

Extensive theoretical efforts have been also devoted to reveal the nature of recently reported $X(3960)$ state. A $D_s \bar{D}_s$ molecular structure has been obtained in effective field theory~\cite{Bayar:2022dqa, Ji:2022uie} and QCD sum rules~\cite{Mutuk:2022ckn}. Within the same configuration, production and strong decay properties of the $X(3960)$ have been studied in Refs.~\cite{Xie:2022lyw, Wu:2023fyh, Hu:2023hrn, Xin:2022bzt, Ji:2022vdj}. A mixture of a conventional $c\bar{c}$ state and $D_s \bar{D}_s$ continuum is found to be the correct mechanism to describe the $X(3960)$ state in an effective field theory study~\cite{Chen:2023eix}. The importance of coupling $D^* \bar{D}^*$ and $D^*_s \bar{D}^*_s$ degree of freedom is highlighted within a one boson-exchange model approach~\cite{Chen:2022dad}. Moreover, a $c\bar{c}s\bar{s}$ compact tetraquark structure is identified as the $X(3960)$ state when using other phenomenological quark models~\cite{Guo:2022crh, Badalian:2023qyi}. 

There are other tetraquark candidates in the $c\bar{c}s\bar{s}$ sector. The $X(4630)$ has been identified as a $D^*_s \bar{D}_s$ molecule in $1^{-+}$ channel by the one boson-exchange model of Ref.~\cite{Yang:2021sue}. This conclusion is also supported by a QCD sum rules study~\cite{Wang:2021ghk}. The $X(3912)$ and $X(4500)$ have been identified as $(cs)(\bar{c}\bar{s})$ diquark-antidiquark structures with $J^{PC}=0^{++}$. The $X(4685)$ has been interpreted as a tetraquark resonance when studying B decays~\cite{Ge:2021sdq} and using QCD sum rules~\cite{Turkan:2021ome, Yang:2022zxe}. Besides, several $D^{(*)}_s \bar{D}^{(*)}_s$ bound or resonance states are claimed in Refs.~\cite{Meng:2020cbk, Wang:2021gml, Wang:2020dya}. Finally, other properties as structure content and decays of charmonium-like states such as the $X(3915)$, $X(4140)$, $X(4160)$, $X(4274)$, $X(4350)$ and $X(4700)$ have been reported in Refs.~\cite{Hao:2019fjg, Hao:2020fsl, Guo:2022zbc, Wang:2022clw, Wang:2020vhq, Li:2022fjf, Ferretti:2020civ, Yang:2010sf, Duan:2020tsx, Giron:2020qpb, Liu:2021xje, Badalian:2022hfu, Li:2022fgd}.

Following our systematic investigation of bottomonium-like tetraquarks~\cite{Yang:2022cut} and complementing our previous study of hidden-charm tetraquarks with strange content~\cite{Yang:2021zhe}, we intend here to close the circle and perform a comprehensive analysis of charmonium-like tetraquarks $c\bar{c}q\bar{q}$ $(q=u,\,d)$ and $c\bar{c}s\bar{s}$, with spin-parity $J^P=0^+$, $1^+$ and $2^+$, and isospin $I=0$ and $1$. We shall employ a QCD-inspired chiral quark model which has been successfully applied before in the description of various multiquark systems, \emph{e.g.} strange-, double- and full-heavy tetraquarks~\cite{Yang:2021izl, gy:2020dht, gy:2020dhts, Yang:2021hrb}, and hidden-, double- and full-heavy pentaquarks~\cite{Yang:2015bmv, Yang:2018oqd, gy:2020dcp, Yang:2022bfu}. The formulation of the phenomenological model in complex-scaling method (CSM) has been discussed in detail in Ref.~\cite{Yang:2020atz}. The CSM shall allow us to distinguish between bound, resonance and scattering states, and thus perform a complete analysis of the scattering singularities within the same formalism. Furthermore, the meson-meson, diquark-antidiquark and K-type configurations, plus their couplings with all possible color structures, shall be considered.

We arrange the manuscript as follows. In Sec.~\ref{sec:model} the theoretical framework is presented, we briefly introduce the complex-range method applied to a chiral quark model and the $c\bar{c}q\bar{q}$ $(q=u,\,d)$ and $c\bar{c}s\bar{s}$ tetraquark wave functions. Section~\ref{sec:results} is devoted to the analysis and discussion of the obtained results. Finally, a summary is presented in Sec.~\ref{sec:summary}.


\section{Theoretical framework}
\label{sec:model}

A throughout review of the theoretical formalism used herein has been published in Ref.~\cite{Yang:2020atz}. We shall then focus on the most relevant features of the phenomenological model and the numerical method concerning the hidden-charm tetraquarks $c\bar{c}q\bar{q}$ $(q=u,\,d,\,s)$.
 
Within the framework of complex-scaling range, the relative coordinate of a two-body interaction is rotated in the complex plane by an angle $\theta$, $\vec{r}_{ij}\to \vec{r}_{ij} e^{i\theta}$. Therefore, the general form of our four-body Hamiltonian reads as
\begin{equation}
H(\theta) = \sum_{i=1}^{4}\left( m_i+\frac{\vec{p\,}^2_i}{2m_i}\right) - T_{\text{CM}} + \sum_{j>i=1}^{4} V(\vec{r}_{ij} e^{i\theta}) \,,
\label{eq:Hamiltonian}
\end{equation}
where $m_{i}$ is the quark mass, $\vec{p}_i$ is the quark's momentum, $T_{\text{CM}}$ is the center-of-mass kinetic energy and the last term is the two-body potential. The complex scaled Schr\"odinger equation:
\begin{equation}\label{CSMSE}
\left[ H(\theta)-E(\theta) \right] \Psi_{JM}(\theta)=0	
\end{equation}
has (complex) eigenvalues which can be classified into three types: bound, resonance and scattering states. In particular, bound-states and resonances are independent of the rotated angle $\theta$, with the first ones placed on the real-axis of the complex energy plane and the second ones located above the continuum threshold with a total decay width $\Gamma=-2\,\text{Im}(E)$.

The dynamics of the $c\bar{c}q\bar{q}$ tetraquark system is driven by a two-body potential,
\begin{equation}
\label{CQMV}
V(\vec{r}_{ij}) = V_{\chi}(\vec{r}_{ij}) + V_{\text{CON}}(\vec{r}_{ij}) + V_{\text{OGE}}(\vec{r}_{ij})  \,,
\end{equation}
which takes into account the most relevant features of QCD at its low energy regime: dynamical chiral symmetry breaking, confinement and perturbative one-gluon exchange interaction. Herein, the low-lying $S$-wave positive parity $c\bar{c}q\bar{q}$ tetraquark states shall be investigated, and thus the central and spin-spin terms of the potential are the only ones needed.

One consequence of the dynamical breaking of chiral symmetry is that Goldstone boson exchange interactions appear between constituent light quarks $u$, $d$ and $s$. Therefore, the chiral interaction can be written as:
\begin{equation}
V_{\chi}(\vec{r}_{ij}) = V_{\pi}(\vec{r}_{ij})+ V_{\sigma}(\vec{r}_{ij}) + V_{K}(\vec{r}_{ij}) + V_{\eta}(\vec{r}_{ij}) \,,
\end{equation}
given by
\begin{align}
&
V_{\pi}\left( \vec{r}_{ij} \right) = \frac{g_{ch}^{2}}{4\pi}
\frac{m_{\pi}^2}{12m_{i}m_{j}} \frac{\Lambda_{\pi}^{2}}{\Lambda_{\pi}^{2}-m_{\pi}
^{2}}m_{\pi} \Bigg[ Y(m_{\pi}r_{ij}) \nonumber \\
&
\hspace*{1.20cm} - \frac{\Lambda_{\pi}^{3}}{m_{\pi}^{3}}
Y(\Lambda_{\pi}r_{ij}) \bigg] (\vec{\sigma}_{i}\cdot\vec{\sigma}_{j})\sum_{a=1}^{3}(\lambda_{i}^{a}
\cdot\lambda_{j}^{a}) \,, \\
& 
V_{\sigma}\left( \vec{r}_{ij} \right) = - \frac{g_{ch}^{2}}{4\pi}
\frac{\Lambda_{\sigma}^{2}}{\Lambda_{\sigma}^{2}-m_{\sigma}^{2}}m_{\sigma} \Bigg[Y(m_{\sigma}r_{ij}) \nonumber \\
&
\hspace*{1.20cm} - \frac{\Lambda_{\sigma}}{m_{\sigma}}Y(\Lambda_{\sigma}r_{ij})
\Bigg] \,,
\end{align}
\begin{align}
& 
V_{K}\left( \vec{r}_{ij} \right)= \frac{g_{ch}^{2}}{4\pi}
\frac{m_{K}^2}{12m_{i}m_{j}}\frac{\Lambda_{K}^{2}}{\Lambda_{K}^{2}-m_{K}^{2}}m_{
K} \Bigg[ Y(m_{K}r_{ij}) \nonumber \\
&
\hspace*{1.20cm} -\frac{\Lambda_{K}^{3}}{m_{K}^{3}}Y(\Lambda_{K}r_{ij}) \Bigg] (\vec{\sigma}_{i}\cdot\vec{\sigma}_{j}) \sum_{a=4}^{7}(\lambda_{i}^{a} \cdot \lambda_{j}^{a}) \,, \\
& 
V_{\eta}\left( \vec{r}_{ij} \right) = \frac{g_{ch}^{2}}{4\pi}
\frac{m_{\eta}^2}{12m_{i}m_{j}} \frac{\Lambda_{\eta}^{2}}{\Lambda_{\eta}^{2}-m_{
\eta}^{2}}m_{\eta} \Bigg[ Y(m_{\eta}r_{ij}) \nonumber \\
&
\hspace*{1.20cm} -\frac{\Lambda_{\eta}^{3}}{m_{\eta}^{3}
}Y(\Lambda_{\eta}r_{ij}) \Bigg] (\vec{\sigma}_{i}\cdot\vec{\sigma}_{j})
\Big[\cos\theta_{p} \left(\lambda_{i}^{8}\cdot\lambda_{j}^{8}
\right) \nonumber \\
&
\hspace*{1.20cm} -\sin\theta_{p} \Big] \,,
\end{align}
where $Y(x)=e^{-x}/x$ is the standard Yukawa function. The physical $\eta$ meson, instead of the octet one, is considered by introducing the angle $\theta_p$. The $\lambda^{a}$ are the SU(3) flavor Gell-Mann matrices. Taken from their experimental values, $m_{\pi}$, $m_{K}$ and $m_{\eta}$ are the masses of the SU(3) Goldstone bosons. The value of $m_{\sigma}$ is determined through the relation $m_{\sigma}^{2}\simeq m_{\pi}^{2}+4m_{u,d}^{2}$~\cite{Scadron:1982eg}. Finally, the chiral coupling constant, $g_{ch}$, is determined from the $\pi NN$ coupling constant through
\begin{equation}
\frac{g_{ch}^{2}}{4\pi}=\frac{9}{25}\frac{g_{\pi NN}^{2}}{4\pi} \frac{m_{u,d}^{2}}{m_{N}^2} \,,
\end{equation}
which assumes that flavor SU(3) is an exact symmetry only broken by the different mass of the strange quark.

Color confinement should be encoded in the non-Abelian character of QCD. It has been demonstrated by lattice-regularized QCD that multi-gluon exchanges produce an attractive linearly rising potential proportional to the distance between infinite-heavy quarks~\cite{Bali:2005fu}. However, the spontaneous creation of light-quark pairs from the QCD vacuum may give rise at the same scale to a breakup of the created color flux-tube~\cite{Bali:2005fu}. These two observations can be described phenomenologically by
\begin{equation}
V_{\text{CON}}(\vec{r}_{ij})=\left[-a_{c}(1-e^{-\mu_{c}r_{ij}})+\Delta \right] 
(\lambda_{i}^{c}\cdot \lambda_{j}^{c}) \,,
\label{eq:conf}
\end{equation}
where $a_{c}$, $\mu_{c}$ and $\Delta$ are model parameters, and the SU(3) color Gell-Mann matrices are denoted as $\lambda^c$. One can see in Eq.~\eqref{eq:conf} that the potential is linear at short inter-quark distances with an effective confinement strength $\sigma = -a_{c} \, \mu_{c} \, (\lambda^{c}_{i}\cdot \lambda^{c}_{j})$, while it becomes constant at large distances, $V_{\text{thr.}} = (\Delta-a_c)(\lambda^{c}_{i}\cdot \lambda^{c}_{j})$.

Beyond the chiral symmetry breaking scale one expects the dynamics to be
governed by QCD perturbative effects. In particular, the one-gluon exchange potential, which includes the so-called Coulomb and color-magnetic interactions, is the leading order contribution:
\begin{align}
&
V_{\text{OGE}}(\vec{r}_{ij}) = \frac{1}{4} \alpha_{s} (\lambda_{i}^{c}\cdot \lambda_{j}^{c}) \Bigg[\frac{1}{r_{ij}} \nonumber \\ 
&
\hspace*{1.60cm} - \frac{1}{6m_{i}m_{j}} (\vec{\sigma}_{i}\cdot\vec{\sigma}_{j}) 
\frac{e^{-r_{ij}/r_{0}(\mu_{ij})}}{r_{ij} r_{0}^{2}(\mu_{ij})} \Bigg] \,,
\end{align}
where $r_{0}(\mu_{ij})=\hat{r}_{0}/\mu_{ij}$ is a regulator which depends on the reduced mass of the $q\bar{q}$ pair, the Pauli matrices are denoted by $\vec{\sigma}$, and the contact term has been regularized as
\begin{equation}
\delta(\vec{r}_{ij}) \sim \frac{1}{4\pi r_{0}^{2}(\mu_{ij})}\frac{e^{-r_{ij} / r_{0}(\mu_{ij})}}{r_{ij}} \,.
\end{equation}

An effective scale-dependent strong coupling constant, $\alpha_s(\mu_{ij})$, provides a consistent description of mesons and baryons from light to heavy quark sectors. We use the frozen coupling constant of Ref.~\cite{Segovia:2013wma},
\begin{equation}
\alpha_{s}(\mu_{ij})=\frac{\alpha_{0}}{\ln\left(\frac{\mu_{ij}^{2}+\mu_{0}^{2}}{\Lambda_{0}^{2}} \right)} \,,
\end{equation}
in which $\alpha_{0}$, $\mu_{0}$ and $\Lambda_{0}$ are parameters of the model.

The model parameters are listed in Table~\ref{tab:model}. Additionally, for later concern, Table~\ref{MesonMass} lists theoretical and experimental (if available) masses of $1S$, $2S$ and $3S$ states of $q\bar{q}$, $c\bar{q}$ $(q=u,\,d,\,s)$ and $c\bar{c}$ mesons.

\begin{table}[!t]
\caption{\label{tab:model} Model parameters.}
\begin{ruledtabular}
\begin{tabular}{llr}
Quark masses     & $m_q\,(q=u,\,d)$ (MeV) & 313 \\
                 & $m_s$ (MeV) &  555 \\
                 & $m_c$ (MeV) & 1752 \\[2ex]
Goldstone bosons & $\Lambda_\pi=\Lambda_\sigma~$ (fm$^{-1}$) &   4.20 \\
                 & $\Lambda_\eta=\Lambda_K$ (fm$^{-1}$)      &   5.20 \\
                 & $g^2_{ch}/(4\pi)$                         &   0.54 \\
                 & $\theta_P(^\circ)$                        & -15 \\[2ex]
Confinement      & $a_c$ (MeV)         & 430 \\
                 & $\mu_c$ (fm$^{-1})$ & 0.70 \\
                 & $\Delta$ (MeV)      & 181.10 \\[2ex]
OGE              & $\alpha_0$              & 2.118 \\
                 & $\Lambda_0~$(fm$^{-1}$) & 0.113 \\
                 & $\mu_0~$(MeV)           & 36.976 \\
                 & $\hat{r}_0~$(MeV~fm)    & 28.170 \\
\end{tabular}
\end{ruledtabular}
\end{table}

\begin{table*}[!t]
\caption{\label{MesonMass} Theoretical and experimental (if available) masses of $nL=1S, 2S$ and $3S$ states of $q\bar{q}$, $q\bar{c}\,(q=u, d, s)$ and $c\bar{c}$ mesons.}
\begin{ruledtabular}
\begin{tabular}{lccclccc}
Meson & $nL$ & $M_{\text{The.}}$ (MeV) & $M_{\text{Exp.}}$ (MeV)  & Meson & $nL$ & $M_{\text{The.}}$ (MeV) & $M_{\text{Exp.}}$ (MeV) \\
\hline
$\pi$ & $1S$ &  $149$ & $140$  & $\eta$ & $1S$ &  $689$ & $548$ \\
    & $2S$ & $1291$ & $1300$    &  & $2S$ & $1443$ & $1295$ \\
    & $3S$ & $1749$ & $1800$    &  & $3S$ & $1888$ & - \\[2ex]
$\rho$ & $1S$ &  $772$ & $770$  & $\omega$ & $1S$ &  $696$ & $782$ \\
    & $2S$ & $1479$ & $1450$    &  & $2S$ & $1449$ & $1420$ \\
    & $3S$ & $1940$ & -    &  & $3S$ & $1897$ & $1650$ \\[2ex]
$K$ & $1S$ &  $481$ & $494$  & $K^*$ & $1S$ &  $907$ & $892$ \\
    & $2S$ & $1468$ & $1460$    &  & $2S$ & $1621$ & $1630$ \\
    & $3S$ & $1898$ & $1830$    &  & $3S$ & $1998$ & - \\[2ex]
$\eta'$ & $1S$ &  $828$ & $958$  & $\phi$ & $1S$ &  $1011$ & $1020$ \\
    & $2S$ & $1639$ & $1760$    &  & $2S$ & $1720$ & $1680$ \\
    & $3S$ & $2057$ & -    &  & $3S$ & $2097$ & $2170$ \\[2ex]
$D$ & $1S$ &  $1897$ & $1870$  & $D^*$ & $1S$ &  $2017$ & $2007$ \\
    & $2S$ & $2648$ & -    &  & $2S$ & $2704$ & - \\
    & $3S$ & $3043$ & -    &  & $3S$ & $3072$ & - \\[2ex]
$D_s$ & $1S$ &  $1989$ & $1968$  & $D^*_s$ & $1S$ &  $2115$ & $2112$ \\
    & $2S$ & $2705$ & -    &  & $2S$ & $2769$ & - \\
    & $3S$ & $3118$ & -    &  & $3S$ & $3165$ & - \\[2ex]
$\eta_c$ & $1S$ &  $2989$ & $2981$  & $J/\psi$ & $1S$ &  $3097$ & $3097$ \\
    & $2S$ & $3627$ & -    &  & $2S$ & $3685$ & - \\
    & $3S$ & $4026$ & -    &  & $3S$ & $4063$ & -
\end{tabular}
\end{ruledtabular}
\end{table*}

\begin{figure}[ht]
\epsfxsize=3.4in \epsfbox{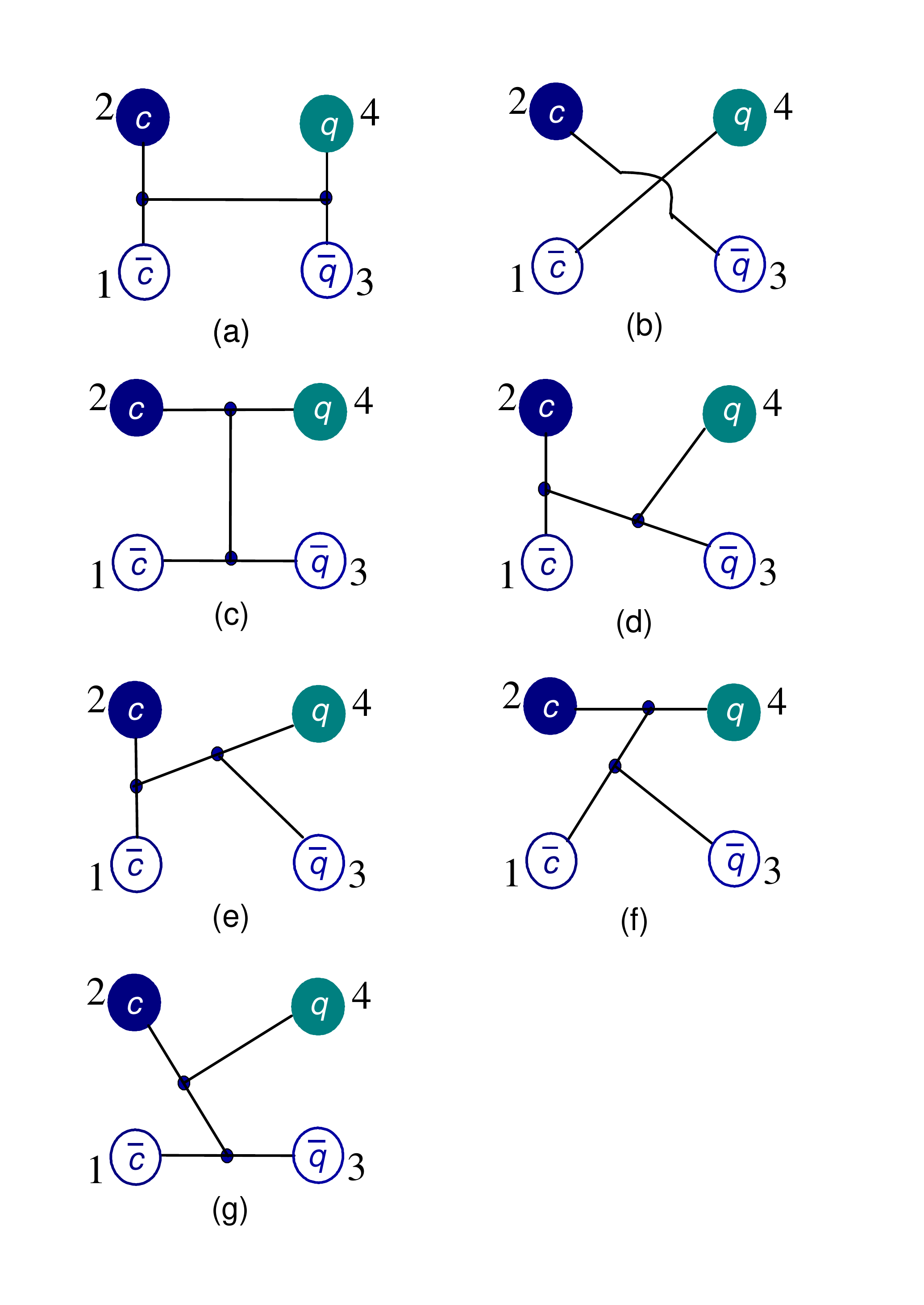}
\caption{\label{QQqq} Seven types of configurations are considered for the $c\bar{c}q\bar{q}$ $(q=u,\,d,\,s)$ tetraquarks. Panels $(a)$ and $(b)$ are meson-meson structures, panel $(c)$ is diquark-antidiquark arrangement and the other K-type configurations are from panel $(d)$ to $(g)$.}
\end{figure}

Complete $S$-wave $c\bar{c}q\bar{q}$ $(q=u,\,d,\,s)$ tetraquark configurations are considered in Figure~\ref{QQqq}. Particularly, Figs.~\ref{QQqq}(a) and (b) are the meson-meson structures, Fig.~\ref{QQqq}(c) is the diquark-antidiquark arrangement, and the other K-type configurations are from panels (d) to (g). All of them, and their couplings, are considered in our investigation. However, for the purpose of solving a manageable $4$-body problem, the K-type configurations are sometimes restricted as in our previous investigations~\cite{Yang:2021zhe, Yang:2022cut}. Furthermore, it is important to note that just one configuration would be enough for the calculation, if all radial and orbital excited states were taken into account; however, this is obviously inefficient and thus an economic way to proceed is the combination of the different mentioned configurations.

The multiquark system's wave function at the quark level is an internal product of color, spin, flavor and space terms. Concerning the color degree-of-freedom, the colorless wave function of a $4$-quark system in meson-meson configuration can be obtained by either two coupled color-singlet clusters, $1\otimes 1$:
\begin{align}
\label{Color1}
\chi^c_1 &= \frac{1}{3}(\bar{r}r+\bar{g}g+\bar{b}b)\times (\bar{r}r+\bar{g}g+\bar{b}b) \,,
\end{align}
or two coupled color-octet clusters, $8\otimes 8$:
\begin{align}
\label{Color2}
\chi^c_2 &= \frac{\sqrt{2}}{12}(3\bar{b}r\bar{r}b+3\bar{g}r\bar{r}g+3\bar{b}g\bar{g}b+3\bar{g}b\bar{b}g+3\bar{r}g\bar{g}r
\nonumber\\
&+3\bar{r}b\bar{b}r+2\bar{r}r\bar{r}r+2\bar{g}g\bar{g}g+2\bar{b}b\bar{b}b-\bar{r}r\bar{g}g
\nonumber\\
&-\bar{g}g\bar{r}r-\bar{b}b\bar{g}g-\bar{b}b\bar{r}r-\bar{g}g\bar{b}b-\bar{r}r\bar{b}b) \,.
\end{align}
The first color state is the so-called color-singlet channel and the second one is the named hidden-color case.

The color wave functions associated to the diquark-antidiquark structure are the coupled color triplet-antitriplet clusters, $3\otimes \bar{3}$:
\begin{align}
\label{Color3}
\chi^c_3 &= \frac{\sqrt{3}}{6}(\bar{r}r\bar{g}g-\bar{g}r\bar{r}g+\bar{g}g\bar{r}r-\bar{r}g\bar{g}r+\bar{r}r\bar{b}b
\nonumber\\
&-\bar{b}r\bar{r}b+\bar{b}b\bar{r}r-\bar{r}b\bar{b}r+\bar{g}g\bar{b}b-\bar{b}g\bar{g}b
\nonumber\\
&+\bar{b}b\bar{g}g-\bar{g}b\bar{b}g) \,,
\end{align}
and the coupled color sextet-antisextet clusters, $6\otimes \bar{6}$:
\begin{align}
\label{Color4}
\chi^c_4 &= \frac{\sqrt{6}}{12}(2\bar{r}r\bar{r}r+2\bar{g}g\bar{g}g+2\bar{b}b\bar{b}b+\bar{r}r\bar{g}g+\bar{g}r\bar{r}g
\nonumber\\
&+\bar{g}g\bar{r}r+\bar{r}g\bar{g}r+\bar{r}r\bar{b}b+\bar{b}r\bar{r}b+\bar{b}b\bar{r}r
\nonumber\\
&+\bar{r}b\bar{b}r+\bar{g}g\bar{b}b+\bar{b}g\bar{g}b+\bar{b}b\bar{g}g+\bar{g}b\bar{b}g) \,.
\end{align}

Meanwhile, the colorless wave functions of the K-type structures are given by
\begin{align}
\label{Color5}
\chi^c_5 &= \frac{1}{6\sqrt{2}}(\bar{r}r\bar{r}r+\bar{g}g\bar{g}g-2\bar{b}b\bar{b}b)+
\nonumber\\
&\frac{1}{2\sqrt{2}}(\bar{r}b\bar{b}r+\bar{r}g\bar{g}r+\bar{g}b\bar{b}g+\bar{g}r\bar{r}g+\bar{b}g\bar{g}b+\bar{b}r\bar{r}b)-
\nonumber\\
&\frac{1}{3\sqrt{2}}(\bar{g}g\bar{r}r+\bar{r}r\bar{g}g)+\frac{1}{6\sqrt{2}}(\bar{b}b\bar{r}r+\bar{b}b\bar{g}g+\bar{r}r\bar{b}b+\bar{g}g\bar{b}b) \,,
\end{align}
\begin{align}
\label{Color6}
\chi^c_6 &= \chi^c_1 \,,
\end{align}
\begin{align}
\label{Color7}
\chi^c_7 &= \chi^c_1 \,,
\end{align}
\begin{align}
\label{Color8}
\chi^c_8 &= \frac{1}{4}(1-\frac{1}{\sqrt{6}})\bar{r}r\bar{g}g-\frac{1}{4}(1+\frac{1}{\sqrt{6}})\bar{g}g\bar{g}g-\frac{1}{4\sqrt{3}}\bar{r}g\bar{g}r+
\nonumber\\
&\frac{1}{2\sqrt{2}}(\bar{r}b\bar{b}r+\bar{g}b\bar{b}g+\bar{b}g\bar{g}b+\bar{g}r\bar{r}g+\bar{b}r\bar{r}b)+
\nonumber\\
&\frac{1}{2\sqrt{6}}(\bar{r}r\bar{b}b-\bar{g}g\bar{b}b+\bar{b}b\bar{g}g+\bar{g}g\bar{r}r-\bar{b}b\bar{r}r) \,,
\end{align}
\begin{align}
\label{Color9}
\chi^c_9 &= \frac{1}{2\sqrt{6}}(\bar{r}b\bar{b}r+\bar{r}r\bar{b}b+\bar{g}b\bar{b}g+\bar{g}g\bar{b}b+\bar{r}g\bar{g}r+\bar{r}r\bar{g}g+
\nonumber\\
&\bar{b}b\bar{g}g+\bar{b}g\bar{g}b+\bar{g}g\bar{r}r+\bar{g}r\bar{r}g+\bar{b}b\bar{r}r+\bar{b}r\bar{r}b)+
\nonumber\\
&\frac{1}{\sqrt{6}}(\bar{r}r\bar{r}r+\bar{g}g\bar{g}g+\bar{b}b\bar{b}b) \,,
\end{align}
\begin{align}
\label{Color10}
\chi^c_{10} &= \frac{1}{2\sqrt{3}}(\bar{r}b\bar{b}r-\bar{r}r\bar{b}b+\bar{g}b\bar{b}g-\bar{g}g\bar{b}b+\bar{r}g\bar{g}r-\bar{r}r\bar{g}g-
\nonumber\\
&\bar{b}b\bar{g}g+\bar{b}g\bar{g}b-\bar{g}g\bar{r}r+\bar{g}r\bar{r}g-\bar{b}b\bar{r}r+\bar{b}r\bar{r}b) \,,
\end{align}
\begin{align}
\label{Color11}
\chi^c_{11} &= \chi^c_9 \,,
\end{align}
\begin{align}
\label{Color12}
\chi^c_{12} &= -\chi^c_{10} \,.
\end{align}

According to the nature of $SU(3)$-flavor symmetry, the $c\bar{c}q\bar{q}$ $(q=u,\,d,\,s)$ tetraquark systems could be categorized into three cases: $c\bar{c}q\bar{q}$ with $(q=u,\,d)$, $c\bar{c}s\bar{s}$ and $c\bar{c}q\bar{s}$ $(q=u,\,d)$. Last case was studied by us in Ref.~\cite{Yang:2021zhe} and thus only the first two tetraquark configurations are considered herein. The flavor wave function is then denoted as $\chi^{f_i}_{I, M_I}$, where the superscript $i=1$ and $2$ will refer to $c\bar{c}q\bar{q}$ and $c\bar{c}s\bar{s}$ systems, respectively. We have isoscalar, $I=0$, and isovector, $I=1$, sectors in $c\bar{c}q\bar{q}$ systems, their flavor wave functions read as
\begin{align}
&
\label{FWF0}
\chi_{0,0}^{f_1} = -\frac{1}{\sqrt{2}}(\bar{c}c\bar{u}u+\bar{c}c\bar{d}d) \,, \\
&
\label{FWF1}
\chi_{1,0}^{f_1} = -\frac{1}{\sqrt{2}}(\bar{c}c\bar{u}u-\bar{c}c\bar{d}d) \,,
\end{align}
where the third component of the isospin, $M_I$, is fixed to be zero for simplicity since the Hamiltonian does not have a flavor-dependent interaction which can distinguish the third component of the isospin quantum number.

We are going to considered $S$-wave ground states with spin ranging from $S=0$ to $2$. Therefore, the spin wave functions, $\chi^{\sigma_i}_{S, M_S}$, are given by ($M_S$ can be set to be equal to $S$ without loss of generality):
\begin{align}
\label{SWF0}
\chi_{0,0}^{\sigma_{u_1}}(4) &= \chi^\sigma_{00}\chi^\sigma_{00} \,, \\
\chi_{0,0}^{\sigma_{u_2}}(4) &= \frac{1}{\sqrt{3}}(\chi^\sigma_{11}\chi^\sigma_{1,-1}-\chi^\sigma_{10}\chi^\sigma_{10}+\chi^\sigma_{1,-1}\chi^\sigma_{11}) \,, \\
\chi_{0,0}^{\sigma_{u_3}}(4) &= \frac{1}{\sqrt{2}}\big((\sqrt{\frac{2}{3}}\chi^\sigma_{11}\chi^\sigma_{\frac{1}{2}, -\frac{1}{2}}-\sqrt{\frac{1}{3}}\chi^\sigma_{10}\chi^\sigma_{\frac{1}{2}, \frac{1}{2}})\chi^\sigma_{\frac{1}{2}, -\frac{1}{2}} \nonumber \\ 
&-(\sqrt{\frac{1}{3}}\chi^\sigma_{10}\chi^\sigma_{\frac{1}{2}, -\frac{1}{2}}-\sqrt{\frac{2}{3}}\chi^\sigma_{1, -1}\chi^\sigma_{\frac{1}{2}, \frac{1}{2}})\chi^\sigma_{\frac{1}{2}, \frac{1}{2}}\big) \,, \\
\chi_{0,0}^{\sigma_{u_4}}(4) &= \frac{1}{\sqrt{2}}(\chi^\sigma_{00}\chi^\sigma_{\frac{1}{2}, \frac{1}{2}}\chi^\sigma_{\frac{1}{2}, -\frac{1}{2}}-\chi^\sigma_{00}\chi^\sigma_{\frac{1}{2}, -\frac{1}{2}}\chi^\sigma_{\frac{1}{2}, \frac{1}{2}}) \,,
\end{align}
for $(S,M_S)=(0,0)$, by 
\begin{align}
\label{SWF1}
\chi_{1,1}^{\sigma_{w_1}}(4) &= \chi^\sigma_{00}\chi^\sigma_{11} \,, \\ 
\chi_{1,1}^{\sigma_{w_2}}(4) &= \chi^\sigma_{11}\chi^\sigma_{00} \,, \\
\chi_{1,1}^{\sigma_{w_3}}(4) &= \frac{1}{\sqrt{2}} (\chi^\sigma_{11} \chi^\sigma_{10}-\chi^\sigma_{10} \chi^\sigma_{11}) \,, \\
\chi_{1,1}^{\sigma_{w_4}}(4) &= \sqrt{\frac{3}{4}}\chi^\sigma_{11}\chi^\sigma_{\frac{1}{2}, \frac{1}{2}}\chi^\sigma_{\frac{1}{2}, -\frac{1}{2}}-\sqrt{\frac{1}{12}}\chi^\sigma_{11}\chi^\sigma_{\frac{1}{2}, -\frac{1}{2}}\chi^\sigma_{\frac{1}{2}, \frac{1}{2}} \nonumber \\ 
&-\sqrt{\frac{1}{6}}\chi^\sigma_{10}\chi^\sigma_{\frac{1}{2}, \frac{1}{2}}\chi^\sigma_{\frac{1}{2}, \frac{1}{2}} \,, \\
\chi_{1,1}^{\sigma_{w_5}}(4) &= (\sqrt{\frac{2}{3}}\chi^\sigma_{11}\chi^\sigma_{\frac{1}{2}, -\frac{1}{2}}-\sqrt{\frac{1}{3}}\chi^\sigma_{10}\chi^\sigma_{\frac{1}{2}, \frac{1}{2}})\chi^\sigma_{\frac{1}{2}, \frac{1}{2}} \,, \\
\chi_{1,1}^{\sigma_{w_6}}(4) &= \chi^\sigma_{00}\chi^\sigma_{\frac{1}{2}, \frac{1}{2}}\chi^\sigma_{\frac{1}{2}, \frac{1}{2}} \,,
\end{align}
for $(S,M_S)=(1,1)$, and by 
\begin{align}
\label{SWF2}
\chi_{2,2}^{\sigma_{1}}(4) &= \chi^\sigma_{11}\chi^\sigma_{11} \,,
\end{align}
for $(S,M_S)=(2,2)$. The superscripts $u_1,\ldots,u_4$ and $w_1,\ldots,w_6$ determine the spin wave function for each configuration of the $c\bar c q\bar q$ $(q=u,\,d,\,s)$ tetraquark system, their specific values are shown in Table~\ref{SpinIndex}. Furthermore, the expressions above are obtained by considering the coupling of two sub-clusters whose spin wave functions are given by trivial SU(2) algebra whose necessary basis reads as
\begin{align}
\label{Spin}
\chi^\sigma_{00} &= \frac{1}{\sqrt{2}}(\chi^\sigma_{\frac{1}{2}, \frac{1}{2}} \chi^\sigma_{\frac{1}{2}, -\frac{1}{2}}-\chi^\sigma_{\frac{1}{2}, -\frac{1}{2}} \chi^\sigma_{\frac{1}{2}, \frac{1}{2}}) \,, \\
\chi^\sigma_{11} &= \chi^\sigma_{\frac{1}{2}, \frac{1}{2}} \chi^\sigma_{\frac{1}{2}, \frac{1}{2}} \,, \\
\chi^\sigma_{1,-1} &= \chi^\sigma_{\frac{1}{2}, -\frac{1}{2}} \chi^\sigma_{\frac{1}{2}, -\frac{1}{2}} \,, \\
\chi^\sigma_{10} &= \frac{1}{\sqrt{2}}(\chi^\sigma_{\frac{1}{2}, \frac{1}{2}} \chi^\sigma_{\frac{1}{2}, -\frac{1}{2}}+\chi^\sigma_{\frac{1}{2}, -\frac{1}{2}} \chi^\sigma_{\frac{1}{2}, \frac{1}{2}}) \,.
\end{align}

\begin{table}[!t]
\caption{\label{SpinIndex} Values of the superscripts $u_1,\ldots,u_4$ and $w_1,\ldots,w_6$ that specify the spin wave function for each configuration of the $c\bar{c}q\bar{q}$ $(q=u,\,d,\,s)$ tetraquark system.}
\begin{ruledtabular}
\begin{tabular}{lcccccc}
& Di-meson & Diquark-antidiquark & $K_1$ & $K_2$ & $K_3$ & $K_4$ \\
\hline
$u_1$ & 1 & 3 & & & & \\
$u_2$ & 2 & 4 & & & & \\
$u_3$ &   &   & 5 & 7 &  9 & 11 \\
$u_4$ &   &   & 6 & 8 & 10 & 12 \\[2ex]
$w_1$ & 1 & 4 & & & & \\
$w_2$ & 2 & 5 & & & & \\
$w_3$ & 3 & 6 & & & & \\
$w_4$ &   &   & 7 & 10 & 13 & 16 \\
$w_5$ &   &   & 8 & 11 & 14 & 17 \\
$w_6$ &   &   & 9 & 12 & 15 & 18
\end{tabular}
\end{ruledtabular}
\end{table}

Among the different methods to solve the Schr\"odinger-like 4-body bound state equation, we use the Rayleigh-Ritz variational principle which is one of the most extended tools to solve eigenvalue problems because its simplicity and flexibility. Moreover, we use the complex-range method and thus the spatial wave function is written as follows:
\begin{equation}
\label{eq:WFexp}
\psi_{LM_L}(\theta)= \left[ \left[ \phi_{n_1l_1}(\vec{\rho}e^{i\theta}\,) \phi_{n_2l_2}(\vec{\lambda}e^{i\theta}\,)\right]_{l} \phi_{n_3l_3}(\vec{R}e^{i\theta}\,) \right]_{L M_L} \,,
\end{equation}
where the internal Jacobi coordinates are defined as
\begin{align}
\vec{\rho} &= \vec{x}_1-\vec{x}_2 \,, \\
\vec{\lambda} &= \vec{x}_3 - \vec{x}_4 \,, \\
\vec{R} &= \frac{m_1 \vec{x}_1 + m_2 \vec{x}_2}{m_1+m_2}- \frac{m_3 \vec{x}_3 + m_4 \vec{x}_4}{m_3+m_4} \,,
\end{align}
for the meson-meson configurations of Figs.~\ref{QQqq}(a) and (b); and as
\begin{align}
\vec{\rho} &= \vec{x}_1-\vec{x}_3 \,, \\
\vec{\lambda} &= \vec{x}_2 - \vec{x}_4 \,, \\
\vec{R} &= \frac{m_1 \vec{x}_1 + m_3 \vec{x}_3}{m_1+m_3}- \frac{m_2 \vec{x}_2 + m_4 \vec{x}_4}{m_2+m_4} \,,
\end{align}
for the diquark-antidiquark structure of Fig.~\ref{QQqq}(c). The remaining K-type configurations shown in Fig.~\ref{QQqq}(d) to \ref{QQqq}(g) are ($i, j, k, l$ take values according to the panels (d) to (g) of Fig.~\ref{QQqq}):
\begin{align}
\vec{\rho} &= \vec{x}_i-\vec{x}_j \,, \\
\vec{\lambda} &= \vec{x}_k- \frac{m_i \vec{x}_i + m_j \vec{x}_j}{m_i+m_j} \,, \\
\vec{R} &= \vec{x}_l- \frac{m_i \vec{x}_i + m_j \vec{x}_j+m_k \vec{x}_k}{m_i+m_j+m_k} \,.
\end{align}
It becomes obvious now that the center-of-mass kinetic term $T_\text{CM}$ can be completely eliminated for a non-relativistic system defined in any of the above sets of relative coordinates.

A crucial aspect of the Rayleigh-Ritz variational method is the basis expansion of the trial wave function. We are going to use the Gaussian expansion method (GEM)~\cite{Hiyama:2003cu} in which each relative coordinate is expanded in terms of Gaussian basis functions whose sizes are taken in geometric progression. This method has proven to be very efficient on solving the bound-state problem of a multiquark systems~\cite{Yang:2015bmv, gy:2020dcp, gy:2020dht} and the details on how the geometric progression is fixed can be found in \emph{e.g} Ref.~\cite{Yang:2015bmv}. Therefore, the form of the orbital wave functions, $\phi$'s, in Eq.~\eqref{eq:WFexp} is 
\begin{align}
&
\phi_{nlm}(\vec{r}e^{i\theta}\,) = N_{nl} \, (re^{i\theta})^{l} \, e^{-\nu_{n} (re^{i\theta})^2} \, Y_{lm}(\hat{r}) \,.
\end{align}
Since only $S$-wave states of $c\bar{c}q\bar{q}$ tetraquarks are investigated in this work, the spherical harmonic function is just a constant, \emph{viz.} $Y_{00}=\sqrt{1/4\pi}$, and thus no laborious Racah algebra is needed while computing matrix elements.

Finally, the complete wave-function that fulfills the Pauli principle is written as
\begin{equation}
\label{TPs}
\Psi_{JM_J,I,i,j,k}(\theta)=\left[ \left[ \psi_{LM_L}(\theta) \chi^{\sigma_i}_{SM_S}(4) \right]_{JM_J} \chi^{f_j}_I \chi^{c}_k \right] \,.
\end{equation}


\section{Results}
\label{sec:results}

In this calculation, we investigate the $S$-wave $c\bar{c}q\bar{q}$ $(q=u,\,d)$ and $c\bar{c}s\bar{s}$ tetraquarks by taking into account meson-meson, diquark-antidiquark and K-type configurations. In our approach, a $4-$body state has positive parity assuming that the angular momenta $l_1$, $l_2$ and $l_3$ in spatial wave function are all equal to zero. Accordingly, the total angular momentum, $J$, coincides with the total spin, $S$, and can take values of $0$, $1$ and $2$. Meanwhile, the value of isospin, $I$, can be either $0$ or $1$ considering the quark content of $c\bar{c}q\bar{q}$ system in the $SU(2)$ flavor symmetry, it is $0$ for $c\bar{c}s\bar{s}$ system.

Tables~\ref{GresultCC1} to~\ref{GresultCC9} list our calculated results of the low-lying hidden-charm tetraquark states. The allowed meson-meson, diquark-antidiquark and K-type configurations are listed in the first column; when possible, the experimental value of the non-interacting meson-meson threshold is labeled in parentheses. An index is assigned to each channel in the second column, which indicates a particular combination of spin ($\chi_J^{\sigma_i}$), flavor ($\chi_I^{f_j}$) and color ($\chi_k^c$) wave functions that are shown explicitly in the third column. The theoretical mass obtained in each channel is shown in the fourth column and the coupled result for each kind of configuration is presented in the last column. Last row of the table indicates the lowest-lying mass obtained when a complete coupled-channels calculation is performed. 

The CSM is employed in the complete coupled-channels calculation, we show in Fig.~\ref{PP1} to~\ref{PP9} the distribution of complex eigenenergies and, therein, the obtained resonance states are indicated inside circles. Some insights about the nature of these resonances are given by computing their sizes and probabilities of the different tetraquark configurations in their wave functions, the results are listed among the Tables~\ref{GresultCC1} to~\ref{GresultCC9}. A summary of our most salient results is presented in Table~\ref{GresultCCT}.

Let us proceed now to describe in detail our theoretical findings for each sector of hidden-charm tetraquarks.

\subsection{The $\mathbf{c\bar{c}q\bar{q}}\,(q=u,\,d)$ tetraquarks}

Several resonances whose masses range from $3.9$ GeV to $4.8$ GeV are obtained in this tetraquark sector, they can be well identified as experimental data on charmonium states. Each iso-scalar and -vector sectors with total spin and parity $J^P=0^+$, $1^+$ and $2^+$ shall be discussed individually below. In particular, due to the equivalence between some K-type $c\bar{c}q\bar{q}$ tetraquark configurations, only $K_1$ and $K_3$ are considered.


\begin{table}[!t]
\caption{\label{GresultCC1} Lowest-lying $c\bar{c}q\bar{q}$ tetraquark states with $I(J^P)=0(0^+)$ calculated within the real range formulation of the chiral quark model.
The allowed meson-meson, diquark-antidiquark and K-type configurations are listed in the first column; when possible, the experimental value of the non-interacting meson-meson threshold is labeled in parentheses. Each channel is assigned an index in the 2nd column, it reflects a particular combination of spin ($\chi_J^{\sigma_i}$), flavor ($\chi_I^{f_j}$) and color ($\chi_k^c$) wave functions that are shown explicitly in the 3rd column. The theoretical mass obtained in each channel is shown in the 4th column and the coupled result for each kind of configuration is presented in the 5th column.
When a complete coupled-channels calculation is performed, last row of the table indicates the calculated lowest-lying mass.
(unit: MeV).}
\begin{ruledtabular}
\begin{tabular}{lcccc}
~~Channel   & Index & $\chi_J^{\sigma_i}$;~$\chi_I^{f_j}$;~$\chi_k^c$ & $M$ & Mixed~~ \\
        &   &$[i; ~j; ~k]$ &  \\[2ex]
$(\eta_c \eta)^1 (3529)$          & 1  & [1;~1;~1]  & $3678$ & \\
$(J/\psi \omega)^1 (3879)$  & 2  & [2;~1;~1]   & $3793$ &  \\
$(D \bar{D})^1 (3740)$          & 3  & [1;~1;~1]  & $3794$ & \\
$(D^* \bar{D}^*)^1 (4014)$  & 4  & [2;~1;~1]   & $4034$ & $3678$ \\[2ex]
$(\eta_c \eta)^8$          & 5  & [1;~1;~2]  & $4395$ & \\
$(J/\psi \omega)^8$  & 6  & [2;~1;~2]   & $4239$ &  \\
$(D \bar{D})^8$          & 7  & [1;~1;~2]  & $4237$ & \\
$(D^* \bar{D}^*)^8$  & 8  & [2;~1;~2]   & $4223$ & $4107$ \\[2ex]
$(cq)(\bar{q}\bar{c})$      & 9   & [3;~1;~3]  & $4207$ & \\
$(cq)(\bar{q}\bar{c})$      & 10   & [3;~1;~4]  & $4294$ & \\
$(cq)^*(\bar{q}\bar{c})^*$  & 11  & [4;~1;~3]   & $4235$ & \\
$(cq)^*(\bar{q}\bar{c})^*$  & 12  & [4;~1;~4]   & $4177$ & $4066$ \\[2ex]
$K_1$  & 13  & [5;~1;~5]   & $4239$ & \\
  & 14  & [6;~1;~5]   & $4399$ & \\
  & 15  & [5;~1;~6]   & $4130$ & \\
  & 16  & [6;~1;~6]   & $4019$ & $4014$ \\[2ex]
$K_3$  & 17  & [9;~1;~9]   & $4181$ & \\
  & 18  & [10;~1;~9]   & $4243$ & \\
  & 19  & [9;~1;~10]   & $4290$ & \\
  & 20  & [10;~1;~10]   & $4206$ & $4057$ \\[2ex]
\multicolumn{4}{c}{Complete coupled-channels:} & $3678$
\end{tabular}
\end{ruledtabular}
\end{table}

\begin{figure}[!t]
\includegraphics[clip, trim={3.0cm 1.9cm 3.0cm 1.0cm}, width=0.45\textwidth]{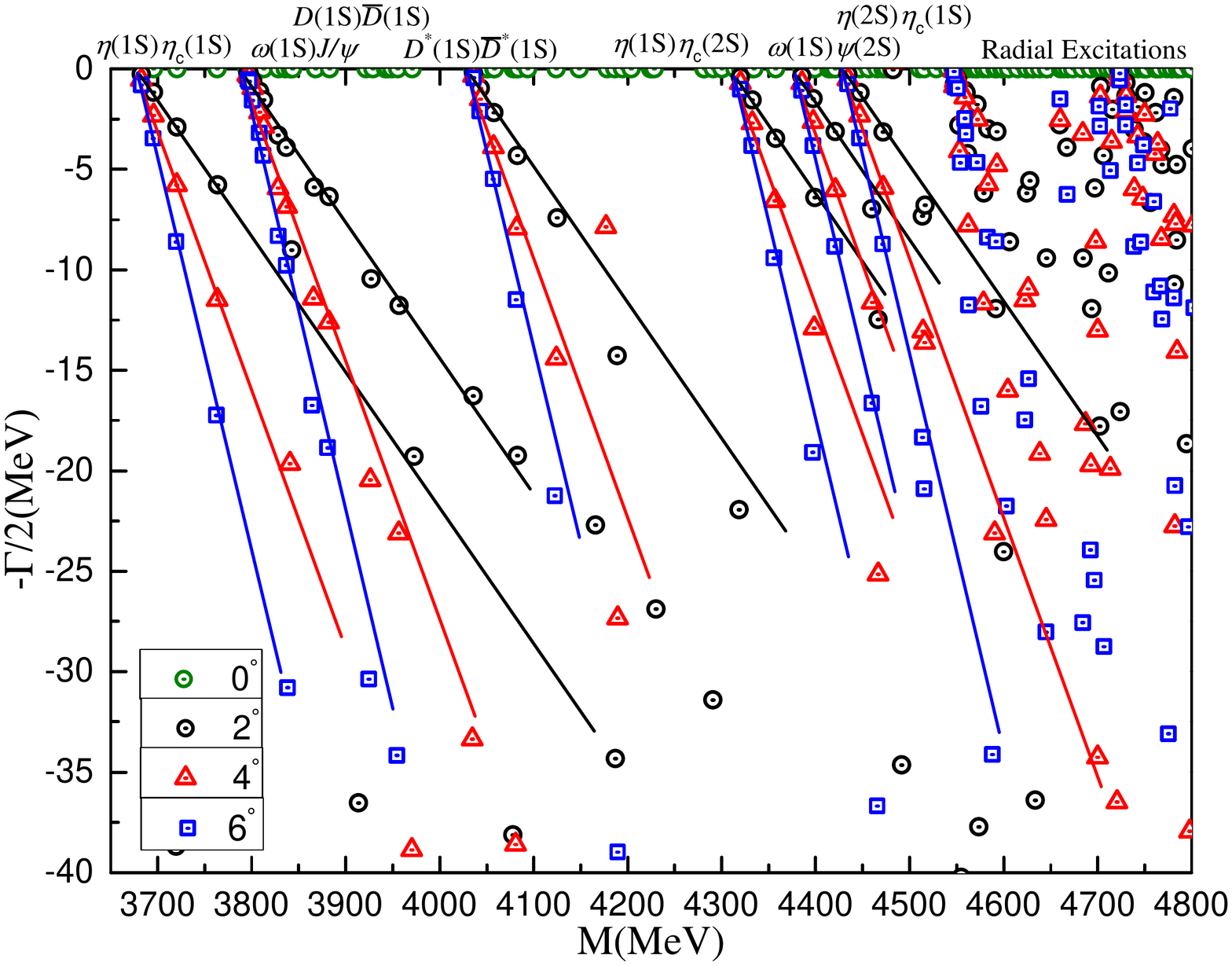} \\
\includegraphics[clip, trim={3.0cm 1.9cm 3.0cm 1.0cm}, width=0.45\textwidth]{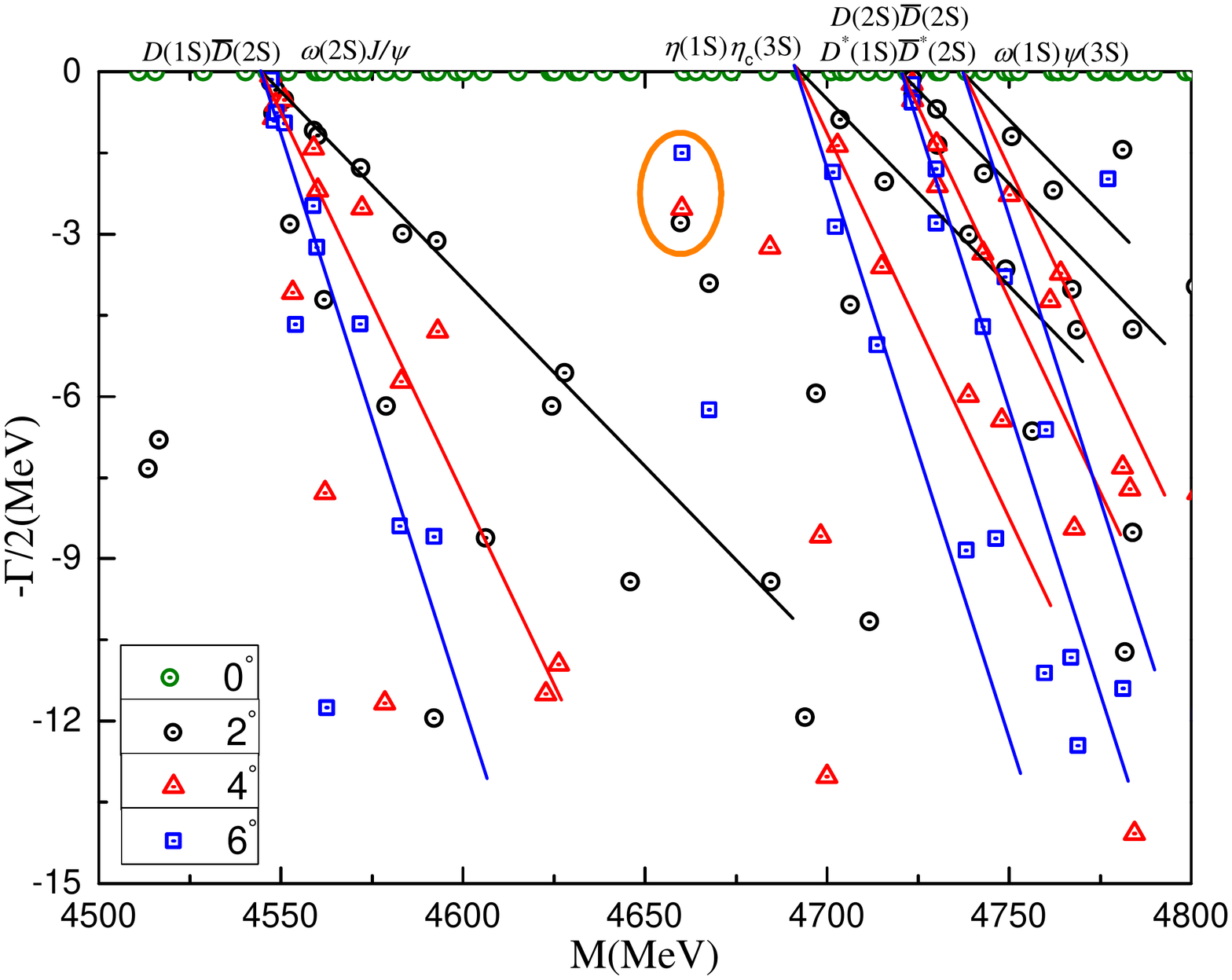}
\caption{\label{PP1} {\it Top panel:} The complete coupled-channels calculation of $c\bar{c}q\bar{q}$ tetraquark system with $I(J^P)=0(0^+)$ quantum numbers. {\it Bottom panel:} Enlarged top panel, with real values of energy ranging from $4.5\,\text{GeV}$ to $4.8\,\text{GeV}$. We use the complex-scaling method of the chiral quark model varying $\theta$ from $0^\circ$ to $6^\circ$.}
\end{figure}

\begin{table}[!t]
\caption{\label{GresultR1} Compositeness of the exotic resonances obtained in a complete coupled-channels analysis by the CSM in $0(0^+)$ state of $c\bar{c}q\bar{q}$ tetraquark. Particularly, the first column is the resonance pole labeled by $M+i\Gamma$, unit in MeV; the second one is the distance between any two quarks or quark-antiquark, unit in fm; and the component of resonance state ($S$: dimeson structure in color-singlet channel; $H$: dimeson structure in hidden-color channel; $Di$: diquark-antiquark configuration; $K$: K-type configuration).}
\begin{ruledtabular}
\begin{tabular}{lccc}
Resonance       & \multicolumn{3}{c}{Structure} \\[2ex]
$4660+i5.6$   & \multicolumn{3}{c}{$r_{c\bar{c}}:0.7$;\,\,\,\,\,\,$r_{cq}:1.3$;\,\,\,\,\,\,$r_{c\bar{q}}:1.2$} \\
& \multicolumn{3}{c}{$r_{q\bar{c}}:1.3$;\,\,\,\,\,\,$r_{\bar{q}\bar{c}}:1.2$;\,\,\,\,\,\,$r_{q\bar{q}}:1.5$} \\
& \multicolumn{3}{c}{$S$: 18.2\%;\, $H$: 28.2\%;\, $Di$: 7.3\%;\, $K$: 46.3\%}
\end{tabular}
\end{ruledtabular}
\end{table}

{\bf The $\bm{I(J^P)=0(0^+)}$ sector:} Four meson-meson channels, $\eta_c \eta$, $J/\psi \omega$, $D \bar{D}$ and $D^* \bar{D}^*$ in both color-singlet and hidden-color configurations, four diquark-antidiquark channels, along with $K_1$- and $K_3$-type configurations are individually studied in Table~\ref{GresultCC1}. A first conclusion can be drawn immediately, a bound state is unavailable in each single channel computation. In particular, the four color-singlet meson-meson states are located above their thresholds with masses at $3.68$ GeV, $3.79$ GeV, $3.79$ GeV and $4.03$ GeV, respectively. The states corresponding to the other three tetraquark configurations, hidden-color meson-meson, diquark-antidiquark, and K-type channels, lie in an energy region which ranges from $4.02$ GeV to $4.40$ GeV. If a coupled-channels study among each type of particular configuration is performed, the $c\bar{c}q\bar{q}$ tetraquark in singlet-color channel remains unbound, with the lowest coupled-mass, $3.68$ GeV, just at the $\eta_c \eta$ theoretical threshold. Meanwhile, the lowest mass is $4.11$ GeV, $4.07$ GeV, $4.01$ GeV and $4.06$ GeV for exotic structures: hidden-color, diquark-antidiquark and K-types, respectively. One can notice that there large mass shifts due to the coupled-channel effect; however, it is still not strong enough to have a stable bound state in a fully coupled-channels calculation. The lowest eigenenergy, $3.68$ GeV, listed in the last row of Table~\ref{GresultCC1}, continues indicating the scattering nature of the $c\bar{c}q\bar{q}$ tetraquark with quantum numbers $I(J^P)=0(0^+)$.

In a further step, introducing a rotated angle $\theta$ in the Hamiltonian, a complete coupled-channels calculation is performed in the complex-range approach. The output is shown in Fig.~\ref{PP1}. Therein, with a rotated angle ranging from $0^\circ$ to $6^\circ$, the scattering states of two mesons are well presented within an interval of $3.7-4.8$ GeV. Particularly, the four ground states, $\eta_c(1S) \eta(1S)$, $J/\psi(1S)\omega(1S)$, $D(1S)\bar{D}(1S)$ and $D^*(1S)\bar{D}^*(1S)$, are located within $3.7-4.2$ GeV. Moreover, their radial excitations are generally located in a region from $4.3$ GeV to $4.8$ GeV. From the top panel of Fig.~\ref{PP1} one can conclude that no stable pole is obtained within $3.7-4.5$ GeV. Hence, the ground and first radial excitation states are all of scattering nature. However, one resonance pole is found in the high energy region of the bottom panel of Fig.~\ref{PP1}. Therein, six radial excitation states, $D(1S)\bar{D}(2S)$,  $J/\psi(1S)\omega(2S)$, $\eta_c(3S) \eta(1S)$, $D(2S)\bar{D}(2S)$, $D^*(1S)\bar{D}^*(2S)$ and $\psi(3S)\omega(1S)$, are generally align along their threshold lines. Other than that, a stable pole is circled whose mass and width is $4660$ MeV and $5.6$ MeV, respectively.

Table~\ref{GresultR1} lists our results on the compositeness of the $(4660+i5.6)$ MeV resonance by calculating its size and component probabilities. In particular, a compact tetraquark structure is obtained with the distance between any two quarks or quark-antiquark less than $1.5$ fm. This nature is also confirmed by the fact that the dominant channel is a exotic one, \emph{i.e.} hidden-color $(28\%)$ and K-type $(46\%)$. Accordingly, the experimental signals $X(4630)$ in $I(J^P)=0(?^?)$ and $X(4700)$ in $I(J^P)=0(0^+)$ could be both identified with our compact $c\bar{c}q\bar{q}$ tetraquark resonance. Meanwhile, the golden two-meson decay channel to find the exotic resonance is suggested to be $J/\psi \omega$, which is the dominant component of color-singlet di-meson structures.


\begin{table}[!t]
\caption{\label{GresultCC2} Lowest-lying $c\bar{c}q\bar{q}$ tetraquark states with $I(J^P)=0(1^+)$ calculated within the real range formulation of the chiral quark model. The results are similarly organized as those in Table~\ref{GresultCC1}.
(unit: MeV).}
\begin{ruledtabular}
\begin{tabular}{lcccc}
~~Channel   & Index & $\chi_J^{\sigma_i}$;~$\chi_I^{f_j}$;~$\chi_k^c$ & $M$ & Mixed~~ \\
        &   &$[i; ~j; ~k]$ &  \\[2ex]
$(\eta_c \omega)^1 (3763)$          & 1  & [1;~1;~1]  & $3685$ & \\
$(J/\psi \eta)^1 (3645)$  & 2  & [2;~1;~1]   & $3786$ &  \\
$(J/\psi \omega)^1 (3879)$  & 3  & [3;~1;~1]   & $3793$ &  \\
$(D \bar{D}^*)^1 (3877)$          & 4  & [1;~1;~1]  & $3914$ & \\
$(D^* \bar{D}^*)^1 (4014)$  & 5  & [3;~1;~1]   & $4034$ & $3685$ \\[2ex]
$(\eta_c \omega)^8$  & 6  & [1;~1;~2]  & $4278$ & \\
$(J/\psi\ \eta)^8$    & 7  & [2;~1;~2]   & $4397$ &  \\
$(J/\psi \omega)^8$  & 8  & [3;~1;~2]   & $4260$ &  \\
$(D \bar{D}^*)^8$          & 9 & [1;~1;~2]  & $4242$ & \\
$(D^* \bar{D}^*)^8$      & 10  & [3;~1;~2]   & $4227$ & $4119$ \\[2ex]
$(cq)(\bar{q}\bar{c})^*$      & 11   & [4;~1;~3]  & $4243$ & \\
$(cq)(\bar{q}\bar{c})^*$      & 12   & [4;~1;~4]  & $4277$ & \\
$(cq)^*(\bar{q}\bar{c})^*$  & 13  & [6;~1;~3]   & $4226$ & \\
$(cq)^*(\bar{q}\bar{c})^*$  & 14  & [6;~1;~4]   & $4189$ & $4094$ \\[2ex]
$K_1$  & 15  & [7;~1;~5]   & $4360$ & \\
  & 16  & [8;~1;~5]   & $4315$ & \\
  & 17  & [9;~1;~5]   & $4281$ & \\
  & 18  & [7;~1;~6]   & $4127$ & \\
  & 19  & [8;~1;~6]   & $4128$ & \\
  & 20  & [9;~1;~6]   & $4022$ & $4021$ \\[2ex]
$K_3$  & 21  & [13;~1;~9]   & $4230$ & \\
  & 22  & [14;~1;~9]   & $4175$ & \\
  & 23  & [15;~1;~9]   & $4231$ & \\
  & 24  & [13;~1;~10]   & $4276$ & \\
  & 25  & [14;~1;~10]   & $4215$ & \\
  & 26  & [15;~1;~10]   & $4235$ & $4077$ \\[2ex]
\multicolumn{4}{c}{Complete coupled-channels:} & $3685$
\end{tabular}
\end{ruledtabular}
\end{table}

\begin{figure}[!t]
\includegraphics[clip, trim={3.0cm 1.9cm 3.0cm 1.0cm}, width=0.45\textwidth]{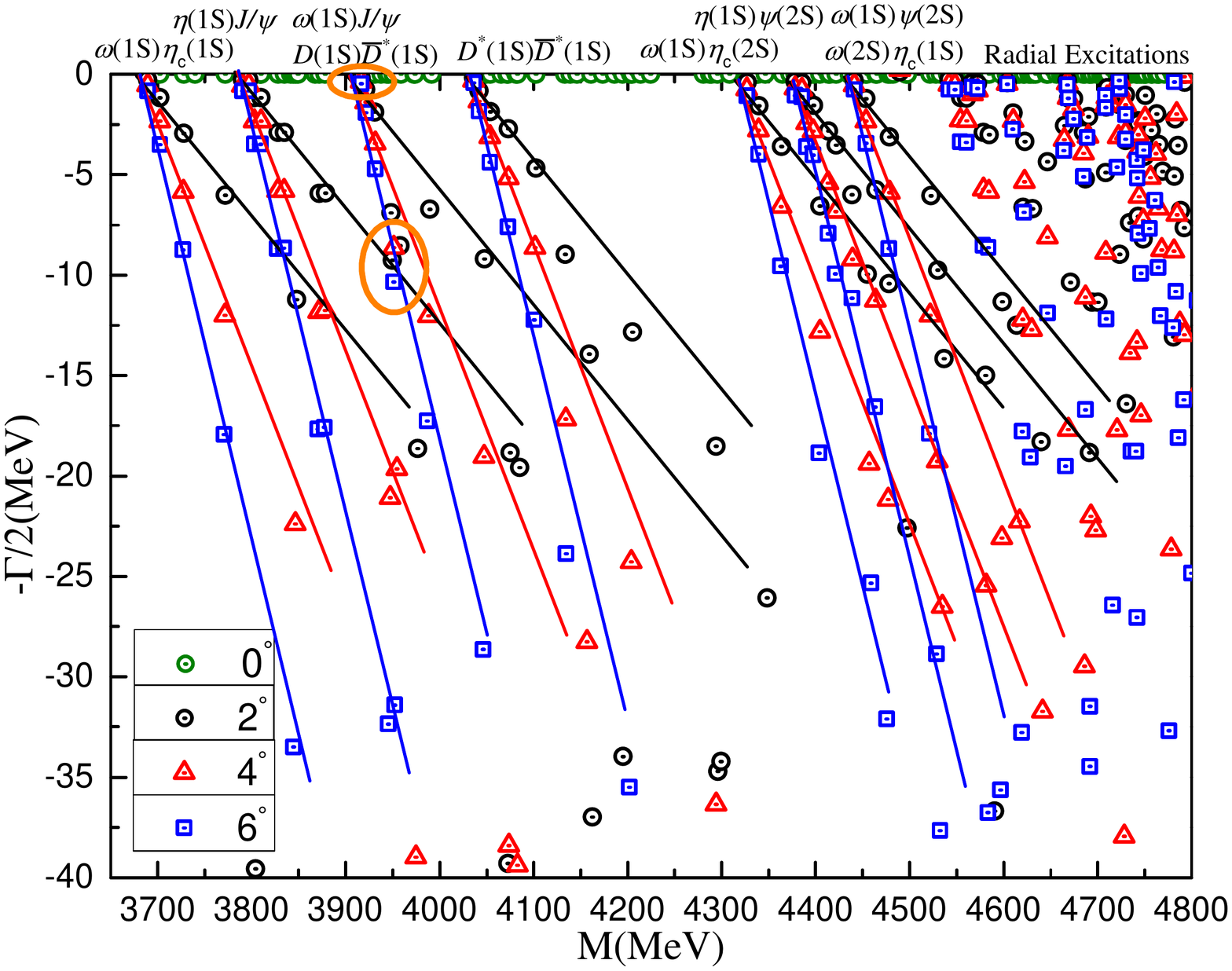} \\
\includegraphics[clip, trim={3.0cm 1.9cm 3.0cm 1.0cm}, width=0.45\textwidth]{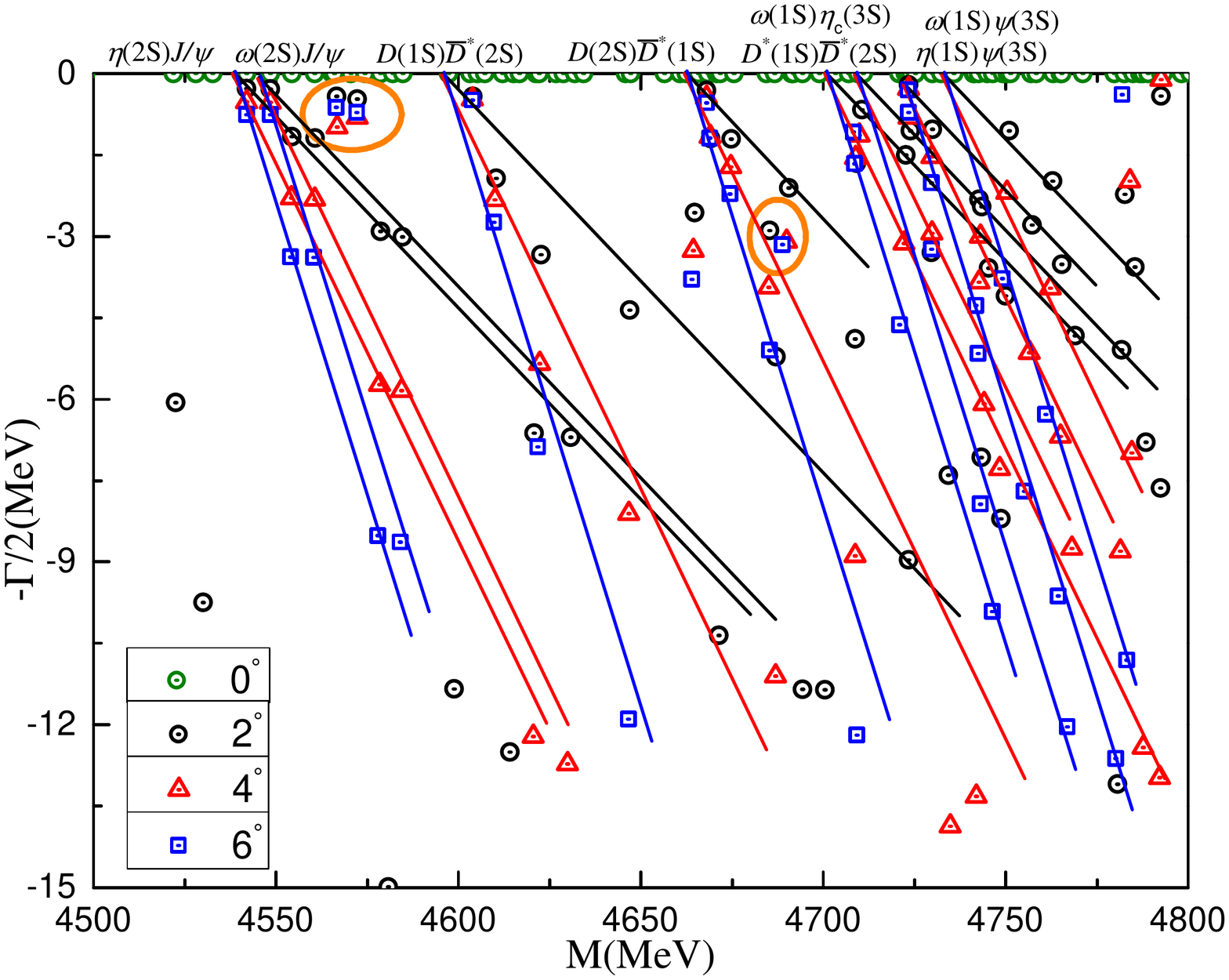}
\caption{\label{PP2} The complete coupled-channels calculation of $c\bar{c}q\bar{q}$ tetraquark system with $I(J^P)=0(1^+)$. {\it Bottom panel:} Enlarged top panel, with real values of energy ranging from $4.5\,\text{GeV}$ to $4.8\,\text{GeV}$..}
\end{figure}

\begin{table}[!t]
\caption{\label{GresultR2} Compositeness of the exotic resonances obtained in a complete coupled-channel analysis by the CSM in $0(1^+)$ state of $c\bar{c}q\bar{q}$ tetraquark. The results are similarly organized as those in Table~\ref{GresultR1}.}
\begin{ruledtabular}
\begin{tabular}{lccc}
Resonance       & \multicolumn{3}{c}{Structure} \\[2ex]
$3916+i0.8$   & \multicolumn{3}{c}{$r_{c\bar{c}}:2.3$;\,\,\,\,\,\,$r_{cq}:2.4$;\,\,\,\,\,\,$r_{c\bar{q}}:0.8$} \\
& \multicolumn{3}{c}{$r_{q\bar{c}}:0.9$;\,\,\,\,\,\,$r_{\bar{q}\bar{c}}:2.4$;\,\,\,\,\,\,$r_{q\bar{q}}:2.4$} \\
& \multicolumn{3}{c}{$S$: 44.7\%;\, $H$: 3.2\%;\, $Di$: 15.2\%;\, $K$: 36.9\%}\\[1.5ex]
$3950+i18.6$   & \multicolumn{3}{c}{$r_{c\bar{c}}:1.3$;\,\,\,\,\,\,$r_{cq}:2.1$;\,\,\,\,\,\,$r_{c\bar{q}}:1.8$} \\
& \multicolumn{3}{c}{$r_{q\bar{c}}:2.3$;\,\,\,\,\,\,$r_{\bar{q}\bar{c}}:1.9$;\,\,\,\,\,\,$r_{q\bar{q}}:2.4$} \\
& \multicolumn{3}{c}{$S$: 8.5\%;\, $H$: 30.3\%;\, $Di$: 14.2\%;\, $K$: 47.0\%}\\[1.5ex]
$4567+i2.0$   & \multicolumn{3}{c}{$r_{c\bar{c}}:0.8$;\,\,\,\,\,\,$r_{cq}:1.8$;\,\,\,\,\,\,$r_{c\bar{q}}:1.7$} \\
& \multicolumn{3}{c}{$r_{q\bar{c}}:1.8$;\,\,\,\,\,\,$r_{\bar{q}\bar{c}}:1.7$;\,\,\,\,\,\,$r_{q\bar{q}}:1.9$} \\
& \multicolumn{3}{c}{$S$: 19.4\%;\, $H$: 21.7\%;\, $Di$: 23.1\%;\, $K$: 35.8\%}\\[1.5ex]
$4572+i1.6$   & \multicolumn{3}{c}{$r_{c\bar{c}}:0.8$;\,\,\,\,\,\,$r_{cq}:1.7$;\,\,\,\,\,\,$r_{c\bar{q}}:1.6$} \\
& \multicolumn{3}{c}{$r_{q\bar{c}}:1.7$;\,\,\,\,\,\,$r_{\bar{q}\bar{c}}:1.7$;\,\,\,\,\,\,$r_{q\bar{q}}:1.9$} \\
& \multicolumn{3}{c}{$S$: 14.0\%;\, $H$: 31.4\%;\, $Di$: 6.3\%;\, $K$: 48.3\%}\\[1.5ex]
$4690+i6.2$   & \multicolumn{3}{c}{$r_{c\bar{c}}:1.4$;\,\,\,\,\,\,$r_{cq}:2.1$;\,\,\,\,\,\,$r_{c\bar{q}}:1.7$} \\
& \multicolumn{3}{c}{$r_{q\bar{c}}:1.8$;\,\,\,\,\,\,$r_{\bar{q}\bar{c}}:2.1$;\,\,\,\,\,\,$r_{q\bar{q}}:1.9$} \\
& \multicolumn{3}{c}{$S$: 30.8\%;\, $H$: 18.2\%;\, $Di$: 16.2\%;\, $K$: 34.8\%}
\end{tabular}
\end{ruledtabular}
\end{table}

{\bf The $\bm{I(J^P)=0(1^+)}$ sector:} There are 26 channels in this sector; particularly, we have 5 color-singlet meson-meson configurations $\eta_c \omega$, $J/\psi \eta$, $J/\psi \omega$, $D \bar{D}^*$ and $D^* \bar{D}^*$, another consequent 5 hidden-color channels, 4 diquark-antidiquark arrangements and 12 $K-$type configurations. All are listed in Table~\ref{GresultCC2}. First of all, no bound state is found in each single channel calculation. The lowest masses of meson-meson channels are locaed at around the theoretical threshold value. Masses of hidden-color, diquark-antidiquark and $K-$type channels are generally situated within the region $4.2-4.3$ GeV. Secondly, when coupled-channels calculation is performed in each particular configuration, the lowest mass on hidden-color, diquark-antidiquark and $K_3-$type is around $4.1$ GeV, which reduces the original masses by $\sim100$ MeV due to a strong coupling effect. However, coupling is extremely weak in color-singlet and $K_1-$type cases, their lowest coupled-masses are $3.69$ GeV and $4.02$ GeV, respectively. All of these results point out an unbound nature of the $c\bar{c}q\bar{q}$ tetraquark lowest-lying state with spin-parity $I(J^P)=0(1^+)$; moreover, this conclusion holds for a fully coupled-channels calculation, where the lowest calculated mass, $3.69$ GeV, is just the theoretical threshold value of $\eta_c \omega$.

Five narrow resonances are obtained in a further complex-range investigation, where all $c\bar{c}q\bar{q}$ channels are considered. The calculated complex energies are plotted in Fig.~\ref{PP2}. Particularly, within a mass region from $3.7$ GeV to $4.8$ GeV, the ground and radial excitation states of $\eta_c \omega$, $J/\psi \eta$, $J/\psi \omega$, $D \bar{D}^*$ and $D^* \bar{D}^*$ are clearly presented in the top panel. Therein, apart from most scattering dots, two stable resonance poles, which are independent of the rotated angle $\theta$, are circled. Their complex energies are $3916+i0.8$ MeV and $3950+i18.6$ MeV, respectively, and can be interpreted as loosely-bound molecular state and a compact tetraquark configuration when focusing on their quark-quark and quark-antiquark distances shown in Table~\ref{GresultR2}. Such a Table also shows that the coupling effect is strong in both resonances, the lower state is $45\%$ color-singlet and $37\%$ $K-$type, whereas the higher one is $30\%$ hidden-color and $47\%$ $K-$type. In our study, the $D\bar D^\ast$ component is dominant in the resonance with a mass of $3.91$ GeV. Therefore, after performing a mass shift of $-37$ MeV, according to the difference between theoretical and experimental threshold of $D \bar{D}^*$, the modified resonant mass is $3879$ MeV. This calculated resonance pole at $3879+i0.8$ MeV may be identified as the exotic state $X(3872)$, and this result is also confirmed in Refs.~\cite{Ding:2020dio, Xu:2021vsi, Wang:2019tlw}. Furthermore, because mass and width of the other resonance is $3.95$ GeV and $18.6$ MeV, the $X(3940)$ state can be identified as a $c\bar{c}q\bar{q}$ compact tetraquark with $I(J^P)=0(1^+)$.

The bottom panel of Fig.~\ref{PP2} focuses on the dense radial excitation region whose energy interval is $4.5-4.8$ GeV. Eight meson-meson scattering states, $J/\psi \eta(2S)$, $J/\psi \omega(2S)$, $D(1S) \bar{D}^*(2S)$, $D(2S) \bar{D}^*(1S)$, $D^*(1S) \bar{D}^*(2S)$, $\eta_c(3S) \omega(1S)$, $\psi(3S) \eta(1S)$ and $\psi(3S) \omega(1S)$, are clearly presented. Nevertheless, three narrow resonances are obtained, and their complex energies read as $4567+i2.0$ MeV, $4572+i1.6$ MeV and $4690+i6.2$ MeV, respectively. There is a common feature of these resonances, their size is of about $2.0$ fm, and their dominant component, which is greater than $70\%$, is of exotic type, combination of hidden-color, diquark-antidiquark and $K-$type channels. After comparing with experimental data, one may conclude that the $X(4685)$ can be explained as a $I(J^P)=0(1^+)$ $c\bar{c}q\bar{q}$ tetraquark state, and the golden observation channel would be $D\bar{D}^*$. Meanwhile, the other two almost degenerate resonance states at $4.57$ GeV could be identified as the $X(4630)$. The dominant meson-meson decay channel in this case would be $J/\psi \omega$.


\begin{table}[!t]
\caption{\label{GresultCC3} Lowest-lying $c\bar{c}q\bar{q}$ tetraquark states with $I(J^P)=0(2^+)$ calculated within the real range formulation of the chiral quark model. The results are similarly organized as those in Table~\ref{GresultCC1}.
(unit: MeV).}
\begin{ruledtabular}
\begin{tabular}{lcccc}
~~Channel   & Index & $\chi_J^{\sigma_i}$;~$\chi_I^{f_j}$;~$\chi_k^c$ & $M$ & Mixed~~ \\
        &   &$[i; ~j; ~k]$ &  \\[2ex]
$(J/\psi \omega)^1 (3879)$  & 1  & [1;~1;~1]   & $3793$ &  \\
$(D^* \bar{D}^*)^1 (4014)$  & 2  & [1;~1;~1]   & $4034$ & $3793$ \\[2ex]
$(J/\psi \omega)^8$  & 3  & [1;~1;~2]   & $4299$ &  \\
$(D^* \bar{D}^*)^8$  & 4  & [1;~1;~2]   & $4234$ & $4159$ \\[2ex]
$(cq)^*(\bar{q}\bar{c})^*$  & 5  & [1;~1;~3]   & $4241$ & \\
$(cq)^*(\bar{q}\bar{c})^*$  & 6  & [1;~1;~4]   & $4209$ & $4132$ \\[2ex]
$K_1$  & 7  & [1;~1;~5]   & $4303$ & \\
  & 8  & [1;~1;~6]   & $4130$ & $4130$ \\[2ex]
$K_3$  & 9  & [1;~1;~9]   & $4212$ & \\
  & 10  & [1;~1;~10]   & $4229$ & $4137$ \\[2ex]
\multicolumn{4}{c}{Complete coupled-channels:} & $3793$
\end{tabular}
\end{ruledtabular}
\end{table}

\begin{figure}[!t]
\includegraphics[width=0.45\textwidth, trim={2.3cm 2.0cm 3.0cm 1.0cm}]{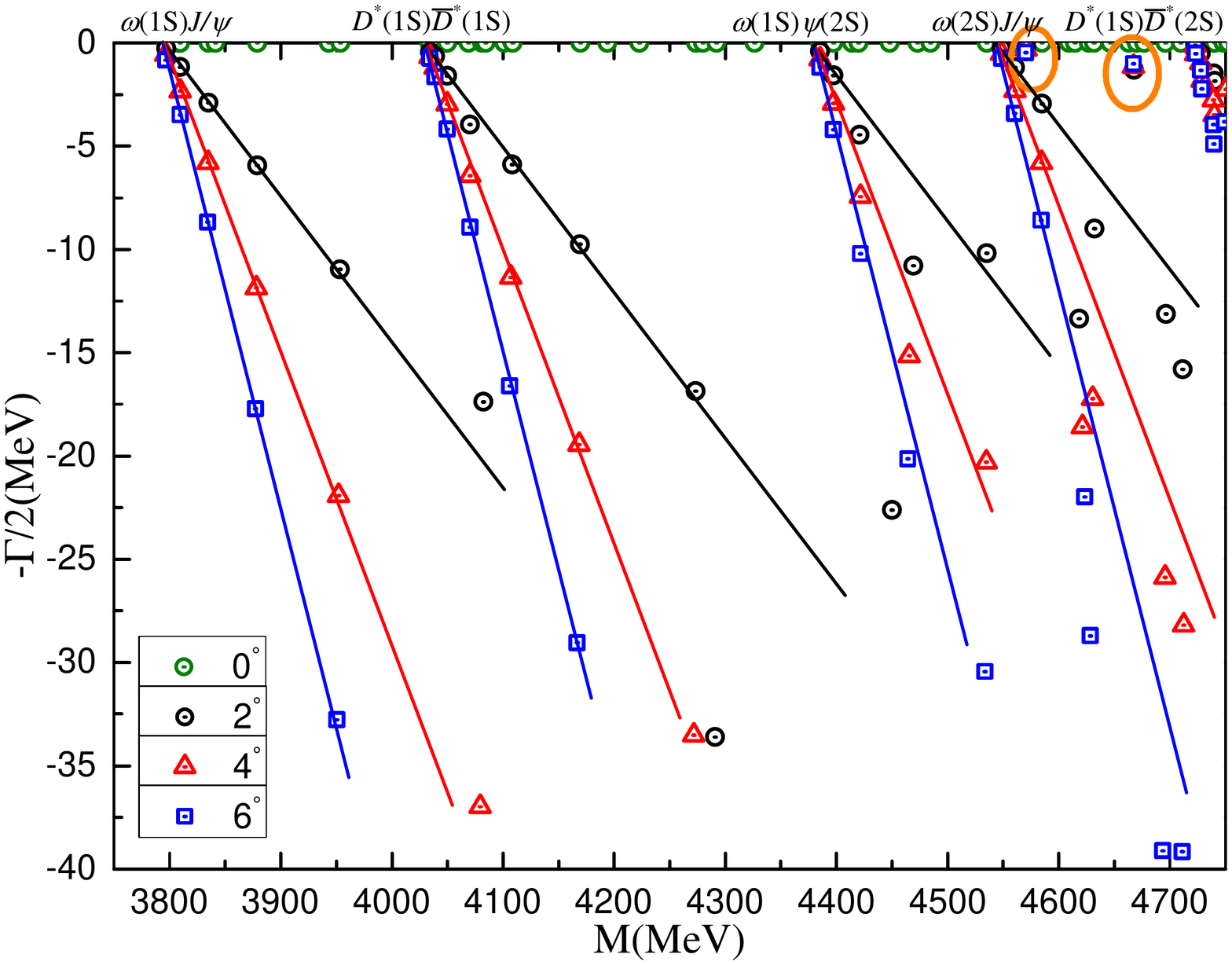}
\caption{\label{PP3} The complete coupled-channels calculation of the $c\bar{c}q\bar{q}$ tetraquark system with $I(J^P)=0(2^+)$ quantum numbers. We use the complex-scaling method of the chiral quark model varying $\theta$ from $0^\circ$ to $6^\circ$.}
\end{figure}

\begin{table}[!t]
\caption{\label{GresultR3} Compositeness of the exotic resonances obtained in a complete coupled-channel analysis by the CSM in $0(2^+)$ state of $c\bar{c}q\bar{q}$ tetraquark. The results are similarly organized as those in Table~\ref{GresultR1}.}
\begin{ruledtabular}
\begin{tabular}{lccc}
Resonance       & \multicolumn{3}{c}{Structure} \\[2ex]
$4570+i0.6$   & \multicolumn{3}{c}{$r_{c\bar{c}}:0.5$;\,\,\,\,\,\,$r_{cq}:1.5$;\,\,\,\,\,\,$r_{c\bar{q}}:1.5$} \\
& \multicolumn{3}{c}{$r_{q\bar{c}}:1.5$;\,\,\,\,\,\,$r_{\bar{q}\bar{c}}:1.5$;\,\,\,\,\,\,$r_{q\bar{q}}:1.6$} \\
& \multicolumn{3}{c}{$S$: 0.2\%;\, $H$: 2.8\%;\, $Di$: 24.9\%;\, $K$: 72.1\%}\\[1.5ex]
$4667+i2.2$   & \multicolumn{3}{c}{$r_{c\bar{c}}:1.2$;\,\,\,\,\,\,$r_{cq}:2.0$;\,\,\,\,\,\,$r_{c\bar{q}}:1.8$} \\
& \multicolumn{3}{c}{$r_{q\bar{c}}:1.9$;\,\,\,\,\,\,$r_{\bar{q}\bar{c}}:1.8$;\,\,\,\,\,\,$r_{q\bar{q}}:2.4$} \\
& \multicolumn{3}{c}{$S$: 0.4\%;\, $H$: 12.9\%;\, $Di$: 38.2\%;\, $K$: 48.5\%}
\end{tabular}
\end{ruledtabular}
\end{table}

{\bf The $\bm{I(J^P)=0(2^+)}$ state:} Table~\ref{GresultCC3} shows that two meson-meson structures, $J/\psi\omega$ and $D^* \bar{D}^*$ (in both color-singlet and hidden-color configurations), two $(cq)^*(\bar{q}\bar{c})^*$ diquark-antidiquark arrangements and four K-type configurations contribute to the $I(J^P)=0(2^+)$ state. First of all, a bound state is still unavailable in each single channel calculation. The lowest masses of the two meson-meson configurations in color-singlet channel are $3.79$ GeV and $4.03$ GeV, respectively. Besides, the other eight exotic channels masses are located at around $4.2$ GeV. Secondly, coupled-channel computations are performed in each configuration. However, the lowest energy remains at $J/\psi \omega$ theoretical threshold value of $3.79$ GeV (this result holds even in a fully coupled case), and masses of the other four configurations are around $4.13$ GeV.

By employing the complex-scaling method in a complete coupled-channel calculation, two extremely narrow resonances are obtained. They are circled in Fig.~\ref{PP3}, where the calculated complex energies are plotted. Therein, one can observe that, within an energy interval of $3.7-4.8$ GeV, the $J/\psi\omega$ and $D^* \bar{D}^*$ in both ground and radial excitation states present a scattering nature. Moreover, the two mentioned resonances appear, their complex energy read as $4570+i0.6$ MeV and $4667+i2.2$ MeV. Table~\ref{GresultR3} lists the structure and component probabilities of these two resonances. Firstly, both resonances appear to be exotic because more than $85\%$ is contributed by diquark-antidiquark and K-type configurations. Meanwhile, their sizes are about $1.5$ fm and $2.0$ fm, respectively. Resonant masses are comparable with that of the $X(4630)$; therefore, it is tentative to assign interpret it as a resonance with $c\bar{c}q\bar{q}$ tetraquark component in $0(2^+)$ channel.


\begin{table}[!t]
\caption{\label{GresultCC4} Lowest-lying $c\bar{c}q\bar{q}$ tetraquark states with $I(J^P)=1(0^+)$ calculated within the real range formulation of the chiral quark model. The results are similarly organized as those in Table~\ref{GresultCC1}.
(unit: MeV).}
\begin{ruledtabular}
\begin{tabular}{lcccc}
~~Channel   & Index & $\chi_J^{\sigma_i}$;~$\chi_I^{f_j}$;~$\chi_k^c$ & $M$ & Mixed~~ \\
        &   &$[i; ~j; ~k]$ &  \\[2ex]
$(\eta_c \pi)^1 (3121)$          & 1  & [1;~1;~1]  & $3138$ & \\
$(J/\psi \rho)^1 (3867)$  & 2  & [2;~1;~1]   & $3869$ &  \\
$(D \bar{D})^1 (3740)$          & 3  & [1;~1;~1]  & $3794$ & \\
$(D^* \bar{D}^*)^1 (4014)$  & 4  & [2;~1;~1]   & $4034$ & $3138$ \\[2ex]
$(\eta_c \pi)^8$          & 5  & [1;~1;~2]  & $4276$ & \\
$(J/\psi \rho)^8$  & 6  & [2;~1;~2]   & $4297$ &  \\
$(D \bar{D})^8$          & 7  & [1;~1;~2]  & $4237$ & \\
$(D^* \bar{D}^*)^8$  & 8  & [2;~1;~2]   & $4154$ & $3984$ \\[2ex]
$(cq)(\bar{q}\bar{c})$      & 9   & [3;~1;~3]  & $4207$ & \\
$(cq)(\bar{q}\bar{c})$      & 10   & [3;~1;~4]  & $4211$ & \\
$(cq)^*(\bar{q}\bar{c})^*$  & 11  & [4;~1;~3]   & $4235$ & \\
$(cq)^*(\bar{q}\bar{c})^*$  & 12  & [4;~1;~4]   & $4090$ & $3932$ \\[2ex]
$K_1$  & 13  & [5;~1;~5]   & $4290$ & \\
  & 14  & [6;~1;~5]   & $4278$ & \\
  & 15  & [5;~1;~6]   & $4161$ & \\
  & 16  & [6;~1;~6]   & $3884$ & $3878$ \\[2ex]
$K_3$  & 17  & [9;~1;~9]   & $4086$ & \\
  & 18  & [10;~1;~9]   & $4243$ & \\
  & 19  & [9;~1;~10]   & $4197$ & \\
  & 20  & [10;~1;~10]   & $4206$ & $3924$ \\[2ex]
\multicolumn{4}{c}{Complete coupled-channels:} & $3138$
\end{tabular}
\end{ruledtabular}
\end{table}

\begin{figure}[!t]
\includegraphics[clip, trim={3.0cm 1.9cm 3.0cm 1.0cm}, width=0.45\textwidth]{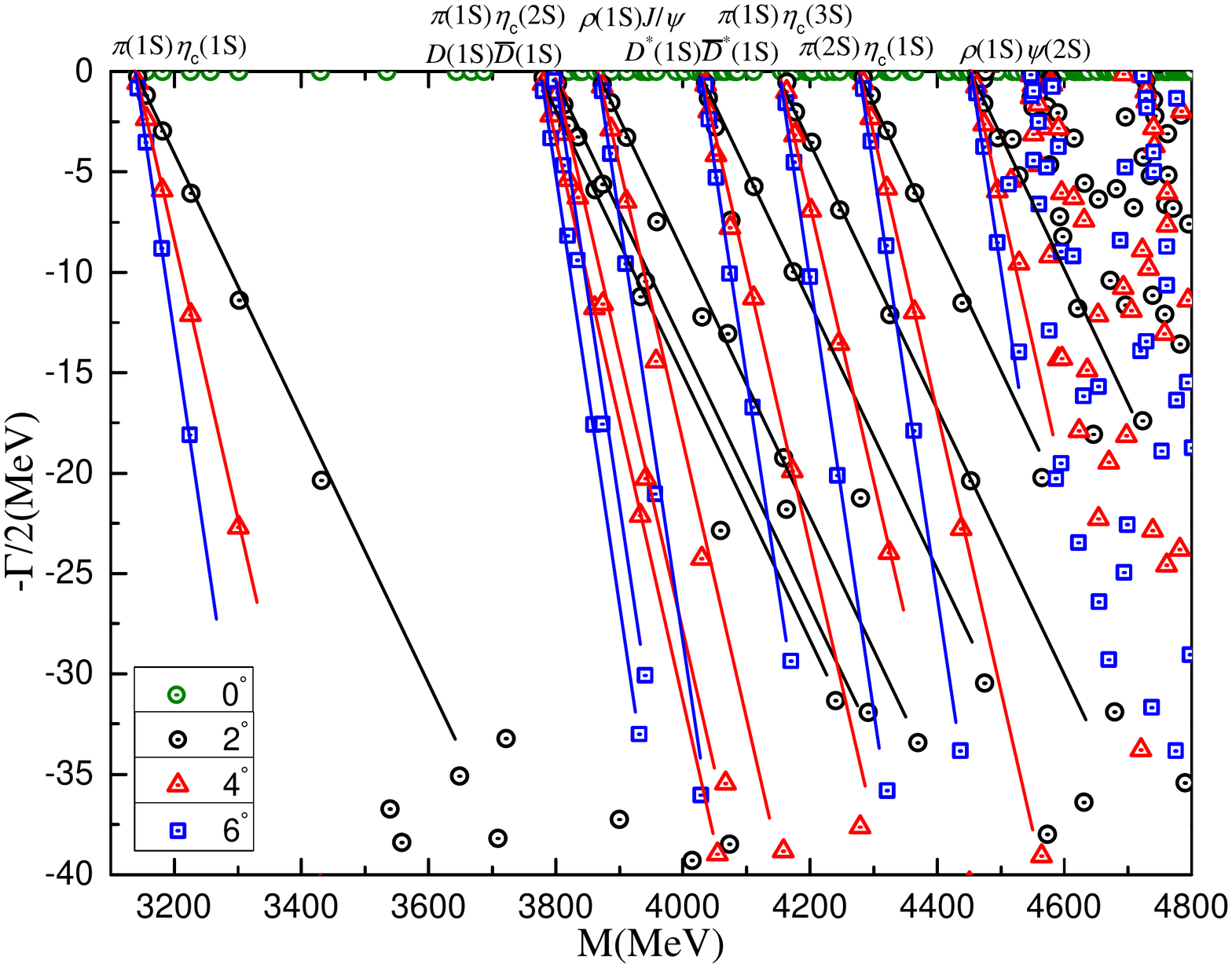} \\
\includegraphics[clip, trim={3.0cm 1.9cm 3.0cm 1.0cm}, width=0.45\textwidth]{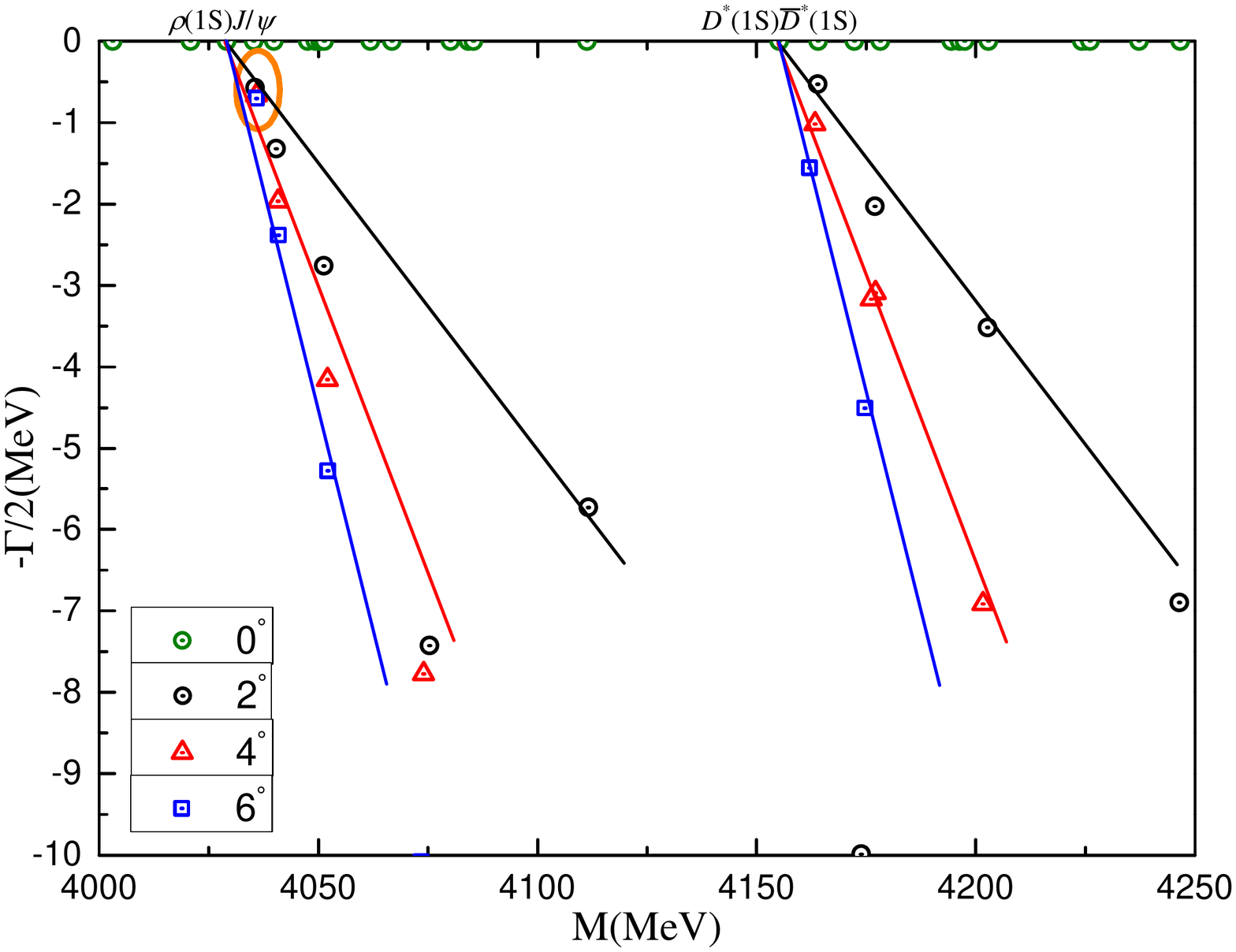}\\
\includegraphics[clip, trim={3.0cm 1.9cm 3.0cm 1.0cm}, width=0.45\textwidth]{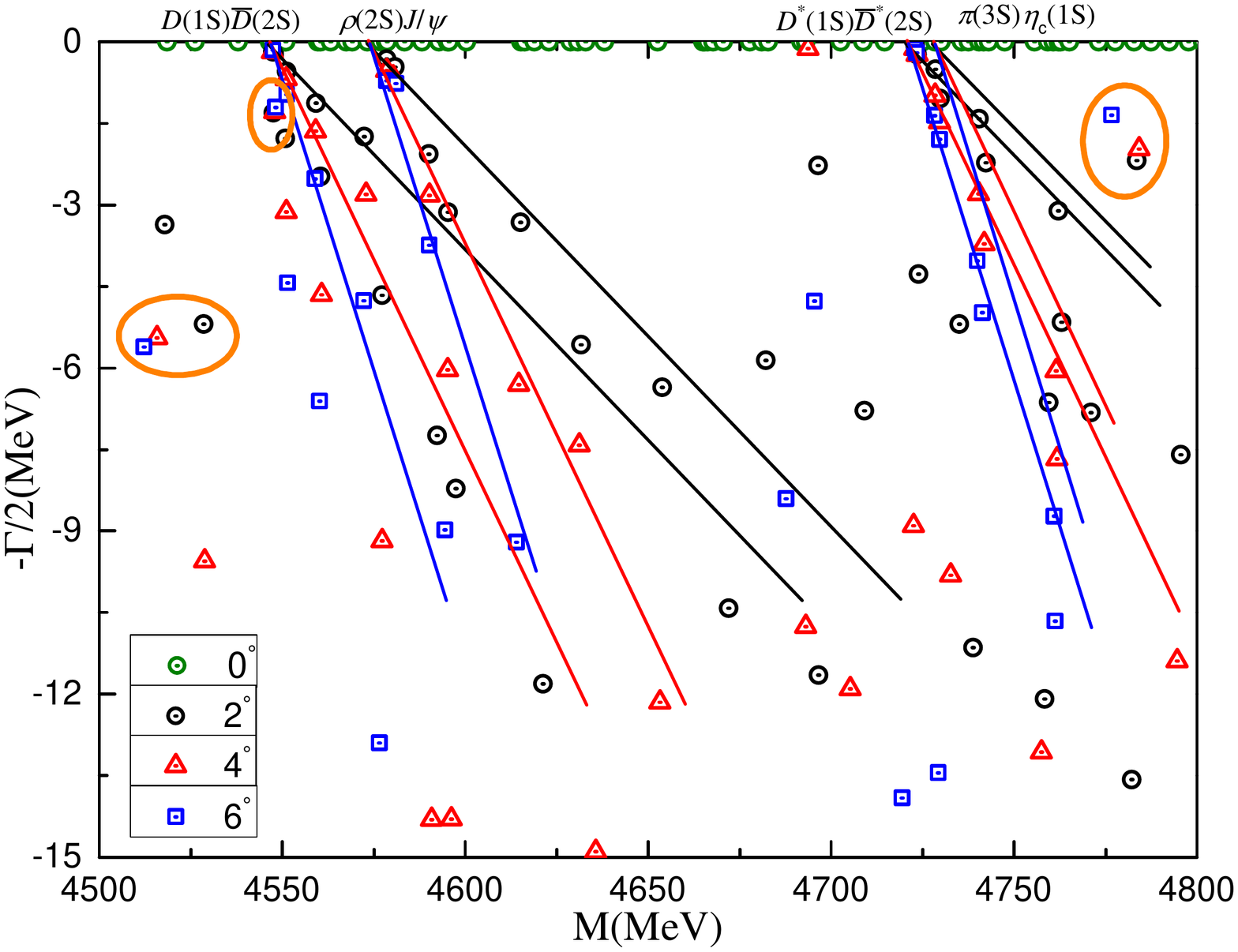}
\caption{\label{PP4} The complete coupled-channels calculation of the $c\bar{c}q\bar{q}$ tetraquark system with $I(J^P)=1(0^+)$ quantum numbers. Particularly, middle and bottom panels are enlarged parts of dense energy region from $4.0\,\text{GeV}$ to $4.25\,\text{GeV}$, and $4.5\,\text{GeV}$ to $4.8\,\text{GeV}$, respectively. We use the complex-scaling method of the chiral quark model varying $\theta$ from $0^\circ$ to $6^\circ$.}
\end{figure}

\begin{table}[!t]
\caption{\label{GresultR4} Compositeness of the exotic resonances obtained in a complete coupled-channel analysis by the CSM in $1(0^+)$ state of $c\bar{c}q\bar{q}$ tetraquark. The results are similarly organized as those in Table~\ref{GresultR1}.}
\begin{ruledtabular}
\begin{tabular}{lccc}
Resonance       & \multicolumn{3}{c}{Structure} \\[2ex]
$4036+i1.4$   & \multicolumn{3}{c}{$r_{c\bar{c}}:2.3$;\,\,\,\,\,\,$r_{cq}:2.4$;\,\,\,\,\,\,$r_{c\bar{q}}:0.9$} \\
   & \multicolumn{3}{c}{$r_{q\bar{c}}:0.9$;\,\,\,\,\,\,$r_{\bar{q}\bar{c}}:2.4$;\,\,\,\,\,\,$r_{q\bar{q}}:2.4$} \\
   & \multicolumn{3}{c}{$S$: 39.8\%;\, $H$: 3.7\%;\, $Di$: 27.9\%;\, $K$: 28.6\%}\\[1.5ex]
$4515+i10.8$   & \multicolumn{3}{c}{$r_{c\bar{c}}:0.9$;\,\,\,\,\,\,$r_{cq}:2.1$;\,\,\,\,\,\,$r_{c\bar{q}}:2.1$} \\
   & \multicolumn{3}{c}{$r_{q\bar{c}}:2.1$;\,\,\,\,\,\,$r_{\bar{q}\bar{c}}:2.1$;\,\,\,\,\,\,$r_{q\bar{q}}:1.0$} \\
   & \multicolumn{3}{c}{$S$: 44.0\%;\, $H$: 1.9\%;\, $Di$: 29.8\%;\, $K$: 24.3\%}\\[1.5ex]
$4548+i2.6$   & \multicolumn{3}{c}{$r_{c\bar{c}}:1.2$;\,\,\,\,\,\,$r_{cq}:1.8$;\,\,\,\,\,\,$r_{c\bar{q}}:1.4$} \\
   & \multicolumn{3}{c}{$r_{q\bar{c}}:1.3$;\,\,\,\,\,\,$r_{\bar{q}\bar{c}}:1.8$;\,\,\,\,\,\,$r_{q\bar{q}}:1.5$} \\
   & \multicolumn{3}{c}{$S$: 43.0\%;\, $H$: 4.1\%;\, $Di$: 30.0\%;\, $K$: 22.9\%}\\[1.5ex]
$4784+i4.0$   & \multicolumn{3}{c}{$r_{c\bar{c}}:0.8$;\,\,\,\,\,\,$r_{cq}:1.7$;\,\,\,\,\,\,$r_{c\bar{q}}:1.7$} \\
   & \multicolumn{3}{c}{$r_{q\bar{c}}:1.7$;\,\,\,\,\,\,$r_{\bar{q}\bar{c}}:1.7$;\,\,\,\,\,\,$r_{q\bar{q}}:2.0$} \\
   & \multicolumn{3}{c}{$S$: 27.3\%;\, $H$: 18.2\%;\, $Di$: 24.3\%;\, $K$: 30.2\%}
\end{tabular}
\end{ruledtabular}
\end{table}

{\bf The $\bm{I(J^P)=1(0^+)}$ sector:} Table~\ref{GresultCC4} shows our results for the isovector $c\bar{c}q\bar{q}$ tetraquark with spin-parity $J^P=0^+$. As in the case of the $I(J^P)=0(0^+)$ state, 20 channels are under investigation and no one shows a bound state. The theoretical mass of the lowest channel $\eta_c \pi$ in color-singlet is $3.13$ GeV, the other three ones, $J/\psi \rho$, $D\bar{D}$ and $D^* \bar{D}^*$, are located at $3.86-4.03$ GeV. Hidden-color, diquark-antiquark and K-type channels are generally in a mass region from $4.1$ GeV to $4.3$ GeV, except the $K_1$-channel whose mass is $3.88$ GeV. The coupling effect is strong in hidden-color, diquark-antidiquark and $K_3$-type configurations, having mass shifts of about $-160$ MeV in each of these structures' coupled-channels calculation. However, the lowest mass in color-singlet and $K_1$-type is almost unchanged in a coupled-channels study. Meanwhile, a scattering state of $\eta_c \pi$ at $3.13$ GeV remains unchanged in a complete coupled-channels calculation.

With the complex-scaling method employed in a fully coupled-channels computation, the calculated complex energies within $3.1-4.8$ GeV are presented in Fig.~\ref{PP4}. Firstly, $\eta_c \pi$, $J/\psi \rho$, $D\bar{D}$ and $D^* \bar{D}^*$ in ground and radial excitation states are of scattering nature. Secondly, no stable pole is found within $3.2-4.0$ GeV. However, a zoom on the energy region from $4.0$ GeV to $4.25$ GeV is plotted in the middle panel of Fig.~\ref{PP4}. Therein, apart from most of the scattering dots of $J/\psi \rho(1S)$ and $D^*(1S) \bar{D}^*(1S)$ states, one stable pole is circled, and its complex energy reads as $4036+i1.4$ MeV. A $D\bar{D}$ molecular structure is identified for this resonance through studying its compositeness in Table~\ref{GresultR4}. In particular, the distances $r_{c\bar{q}}$ and $r_{q\bar{c}}$ are both $0.9$ fm, while the others are $\sim 2.4$ fm. The dominant meson-meson structure in color-singlet channels are $J/\psi \rho (12.4\%)$ and $D\bar{D} (24.8\%)$. Besides, the diquark-antidiquark and K-type components are both around $28\%$. There are four isovector states, $X(4020)$, $X(4050)$, $X(4055)$ and $X(4100)$, the calculated resonance at $4036+i1.4$ MeV is compatible with them, and their spin-parity is suggest to be $0^+$.

In a further step, analyzing the highest energy range $4.5-4.8$ GeV, three more resonances are obtained and shown in Fig.~\ref{PP4}. That is to say, apart from the scattering states of $D(1S)\bar{D}(2S)$, $J/\psi \rho(2S)$, $D^*(1S)\bar{D}^*(2S)$ and $\eta_c(1S) \pi(3S)$ which are clearly visible, three stable poles are shown and their complex energies read $4515+i10.8$ MeV, $4548+i2.6$ MeV and $4784+i4.0$ MeV, respectively. Moreover, we deliver sizes and probability components of these states in Table~\ref{GresultR4}. In particular, their sizes are all less than $2.1$ fm. The resonance at $4.51$ GeV looks like a $J/\psi \rho$ molecule, since $r_{c\bar{c}}$ and $r_{q\bar{q}}$ are $\sim$ 0.9 fm, and the other quark distances are $2.1$ fm. Besides, the $44\%$ color-singlet channel of di-meson configuration consists mainly of $25\%$ $J/\psi \rho$ and $14\%$ $D\bar{D}$. There is also considerable diquark-antidiquark and K-type components, which are around $25\%$. The resonance at $4.54$ GeV is more compact with size less than $1.8$ fm. Meanwhile, the coupling between color-singlet, diquark-antidiquark and K-type is strong, with proportions $43\%$, $30\%$ and $23\%$, respectively. In particular, $12\%$ $J/\psi \rho$ and $27\%$ $D\bar{D}$ contributes to the meson-meson configuration of color-singlet channel. The resonance at $4.78$ GeV evoke a hadro-quarkonium configuration with a distance of $0.8$ fm between the $c\bar{c}$ pair, and much larger sizes for the others, $\sim 1.7$ fm. Furthermore, the wave function content of the four configurations listed in Table~\ref{GresultR4} are similar. Accordingly, the three higher excited resonances, which are located at around $4.5$ GeV and $4.7$ GeV, are expected to be confirmed experimentally.

\begin{table}[!t]
\caption{\label{GresultCC5} Lowest-lying $c\bar{c}q\bar{q}$ tetraquark states with $I(J^P)=1(1^+)$ calculated within the real range formulation of the chiral quark model. The results are similarly organized as those in Table~\ref{GresultCC1}.
(unit: MeV).}
\begin{ruledtabular}
\begin{tabular}{lcccc}
~~Channel   & Index & $\chi_J^{\sigma_i}$;~$\chi_I^{f_j}$;~$\chi_k^c$ & $M$ & Mixed~~ \\
        &   &$[i; ~j; ~k]$ &  \\[2ex]
$(\eta_c \rho)^1 (3751)$          & 1  & [1;~1;~1]  & $3761$ & \\
$(J/\psi \pi)^1 (3237)$  & 2  & [2;~1;~1]   & $3246$ &  \\
$(J/\psi \rho)^1 (3867)$  & 3  & [3;~1;~1]   & $3869$ &  \\
$(D \bar{D}^*)^1 (3877)$          & 4  & [1;~1;~1]  & $3914$ & \\
$(D^* \bar{D}^*)^1 (4014)$  & 5  & [3;~1;~1]   & $4034$ & $3246$ \\[2ex]
$(\eta_c \rho)^8$  & 6  & [1;~1;~2]  & $4332$ & \\
$(J/\psi \pi)^8$    & 7  & [2;~1;~2]   & $4278$ &  \\
$(J/\psi \rho)^8$  & 8  & [3;~1;~2]   & $4316$ &  \\
$(D \bar{D}^*)^8$          & 9 & [1;~1;~2]  & $4242$ & \\
$(D^* \bar{D}^*)^8$      & 10  & [3;~1;~2]   & $4195$ & $4067$ \\[2ex]
$(cq)(\bar{q}\bar{c})^*$      & 11   & [4;~1;~3]  & $4243$ & \\
$(cq)^*(\bar{q}\bar{c})$      & 12   & [4;~1;~4]  & $4236$ & \\
$(cq)^*(\bar{q}\bar{c})^*$  & 13  & [6;~1;~3]   & $4226$ & \\
$(cq)^*(\bar{q}\bar{c})^*$  & 14  & [6;~1;~4]   & $4148$ & $4045$ \\[2ex]
$K_1$  & 15  & [7;~1;~5]   & $4289$ & \\
  & 16  & [8;~1;~5]   & $4300$ & \\
  & 17  & [9;~1;~5]   & $4327$ & \\
  & 18  & [7;~1;~6]   & $4062$ & \\
  & 19  & [8;~1;~6]   & $4117$ & \\
  & 20  & [9;~1;~6]   & $4054$ & $3989$ \\[2ex]
$K_3$  & 21  & [13;~1;~9]   & $4160$ & \\
  & 22  & [14;~1;~9]   & $4202$ & \\
  & 23  & [15;~1;~9]   & $4231$ & \\
  & 24  & [13;~1;~10]   & $4197$ & \\
  & 25  & [14;~1;~10]   & $4247$ & \\
  & 26  & [15;~1;~10]   & $4235$ & $3997$ \\[2ex]
\multicolumn{4}{c}{Complete coupled-channels:} & $3246$
\end{tabular}
\end{ruledtabular}
\end{table}

\begin{figure}[!t]
\includegraphics[clip, trim={3.0cm 1.9cm 3.0cm 1.0cm}, width=0.45\textwidth]{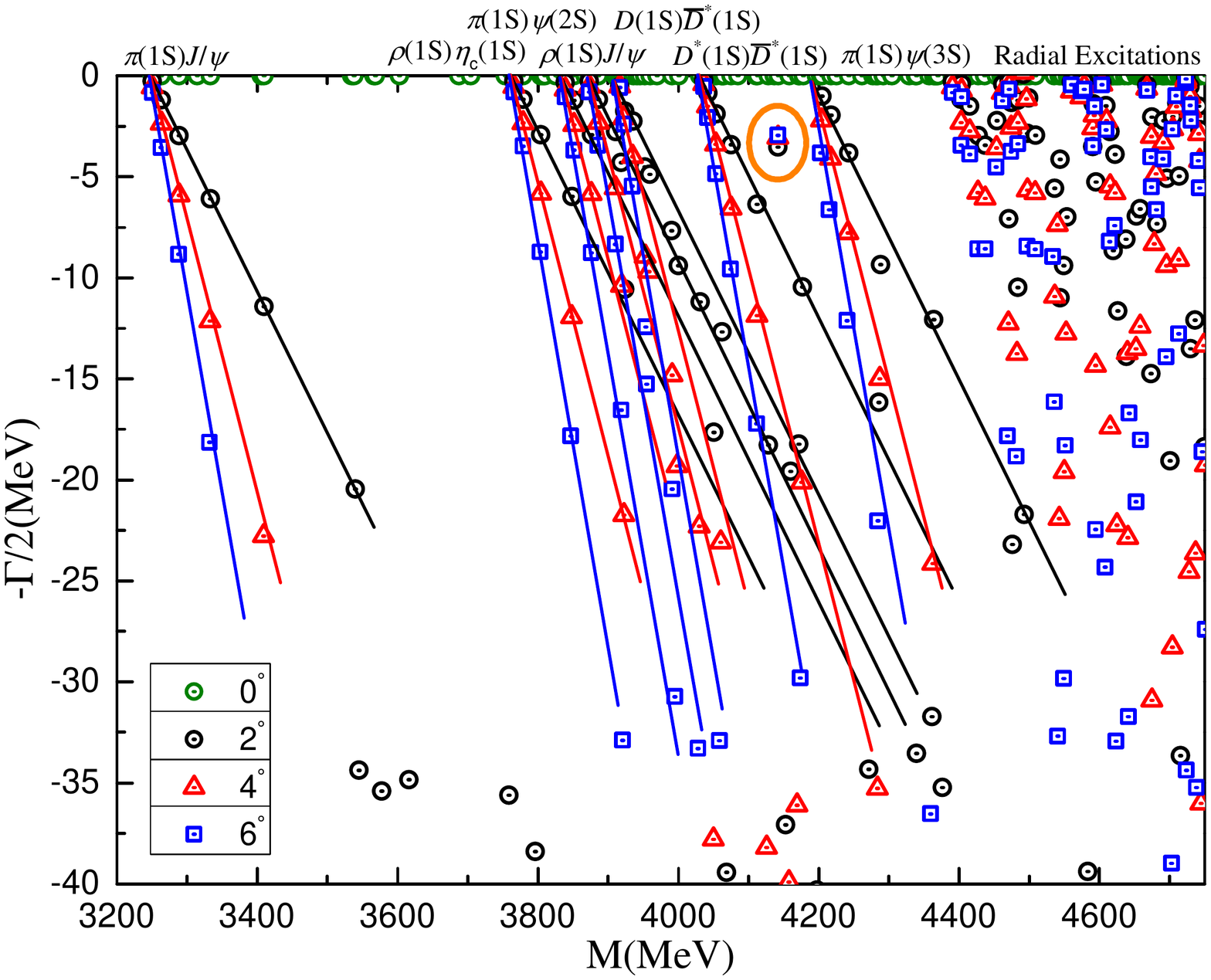} \\[1ex]
\includegraphics[clip, trim={3.0cm 1.9cm 3.0cm 1.0cm}, width=0.45\textwidth]{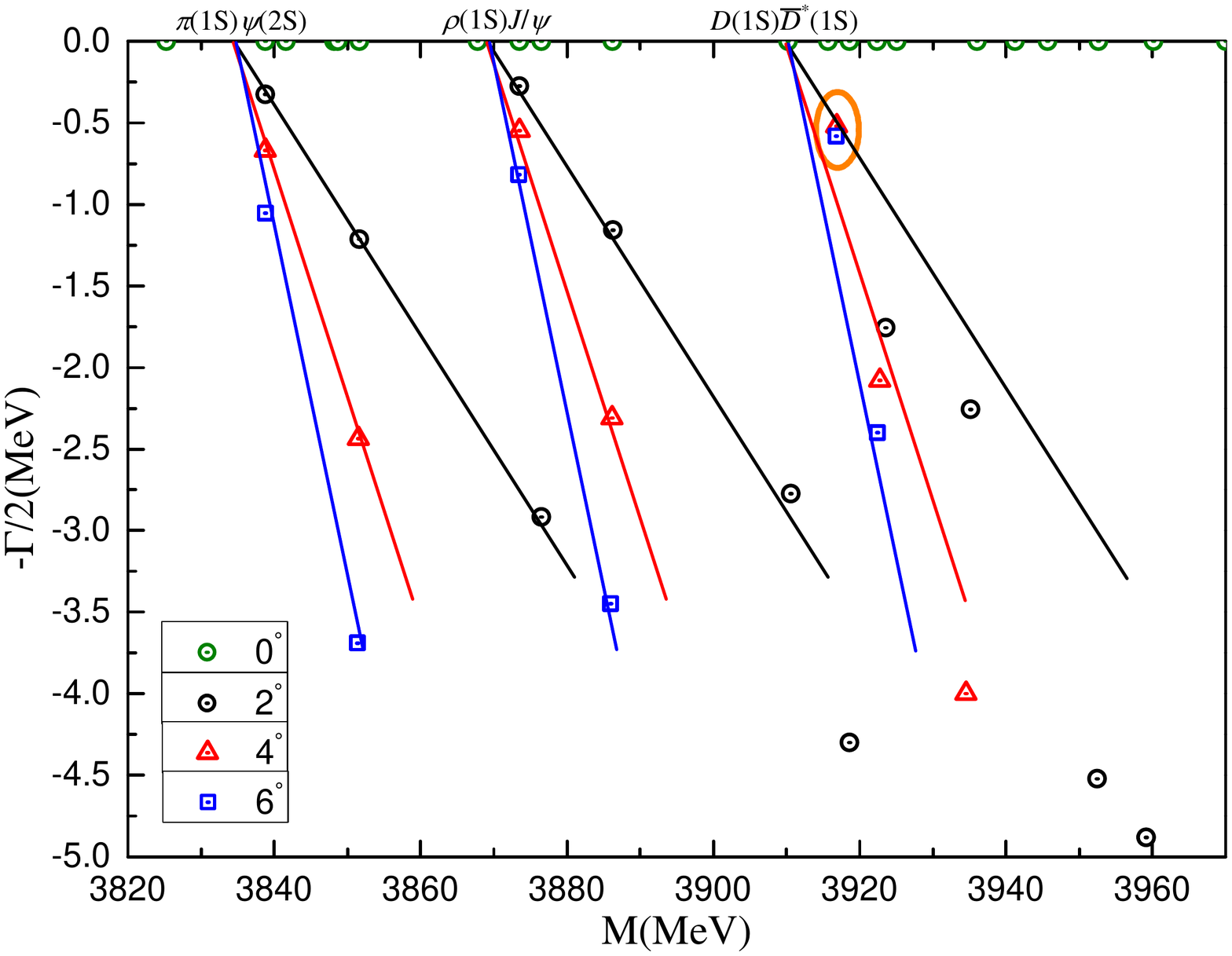}\\[1ex]
\includegraphics[clip, trim={3.0cm 1.9cm 3.0cm 1.0cm}, width=0.45\textwidth]{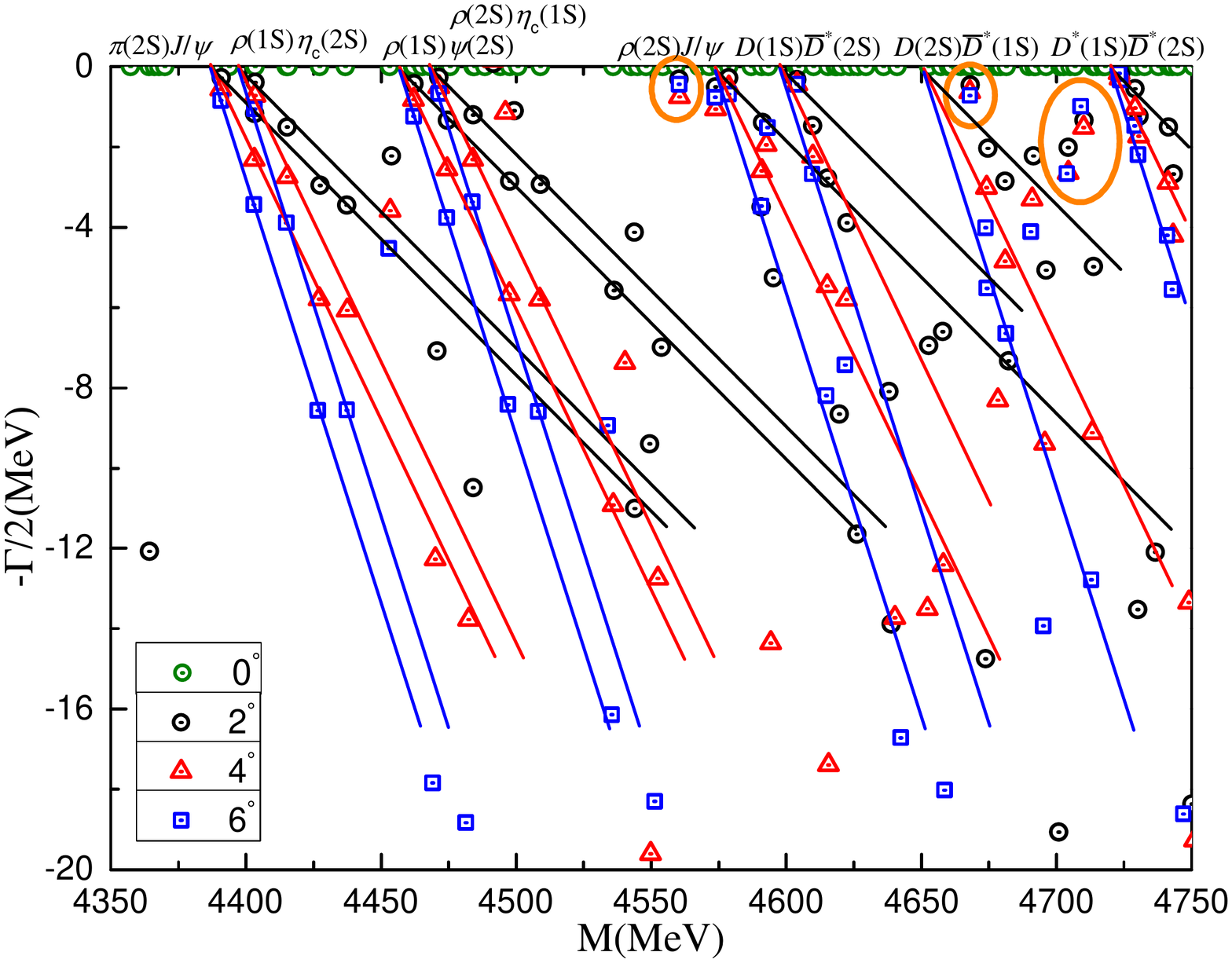}
\caption{\label{PP5} {\it Top panel:} The complete coupled-channels calculation of the $c\bar{c}q\bar{q}$ tetraquark system with $I(J^P)=1(1^+)$ quantum numbers. Particularly, middle and bottom panels are enlarged parts of dense energy region from $3.82\,\text{GeV}$ to $3.97\,\text{GeV}$, and $4.35\,\text{GeV}$ to $4.75\,\text{GeV}$, respectively. We use the complex-scaling method of the chiral quark model varying $\theta$ from $0^\circ$ to $6^\circ$.}
\end{figure}

\begin{table}[!t]
\caption{\label{GresultR5} Compositeness of the exotic resonances obtained in a complete coupled-channel analysis by the CSM in $1(1^+)$ state of $c\bar{c}q\bar{q}$ tetraquark. The results are similarly organized as those in Table~\ref{GresultR1}.}
\begin{ruledtabular}
\begin{tabular}{lccc}
Resonance       & \multicolumn{3}{c}{Structure} \\[2ex]
$3917+i1.1$   & \multicolumn{3}{c}{$r_{c\bar{c}}:2.1$;\,\,\,\,\,\,$r_{cq}:2.2$;\,\,\,\,\,\,$r_{c\bar{q}}:1.1$} \\
   & \multicolumn{3}{c}{$r_{q\bar{c}}:1.2$;\,\,\,\,\,\,$r_{\bar{q}\bar{c}}:2.2$;\,\,\,\,\,\,$r_{q\bar{q}}:2.2$} \\
   & \multicolumn{3}{c}{$S$: 43.6\%;\, $H$: 1.6\%;\, $Di$: 20.4\%;\, $K$: 34.4\%}\\[1.5ex]
$4142+i6.2$   & \multicolumn{3}{c}{$r_{c\bar{c}}:1.3$;\,\,\,\,\,\,$r_{cq}:1.4$;\,\,\,\,\,\,$r_{c\bar{q}}:0.7$} \\
   & \multicolumn{3}{c}{$r_{q\bar{c}}:1.0$;\,\,\,\,\,\,$r_{\bar{q}\bar{c}}:1.4$;\,\,\,\,\,\,$r_{q\bar{q}}:1.3$} \\
   & \multicolumn{3}{c}{$S$: 44.0\%;\, $H$: 4.8\%;\, $Di$: 30.4\%;\, $K$: 20.8\%}\\[1.5ex]
$4560+i1.6$   & \multicolumn{3}{c}{$r_{c\bar{c}}:0.4$;\,\,\,\,\,\,$r_{cq}:1.5$;\,\,\,\,\,\,$r_{c\bar{q}}:1.4$} \\
   & \multicolumn{3}{c}{$r_{q\bar{c}}:1.5$;\,\,\,\,\,\,$r_{\bar{q}\bar{c}}:1.4$;\,\,\,\,\,\,$r_{q\bar{q}}:1.5$} \\
   & \multicolumn{3}{c}{$S$: 28.9\%;\, $H$: 12.1\%;\, $Di$: 16.2\%;\, $K$: 42.8\%}\\[1.5ex]
$4668+i1.2$   & \multicolumn{3}{c}{$r_{c\bar{c}}:2.2$;\,\,\,\,\,\,$r_{cq}:2.3$;\,\,\,\,\,\,$r_{c\bar{q}}:1.4$} \\
   & \multicolumn{3}{c}{$r_{q\bar{c}}:1.0$;\,\,\,\,\,\,$r_{\bar{q}\bar{c}}:2.3$;\,\,\,\,\,\,$r_{q\bar{q}}:2.3$} \\
   & \multicolumn{3}{c}{$S$: 42.6\%;\, $H$: 4.4\%;\, $Di$: 42.8\%;\, $K$: 10.2\%}\\[1.5ex]
$4704+i5.2$   & \multicolumn{3}{c}{$r_{c\bar{c}}:1.8$;\,\,\,\,\,\,$r_{cq}:1.9$;\,\,\,\,\,\,$r_{c\bar{q}}:1.1$} \\
   & \multicolumn{3}{c}{$r_{q\bar{c}}:1.2$;\,\,\,\,\,\,$r_{\bar{q}\bar{c}}:1.9$;\,\,\,\,\,\,$r_{q\bar{q}}:1.9$} \\
   & \multicolumn{3}{c}{$S$: 44.1\%;\, $H$: 3.9\%;\, $Di$: 32.0\%;\, $K$: 20.0\%}\\[1.5ex]
$4710+i3.0$   & \multicolumn{3}{c}{$r_{c\bar{c}}:1.1$;\,\,\,\,\,\,$r_{cq}:1.7$;\,\,\,\,\,\,$r_{c\bar{q}}:1.3$} \\
   & \multicolumn{3}{c}{$r_{q\bar{c}}:1.4$;\,\,\,\,\,\,$r_{\bar{q}\bar{c}}:1.6$;\,\,\,\,\,\,$r_{q\bar{q}}:1.7$} \\
   & \multicolumn{3}{c}{$S$: 28.1\%;\, $H$: 18.4\%;\, $Di$: 18.7\%;\, $K$: 34.8\%}
\end{tabular}
\end{ruledtabular}
\end{table}

{\bf The $\bm{I(J^P)=1(1^+)}$ sector:} The numerical analysis of this case is similar to the $I(J^P)=0(1^+)$ because 26 channels must be explored, they are shown in Table~\ref{GresultCC5}. Firstly, in computations that include single channel, configuration coupled-channels and fully coupled-channels, no bound state is found, and the lowest mass ($3246$ MeV) is just the theoretical threshold value of $J/\psi \pi$. The masses of each channel in exotic configurations, which include hidden-color, diquark-antidiquark and K-type, are located at an energy region from $4.05$ GeV to $4.33$ GeV. Besides, the lowest mass in coupled-channels computation considering only one particular exotic configuration, is about $4.0$ GeV.

If one performs a complex-range study of the complete coupled-channels calculation, where the rotated angle $\theta$ is varied from $0^\circ$ to $6^\circ$, the complex energies are shown in Fig.~\ref{PP5}. Particularly, within $3.2-4.7$ GeV, scattering states of $J/\psi \pi$, $J/\psi \rho$, $\eta_c \rho$, $D\bar{D}^*$ and $D^{(*)} \bar{D}^{(*)}$ are well presented in the top panel. However, one stable resonance pole is clearly obtained at around $4.2$ GeV, and the complex energy is $4142+i6.2$ MeV. From Table~\ref{GresultR5} one can conclude that it is a compact tetraquark configuration, and the size is less than $1.4$ fm. Coupling is strong among color-singlet ($44\%$), diquark-antidiquark ($30\%$) and K-type channels ($21\%$). Particularly, the dominant two mesons decay channels are the $D\bar{D}^*$ ($25\%$) and $D^* \bar{D}^*$ ($12\%$). After revising the experimental data on exotic states at $\sim 4.1$ GeV within the isovector sector, the $X(4100)$, $X(4160)$ and $Z_c(4200)$ can be identified as the obtained resonance. Their golden experimental channels to analyze them are suggested to be $D\bar{D}^*$ and $D^* \bar{D}^*$.

An enlarged part of the energy region from $3.82$ GeV to $3.97$ GeV is plotted in the middle panel of Fig.~\ref{PP5}. Therein, scattering states of $\psi(2S) \pi(1S)$, $J/\psi \rho(1S)$ and $D(1S)\bar{D}^*(1S)$ are clearly shown. Meanwhile, one narrow resonance is obtained, and the complex energy is $3917+i1.1$ MeV. Accordingly, it is consistent with several other theoretical works~\cite{Ding:2020dio, Wang:2020dko, Wang:2022ztm, Chen:2021erj, Wang:2022fdu, Chen:2022ddj}, the $Z_c(3900)$ state can be well identified as a $D\bar{D}^*$ resonance, whose molecular structure is presented in Table~\ref{GresultR5}. Particularly, the dominant component is about $44\%$ $D\bar{D}^*$ state in color-singlet channel, and there is also a strong coupling between diquark-antidiquark ($20\%$) and K-type ($34\%$) configurations. The calculated size is $\sim 2.2$ fm, while it is $1.1$ fm for sub-clusters of $(c\bar{q})$ and $(q\bar{c})$.

Furthermore, four narrow resonances are obtained in the bottom panel of Fig.~\ref{PP5}, which shows an enlarged part of dense energy region from $4.35$ GeV to $4.75$ GeV. First, the $J/\psi \pi$, $J/\psi \rho$, $\eta_c \rho$, $D\bar{D}^*$ and $D^* \bar{D}^*$ scattering states are presented. Besides, the circled stable resonances have complex energies $4560+i1.6$ MeV, $4668+i1.2$ MeV, $4704+i5.2$ MeV and $4710+i3.0$ MeV. Their nature is analyzed in Table~\ref{GresultR5}. In particular, a molecular structure can be distinguish for the resonances at $4.67$ GeV and $4.70$ GeV. The dominant meson-meson channel of them is similar, a combination of $D\bar{D}^* (27\%)$ and $D^* \bar{D}^* (14\%)$. The other two resonances at $4.56$ GeV and $4.71$ GeV are compact $c\bar{c}q\bar{q}$ tetraquarks, whose size is less than $1.7$ fm. Meanwhile, their exotic components, hidden-color, diquark-antidiquark and K-type channels, sum up $\sim 72\%$. The dominant di-meson decay channel of the lower and higher resonance is $J/\psi \rho$ and $D\bar{D}^*$, respectively. These resonances above $4.5$ GeV are also expected to be confirmed in further experiments.


\begin{table}[!t]
\caption{\label{GresultCC6} Lowest-lying $c\bar{c}q\bar{q}$ tetraquark states with $I(J^P)=1(2^+)$ calculated within the real range formulation of the chiral quark model. The results are similarly organized as those in Table~\ref{GresultCC1}.
(unit: MeV).}
\begin{ruledtabular}
\begin{tabular}{lcccc}
~~Channel   & Index & $\chi_J^{\sigma_i}$;~$\chi_I^{f_j}$;~$\chi_k^c$ & $M$ & Mixed~~ \\
        &   &$[i; ~j; ~k]$ &  \\[2ex]
$(J/\psi \rho)^1 (3867)$  & 1  & [1;~1;~1]   & $3869$ &  \\
$(D^* \bar{D}^*)^1 (4014)$  & 2  & [1;~1;~1]   & $4034$ & $3869$ \\[2ex]
$(J/\psi \rho)^8$  & 3  & [1;~1;~2]   & $4351$ &  \\
$(D^* \bar{D}^*)^8$  & 4  & [1;~1;~2]   & $4263$ & $4192$ \\[2ex]
$(cq)^*(\bar{q}\bar{c})^*$  & 5  & [1;~1;~3]   & $4282$ & \\
$(cq)^*(\bar{q}\bar{c})^*$  & 6  & [1;~1;~4]   & $4244$ & $4166$ \\[2ex]
$K_1$  & 7  & [1;~1;~5]   & $4347$ & \\
  & 8  & [1;~1;~6]   & $4161$ & $4161$ \\[2ex]
$K_3$  & 9  & [1;~1;~9]   & $4250$ & \\
  & 10  & [1;~1;~10]   & $4275$ & $4171$ \\[2ex]
\multicolumn{4}{c}{Complete coupled-channels:} & $3869$
\end{tabular}
\end{ruledtabular}
\end{table}

\begin{figure}[!t]
\includegraphics[width=0.49\textwidth, trim={2.3cm 2.0cm 2.0cm 1.0cm}]{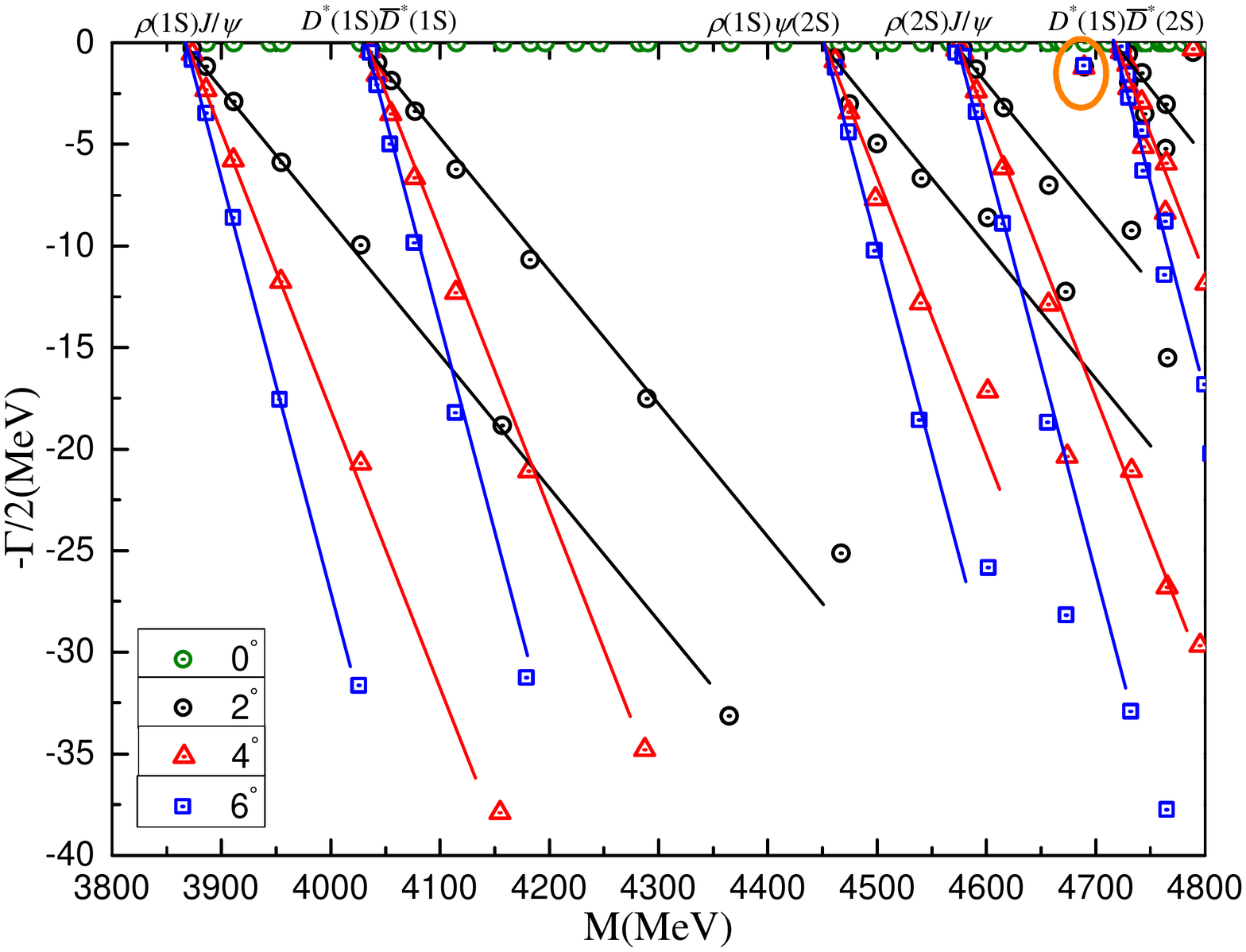}
\caption{\label{PP6} The complete coupled-channels calculation of the $c\bar{c}q\bar{q}$ tetraquark system with $I(J^P)=1(2^+)$ quantum numbers. We use the complex-scaling method of the chiral quark model varying $\theta$ from $0^\circ$ to $6^\circ$.}
\end{figure}

\begin{table}[!t]
\caption{\label{GresultR6} Compositeness of the exotic resonances obtained in a complete coupled-channel analysis by the CSM in $1(2^+)$ state of $c\bar{c}q\bar{q}$ tetraquark. The results are similarly organized as those in Table~\ref{GresultR1}.}
\begin{ruledtabular}
\begin{tabular}{lccc}
Resonance       & \multicolumn{3}{c}{Structure} \\[2ex]
$4689+i2.4$   & \multicolumn{3}{c}{$r_{c\bar{c}}:0.7$;\,\,\,\,\,\,$r_{cq}:1.2$;\,\,\,\,\,\,$r_{c\bar{q}}:1.2$} \\
& \multicolumn{3}{c}{$r_{q\bar{c}}:1.2$;\,\,\,\,\,\,$r_{\bar{q}\bar{c}}:1.2$;\,\,\,\,\,\,$r_{q\bar{q}}:1.3$} \\
& \multicolumn{3}{c}{$S$: 0.2\%;\, $H$: 3.2\%;\, $Di$: 30.9\%;\, $K$: 65.7\%}
\end{tabular}
\end{ruledtabular}
\end{table}

{\bf The $\bm{I(J^P)=1(2^+)}$ sector:} The results for the highest spin and isospin $c\bar{c}q\bar{q}$ state are shown in Table~\ref{GresultCC6} and Fig.~\ref{PP6}. The $J/\psi \rho$ and $D^* \bar{D}^*$ are both scattering states. Furthermore, states of hidden-color, diquark-antidiquark and K-type configurations are generally located in the energy range $4.16-4.35$ GeV. In each coupled-channel calculation, their masses are $\sim 4.16$ GeV, and the scattering nature of lowest channel $J/\psi \rho$ remains at $3869$ MeV.

In Fig.~\ref{PP6}, the distributions of complex energies of $J/\psi \rho$ and $D^* \bar{D}^*$ are well presented. Besides, within the energy region from $3.8$ GeV to $4.8$ GeV, a resonance state is found with complex energy $4689+i2.4$ MeV. A compact $c\bar{c}q\bar{q}$ tetraquark structure is presented in Table~\ref{GresultR6} for this state. Particularly, the distances between two quarks, two antiquarks and quark-antiquark are less than $1.3$ fm. This compact configuration is also compatible with a fact that the dominant component of resonance is made up by diquark-antidiquark $(31\%)$ and K-type $(66\%)$ channels. Hence, this resonance at $4.69$ GeV is expected to be confirmed experimentally, and it can be studied in the $J/\psi \rho$ and $D^* \bar{D}^*$ decay channels.


\subsection{The $\mathbf{c\bar{c}s\bar{s}}$ tetraquarks}

Three spin-parity states, $J^P=0^+$, $1^+$ and $2^+$, with isospin $I=0$, are investigated. Several narrow resonances, which may be identified as experimentally reported states, are only obtained in $J^P=0^+$ and $1^+$ channels. Details of the calculation as well as the related discussion can be found below.


\begin{table}[!t]
\caption{\label{GresultCC7} Lowest-lying $c\bar{c}s\bar{s}$ tetraquark states with $I(J^P)=0(0^+)$ calculated within the real range formulation of the chiral quark model. The results are similarly organized as those in Table~\ref{GresultCC1}.
(unit: MeV).}
\begin{ruledtabular}
\begin{tabular}{lcccc}
~~Channel   & Index & $\chi_J^{\sigma_i}$;~$\chi_I^{f_j}$;~$\chi_k^c$ & $M$ & Mixed~~ \\
        &   &$[i; ~j; ~k]$ &  \\[2ex]
$(\eta_c \eta')^1 (3939)$          & 1  & [1;~2;~1]  & $3817$ & \\
$(J/\psi \phi)^1 (4117)$  & 2  & [2;~2;~1]   & $4108$ &  \\
$(D_s \bar{D}_s)^1 (3936)$          & 3  & [1;~2;~1]  & $3978$ & \\
$(D^*_s \bar{D}^*_s)^1 (4224)$  & 4  & [2;~2;~1]   & $4230$ & $3817$ \\[2ex]
$(\eta_c \eta')^8$          & 5  & [1;~2;~2]  & $4570$ & \\
$(J/\psi \phi)^8$  & 6  & [2;~2;~2]   & $4521$ &  \\
$(D_s \bar{D}_s)^8$          & 7  & [1;~2;~2]  & $4542$ & \\
$(D^*_s \bar{D}^*_s)^8$  & 8  & [2;~2;~2]   & $4494$ & $4333$ \\[2ex]
$(cs)(\bar{s}\bar{c})$      & 9   & [3;~2;~3]  & $4429$ & \\
$(cs)(\bar{s}\bar{c})$      & 10   & [3;~2;~4]  & $4456$ & \\
$(cs)^*(\bar{s}\bar{c})^*$  & 11  & [4;~2;~3]   & $4454$ & \\
$(cs)^*(\bar{s}\bar{c})^*$  & 12  & [4;~2;~4]   & $4354$ & $4290$ \\[2ex]
$K_1$  & 13  & [5;~2;~5]   & $4517$ & \\
  & 14  & [6;~2;~5]   & $4567$ & \\
  & 15  & [5;~2;~6]   & $4405$ & \\
  & 16  & [6;~2;~6]   & $4246$ & $4241$ \\[2ex]
$K_3$  & 17  & [9;~2;~9]   & $4368$ & \\
  & 18  & [10;~2;~9]   & $4475$ & \\
  & 19  & [9;~2;~10]   & $4450$ & \\
  & 20  & [10;~2;~10]   & $4434$ & $4292$ \\[2ex]
\multicolumn{4}{c}{Complete coupled-channels:} & $3817$
\end{tabular}
\end{ruledtabular}
\end{table}

\begin{figure}[!t]
\includegraphics[clip, trim={3.0cm 1.9cm 3.0cm 1.0cm}, width=0.45\textwidth]{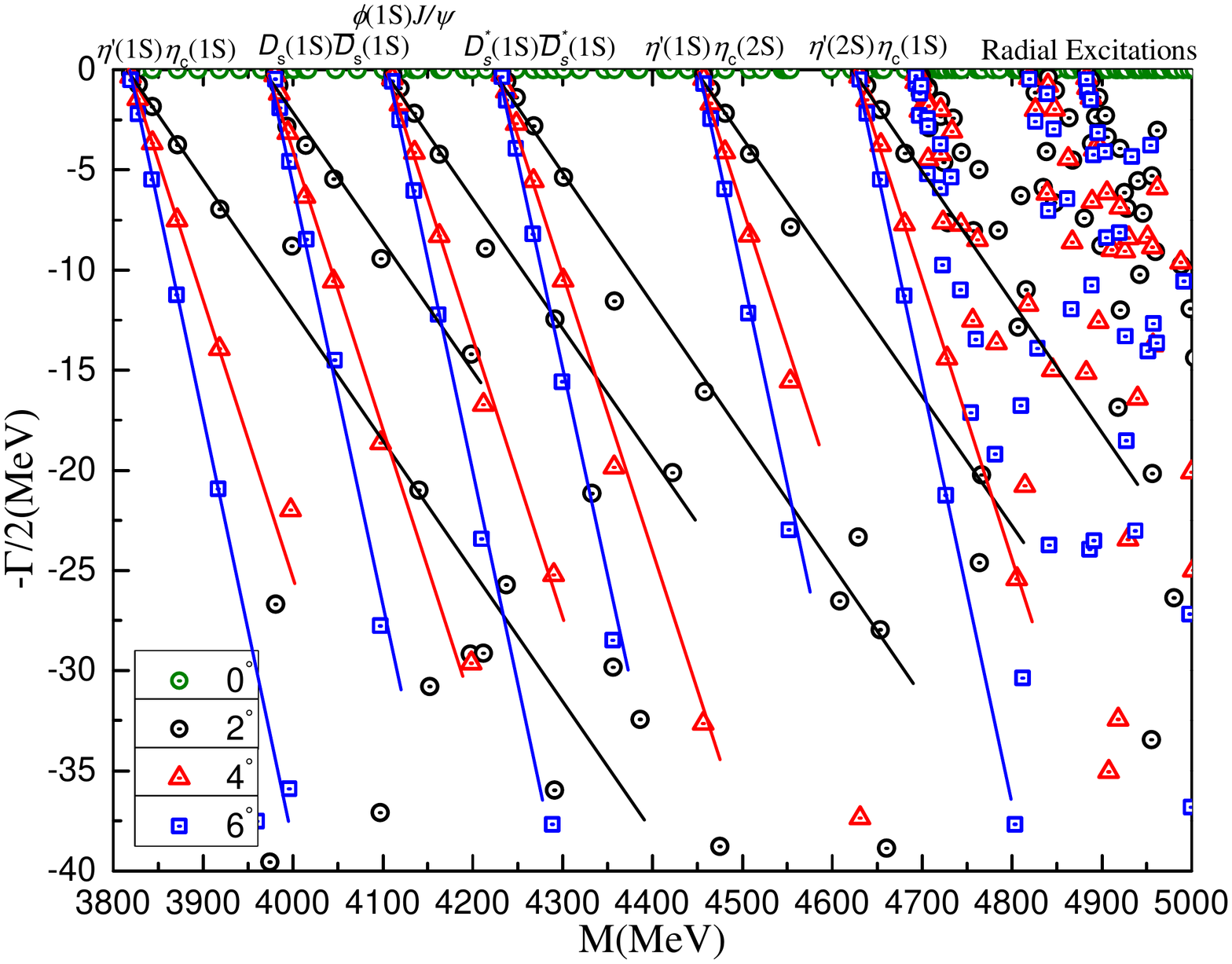} \\
\includegraphics[clip, trim={3.0cm 1.9cm 3.0cm 1.0cm}, width=0.45\textwidth]{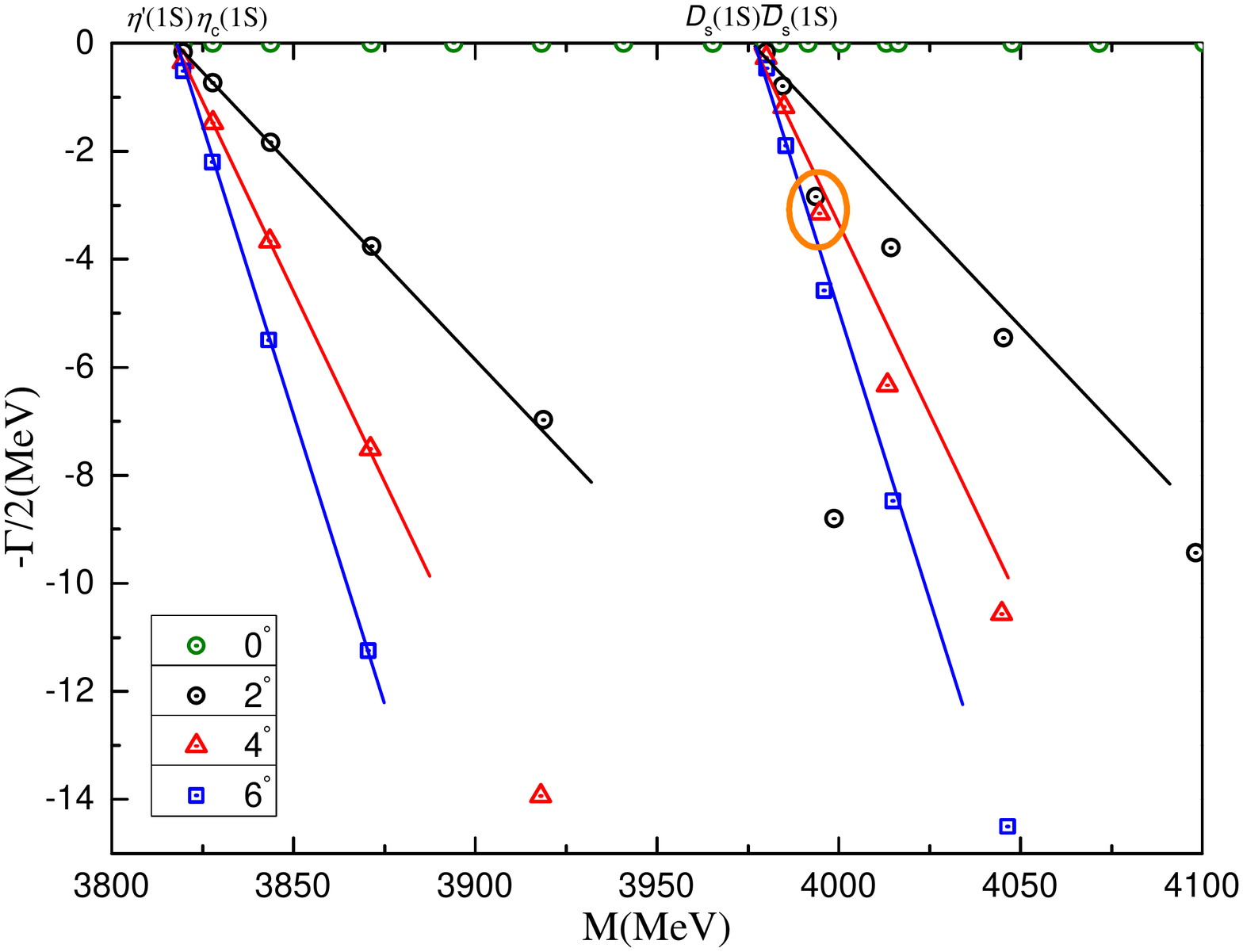}\\
\includegraphics[clip, trim={3.0cm 1.9cm 3.0cm 1.0cm}, width=0.45\textwidth]{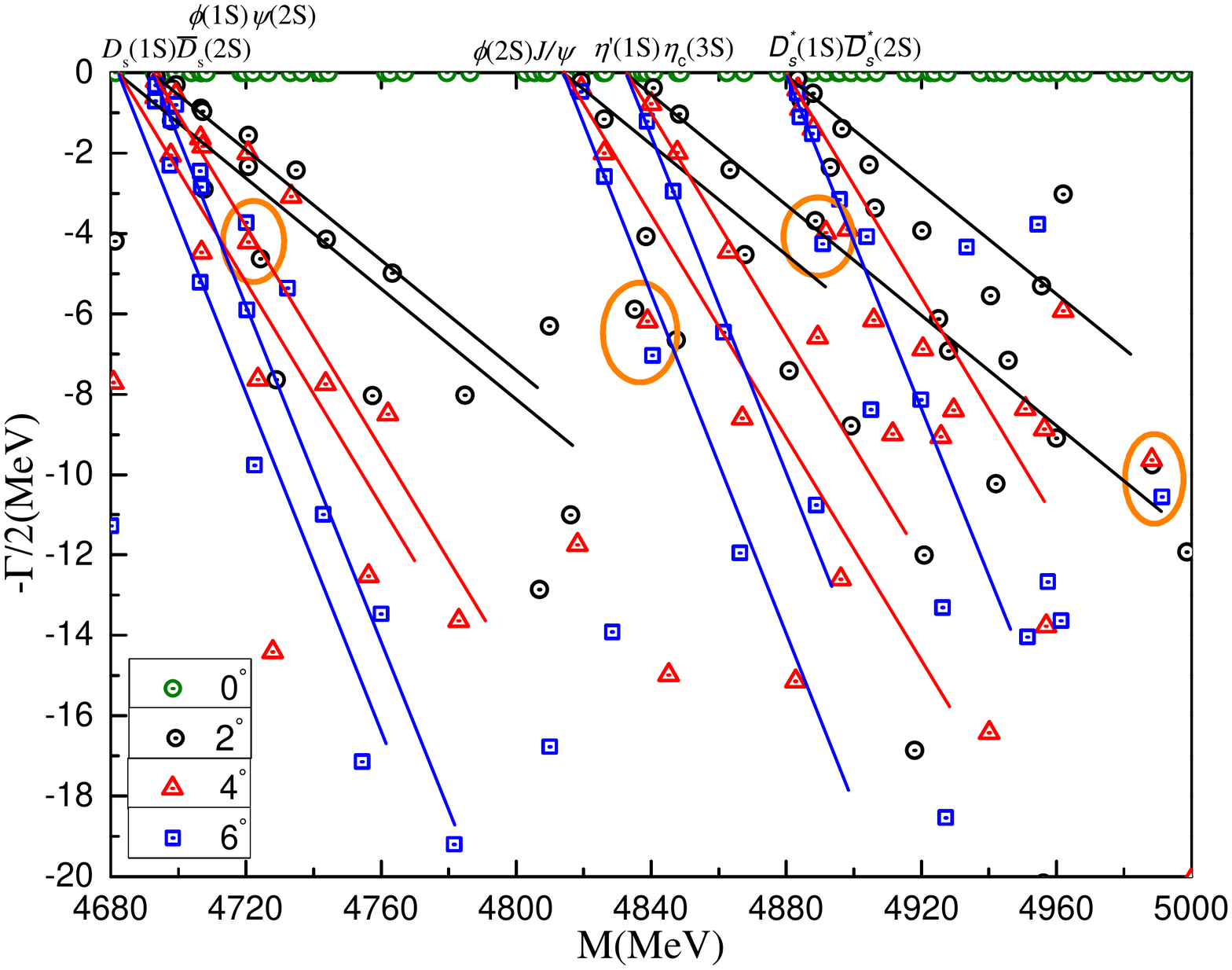}
\caption{\label{PP7} {\it Top panel:} The complete coupled-channels calculation of the $c\bar{c}s\bar{s}$ tetraquark system with $I(J^P)=0(0^+)$ quantum numbers. Particularly, middle and bottom panels are enlarged parts of dense energy region from $3.8\,\text{GeV}$ to $4.1\,\text{GeV}$, and $4.68\,\text{GeV}$ to $5.0\,\text{GeV}$, respectively. We use the complex-scaling method of the chiral quark model varying $\theta$ from $0^\circ$ to $6^\circ$.}
\end{figure}

\begin{table}[!t]
\caption{\label{GresultR7} Compositeness of the exotic resonances obtained in a complete coupled-channel analysis by the CSM in $0(0^+)$ state of $c\bar{c}s\bar{s}$ tetraquark. The results are similarly organized as those in Table~\ref{GresultR1}.}
\begin{ruledtabular}
\begin{tabular}{lccc}
Resonance       & \multicolumn{3}{c}{Structure} \\[2ex]
$3995+i6.4$   & \multicolumn{3}{c}{$r_{c\bar{c}}:2.0$;\,\,\,\,\,\,$r_{cs}:2.1$;\,\,\,\,\,\,$r_{c\bar{s}}:1.4$} \\
   & \multicolumn{3}{c}{$r_{s\bar{c}}:1.4$;\,\,\,\,\,\,$r_{\bar{s}\bar{c}}:2.0$;\,\,\,\,\,\,$r_{s\bar{s}}:2.1$} \\
   & \multicolumn{3}{c}{$S$: 41.3\%;\, $H$: 2.8\%;\, $Di$: 41.7\%;\, $K$: 14.2\%}\\[1.5ex]
$4720+i8.4$   & \multicolumn{3}{c}{$r_{c\bar{c}}:2.0$;\,\,\,\,\,\,$r_{cs}:2.1$;\,\,\,\,\,\,$r_{c\bar{s}}:1.0$} \\
   & \multicolumn{3}{c}{$r_{s\bar{c}}:1.1$;\,\,\,\,\,\,$r_{\bar{s}\bar{c}}:2.1$;\,\,\,\,\,\,$r_{s\bar{s}}:2.1$} \\
   & \multicolumn{3}{c}{$S$: 46.8\%;\, $H$: 1.8\%;\, $Di$: 46.2\%;\, $K$: 5.2\%}\\[1.5ex]
$4838+i12.4$   & \multicolumn{3}{c}{$r_{c\bar{c}}:1.7$;\,\,\,\,\,\,$r_{cs}:2.2$;\,\,\,\,\,\,$r_{c\bar{s}}:1.4$} \\
   & \multicolumn{3}{c}{$r_{s\bar{c}}:1.5$;\,\,\,\,\,\,$r_{\bar{s}\bar{c}}:2.1$;\,\,\,\,\,\,$r_{s\bar{s}}:2.0$} \\
   & \multicolumn{3}{c}{$S$: 29.5\%;\, $H$: 11.9\%;\, $Di$: 34.2\%;\, $K$: 24.4\%}\\[1.5ex]
$4891+i8.0$   & \multicolumn{3}{c}{$r_{c\bar{c}}:1.2$;\,\,\,\,\,\,$r_{cs}:2.0$;\,\,\,\,\,\,$r_{c\bar{s}}:2.0$} \\
   & \multicolumn{3}{c}{$r_{s\bar{c}}:2.1$;\,\,\,\,\,\,$r_{\bar{s}\bar{c}}:2.0$;\,\,\,\,\,\,$r_{q\bar{s}}:2.2$} \\
   & \multicolumn{3}{c}{$S$: 17.2\%;\, $H$: 27.4\%;\, $Di$: 15.2\%;\, $K$: 40.2\%}\\[1.5ex]
$4988+i19.6$   & \multicolumn{3}{c}{$r_{c\bar{c}}:1.8$;\,\,\,\,\,\,$r_{cs}:2.1$;\,\,\,\,\,\,$r_{c\bar{s}}:1.2$} \\
   & \multicolumn{3}{c}{$r_{s\bar{c}}:1.2$;\,\,\,\,\,\,$r_{\bar{s}\bar{c}}:2.1$;\,\,\,\,\,\,$r_{s\bar{s}}:2.1$} \\
   & \multicolumn{3}{c}{$S$: 44.1\%;\, $H$: 3.6\%;\, $Di$: 40.9\%;\, $K$: 11.4\%}
\end{tabular}
\end{ruledtabular}
\end{table}

{\bf The $\bm{I(J^P)=0(0^+)}$ sector:} Table~\ref{GresultCC7} lists the calculated masses. Firstly, no bound state is found in each single channel and in the different variants of coupled-channels computations. The lowest mass is $3817$ MeV, which is the theoretical value of $\eta_c \eta'$ threshold. Besides, we find the following masses $4108$ MeV, $3978$ MeV and $4230$ MeV for the other three di-meson channels in color-singlet configuration, \emph{i.e.} $J/\psi \phi$, $D_s \bar{D}_s$ and $D^*_s \bar{D}^*_s$. The remaining channels, considered exotic configurations, are generally located within an energy interval $4.25-4.57$ GeV.

When a complete coupled-channels computation is performed by using the complex-scaling method, the scattering nature of $\eta_c \eta'$, $J/\psi\phi$, $D_s \bar{D}_s$ and $D^*_s \bar{D}^*_s$ are still kept and shown in the top panel of Fig.~\ref{PP7} (the energy region drawn is from $3.8$ GeV to $5.0$ GeV). No clear resonances are found; however, there is a hint in the middle panel. In particular, within an enlarged part of energy interval $3.8-4.1$ GeV, one stable resonance is obtained at $3995+i6.4$ MeV. A molecular nature of this state can be guess from Table~\ref{GresultR7}. Therein, the distances of $r_{c\bar{s}}$ and $r_{s\bar{c}}$ are both $1.4$ fm, and the other cases are around $2.0$ fm. This $(c\bar{s})-(s\bar{c})$ molecule is confirmed by the wave function probabilities of each channel. Particularly, there is a strong coupling between color-singlet channels $(41\%)$ and diquark-antidiquark ones $(42\%)$; moreover, the dominant meson-meson channels are $D_s \bar{D}_s$ $(19\%)$ and $D^*_s \bar{D}^*_s$ $(18\%)$ states. After performing a mass shift of $-42$ MeV, which is the difference between the theoretical and experimental values of the $D_s \bar{D}_s$ thresholds, the resonance's modified mass is $3953$ MeV. Hence, the recently reported $X(3960)$ state, whose quantum numbers are $I(J^P)=0(0^+)$, can be well assigned in our theoretical framework as a molecule whose dominant meson-meson components are the $D_s \bar{D}_s$ and $D^*_s \bar{D}^*_s$. In particular, the $D_s \bar{D}_s$ molecular resonance is also indicated in Refs.~\cite{Bayar:2022dqa, Ji:2022uie, Mutuk:2022ckn}.

In a further step, we focus on the dense energy region from $4.68$ GeV to $5.0$ GeV, which is plotted in the bottom panel of Fig.~\ref{PP7}. Therein, apart from the radial excitations of $\eta_c \eta'$, $J/\psi\phi$, $D_s \bar{D}_s$ and $D^*_s \bar{D}^*_s$, four narrow resonances are circled, and their complex energies read $4720+i8.4$ MeV, $4838+i12.4$ MeV, $4891+i8.0$ MeV and $4988+i19.6$ MeV. The nature of these states can also be guess by analyzing the data in Table~\ref{GresultR7}. Firstly, the size of these four resonances is around $2.0$ fm, and it seems that they are molecules except the resonance at $4.89$ GeV. Their dominant meson-meson channel is $D^*_s \bar{D}^*_s$; besides, there are also considerable components of diquark-antidiquark and K-type channels. There is only one exotic state, $X(4700)$, reported experimentally within the energy region of interest. On one hand, it can be well identified as a $D^*_s \bar{D}^*_s$ resonance with $I(J^P)=0(0^+)$, on the other hand, the resonances with $c\bar{c}s\bar{s}$ tetraquark content at $4.8-4.9$ GeV are expected to be confirmed in further experimental investigations.


\begin{table}[!t]
\caption{\label{GresultCC8} Lowest-lying $c\bar{c}s\bar{s}$ tetraquark states with $I(J^P)=0(1^+)$ calculated within the real range formulation of the chiral quark model. The results are similarly organized as those in Table~\ref{GresultCC1}.
(unit: MeV).}
\begin{ruledtabular}
\begin{tabular}{lcccc}
~~Channel   & Index & $\chi_J^{\sigma_i}$;~$\chi_I^{f_j}$;~$\chi_k^c$ & $M$ & Mixed~~ \\
        &   &$[i; ~j; ~k]$ &  \\[2ex]
$(\eta_c \phi)^1 (4001)$          & 1  & [1;~2;~1]  & $4000$ & \\
$(J/\psi \eta')^1 (4055)$  & 2  & [2;~2;~1]   & $3925$ &  \\
$(J/\psi \phi)^1 (4117)$  & 3  & [3;~2;~1]   & $4108$ &  \\
$(D_s \bar{D}^*_s)^1 (4080)$          & 4  & [1;~2;~1]  & $4104$ & \\
$(D^*_s \bar{D}^*_s)^1 (4224)$  & 5  & [3;~2;~1]   & $4230$ & $3925$ \\[2ex]
$(\eta_c \phi)^8$  & 6  & [1;~2;~2]  & $4553$ & \\
$(J/\psi \eta')^8$    & 7  & [2;~2;~2]   & $4572$ &  \\
$(J/\psi \phi)^8$  & 8  & [3;~2;~2]   & $4539$ &  \\
$(D_s \bar{D}^*_s)^8$          & 9 & [1;~2;~2]  & $4546$ & \\
$(D^*_s \bar{D}^*_s)^8$      & 10  & [3;~2;~2]   & $4510$ & $4367$ \\[2ex]
$(cs)(\bar{s}\bar{c})^*$      & 11   & [4;~2;~3]  & $4463$ & \\
$(cs)(\bar{s}\bar{c})^*$      & 12   & [4;~2;~4]  & $4461$ & \\
$(cs)^*(\bar{s}\bar{c})^*$  & 13  & [6;~2;~3]   & $4447$ & \\
$(cs)^*(\bar{s}\bar{c})^*$  & 14  & [6;~2;~4]   & $4384$ & $4354$ \\[2ex]
$K_1$  & 15  & [7;~2;~5]   & $4558$ & \\
  & 16  & [8;~2;~5]   & $4546$ & \\
  & 17  & [9;~2;~5]   & $4550$ & \\
  & 18  & [7;~2;~6]   & $4372$ & \\
  & 19  & [8;~2;~6]   & $4389$ & \\
  & 20  & [9;~2;~6]   & $4298$ & $4296$ \\[2ex]
$K_3$  & 21  & [13;~2;~9]   & $4426$ & \\
  & 22  & [14;~2;~9]   & $4413$ & \\
  & 23  & [15;~2;~9]   & $4464$ & \\
  & 24  & [13;~2;~10]   & $4445$ & \\
  & 25  & [14;~2;~10]   & $4446$ & \\
  & 26  & [15;~2;~10]   & $4459$ & $4358$ \\[2ex]
\multicolumn{4}{c}{Complete coupled-channels:} & $3925$
\end{tabular}
\end{ruledtabular}
\end{table}

\begin{figure}[!t]
\includegraphics[clip, trim={3.0cm 1.9cm 3.0cm 1.0cm}, width=0.45\textwidth]{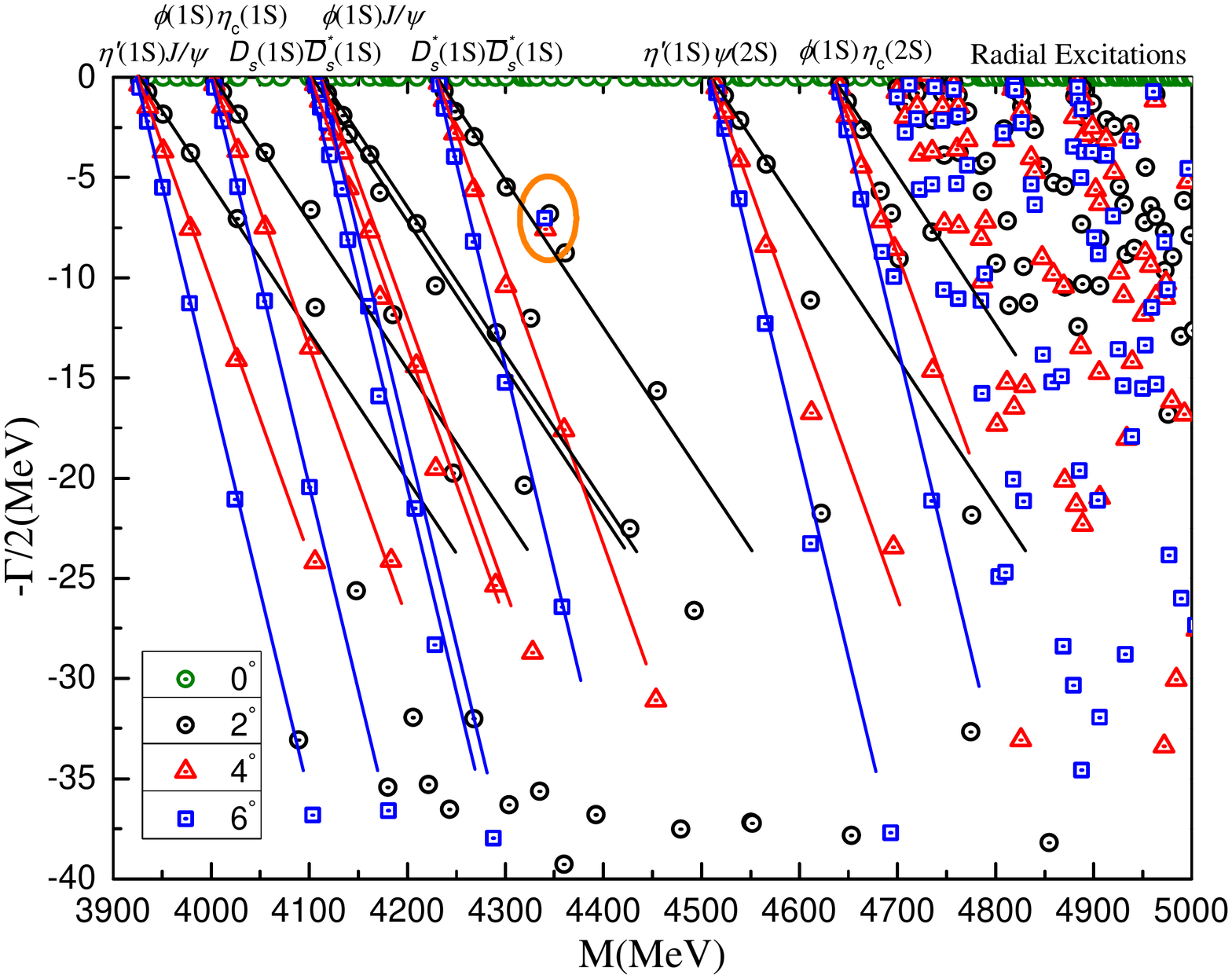}\\
\includegraphics[clip, trim={3.0cm 1.9cm 3.0cm 1.0cm}, width=0.45\textwidth]{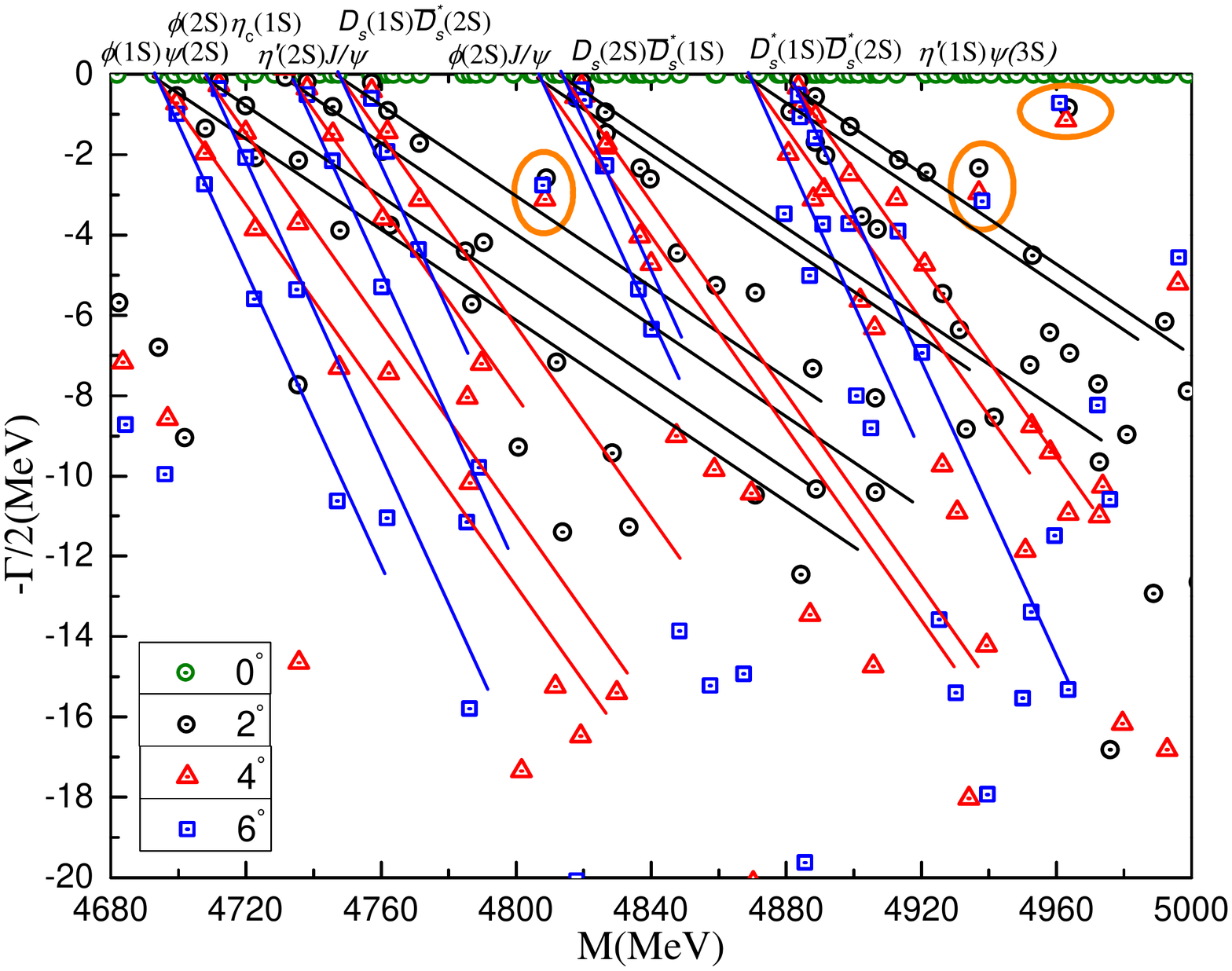}
\caption{\label{PP8} The complete coupled-channels calculation of the $c\bar{c}s\bar{s}$ tetraquark system with $I(J^P)=0(1^+)$ quantum numbers. {\it Bottom panel:} Enlarged top panel, with real values of energy ranging from $4.68\,\text{GeV}$ to $5.0\,\text{GeV}$. We use the complex-scaling method of the chiral quark model varying $\theta$ from $0^\circ$ to $6^\circ$.}
\end{figure}

\begin{table}[!t]
\caption{\label{GresultR8} Compositeness of the exotic resonances obtained in a complete coupled-channel analysis by the CSM in $0(1^+)$ state of $c\bar{c}s\bar{s}$ tetraquark. The results are similarly organized as those in Table~\ref{GresultR1}.}
\begin{ruledtabular}
\begin{tabular}{lccc}
Resonance       & \multicolumn{3}{c}{Structure} \\[2ex]
$4340+i15.2$   & \multicolumn{3}{c}{$r_{c\bar{c}}:1.5$;\,\,\,\,\,\,$r_{cs}:1.7$;\,\,\,\,\,\,$r_{c\bar{s}}:0.9$} \\
   & \multicolumn{3}{c}{$r_{s\bar{c}}:0.9$;\,\,\,\,\,\,$r_{\bar{s}\bar{c}}:1.7$;\,\,\,\,\,\,$r_{s\bar{s}}:1.6$} \\
   & \multicolumn{3}{c}{$S$: 30.7\%;\, $H$: 8.2\%;\, $Di$: 26.1\%;\, $K$: 35.0\%}\\[1.5ex]
$4808+i6.2$   & \multicolumn{3}{c}{$r_{c\bar{c}}:1.9$;\,\,\,\,\,\,$r_{cs}:2.1$;\,\,\,\,\,\,$r_{c\bar{s}}:1.1$} \\
   & \multicolumn{3}{c}{$r_{s\bar{c}}:1.5$;\,\,\,\,\,\,$r_{\bar{s}\bar{c}}:2.0$;\,\,\,\,\,\,$r_{s\bar{s}}:2.0$} \\
   & \multicolumn{3}{c}{$S$: 35.3\%;\, $H$: 2.8\%;\, $Di$: 39.1\%;\, $K$: 22.8\%}\\[1.5ex]
$4937+i6.0$   & \multicolumn{3}{c}{$r_{c\bar{c}}:1.7$;\,\,\,\,\,\,$r_{cs}:2.1$;\,\,\,\,\,\,$r_{c\bar{s}}:1.2$} \\
   & \multicolumn{3}{c}{$r_{s\bar{c}}:1.2$;\,\,\,\,\,\,$r_{\bar{s}\bar{c}}:2.1$;\,\,\,\,\,\,$r_{s\bar{s}}:1.8$} \\
   & \multicolumn{3}{c}{$S$: 48.7\%;\, $H$: 1.0\%;\, $Di$: 48.4\%;\, $K$: 1.9\%}\\[1.5ex]
$4962+i2.4$   & \multicolumn{3}{c}{$r_{c\bar{c}}:1.8$;\,\,\,\,\,\,$r_{cs}:2.1$;\,\,\,\,\,\,$r_{c\bar{s}}:1.3$} \\
   & \multicolumn{3}{c}{$r_{s\bar{c}}:1.3$;\,\,\,\,\,\,$r_{\bar{s}\bar{c}}:2.1$;\,\,\,\,\,\,$r_{s\bar{s}}:2.1$} \\
   & \multicolumn{3}{c}{$S$: 42.3\%;\, $H$: 4.9\%;\, $Di$: 40.7\%;\, K: 12.1\%}
\end{tabular}
\end{ruledtabular}
\end{table}

{\bf The $\bm{I(J^P)=0(1^+)}$ sector:} 26 channels listed in Table~\ref{GresultCC8} contribute. Particularly, five meson-meson channels, $\eta_c \phi$, $J/\psi \eta'$, $J/\psi \phi$, $D_s \bar{D}^*_s$ and $D^*_s \bar{D}^*_s$, are all scattering states, and the lowest mass $3925$ MeV is just the value of the non-interacting $J/\psi \eta'$ theoretical threshold. The other exotic quark arrangements are generally located in energy range $4.30-4.57$ GeV. No bound state is found in any coupled-channels calculation, \emph{i.e.} hidden-color, diquark-antidiquark, K-type and full cases, the lowest masses are around $4.3$ GeV.

When the fully coupled-channels computation is performed within a complex-range formalism, Fig.~\ref{PP8} shows the distribution of complex energies. In particular, the top panel focuses on the scattering states of $\eta_c \phi$, $J/\psi \eta'$, $J/\psi \phi$, $D_s \bar{D}^*_s$ and $D^*_s \bar{D}^*_s$, which are located in a mass region from $3.9$ GeV to $5.0$ GeV. Therein, a stable resonance pole is obtained, whose complex energy is $4340+i15.2$ MeV. This state may be compatible with the $X(4350)$ assignment whose isospin is $0$ but its spin and parity are unknown experimentally. The nature of the theoretical state can be guess by looking at Table~\ref{GresultR8}; therein, one can find that it is of molecular type, with a dominant $D_s \bar{D}^*_s$ meson-meson channel whose wave function probability is $\sim 30\%$. It is worth mentioning that there are also equivalent components of diquark-antidiquark $(26\%)$ and K-type $(35\%)$ configurations.

The energy region from $4.68$ GeV to $5.0$ GeV is enlarged in the bottom panel of Fig.~\ref{PP8}. Generally, radial excitations of $\eta_c \phi$, $J/\psi \eta'$, $J/\psi \phi$, $D_s \bar{D}^*_s$ and $D^*_s \bar{D}^*_s$ are presented. Besides, three narrow resonances are obtained above $4.8$ GeV. Their complex energies are $4808+i6.2$ MeV, $4937+i6.0$ MeV and $4962+i2.4$ MeV. By calculating some structure properties in Table~\ref{GresultR8}, molecular nature is better assigned fort them, with sizes around $2.0$ fm. Furthermore, their dominant meson-meson color-singlet channel is a combination of $D_s \bar{D}^*_s$ $(\sim 24\%)$ and $D^*_s \bar{D}^*_s$ $(\sim 11\%)$. All of these three resonances above $4.8$ GeV are expected to be confirmed experimentally.


\begin{table}[!t]
\caption{\label{GresultCC9} Lowest-lying $c\bar{c}s\bar{s}$ tetraquark states with $I(J^P)=0(2^+)$ calculated within the real range formulation of the chiral quark model. The results are similarly organized as those in Table~\ref{GresultCC1}.
(unit: MeV).}
\begin{ruledtabular}
\begin{tabular}{lcccc}
~~Channel   & Index & $\chi_J^{\sigma_i}$;~$\chi_I^{f_j}$;~$\chi_k^c$ & $M$ & Mixed~~ \\
        &   &$[i; ~j; ~k]$ &  \\[2ex]
$(J/\psi \phi)^1 (4117)$  & 1  & [1;~2;~1]   & $4108$ &  \\
$(D^*_s \bar{D}^*_s)^1 (4224)$  & 2  & [1;~2;~1]   & $4230$ & $4108$ \\[2ex]
$(J/\psi \phi)^8$  & 3  & [1;~2;~2]   & $4570$ &  \\
$(D^*_s \bar{D}^*_s)^8$  & 4  & [1;~2;~2]   & $4539$ & $4412$ \\[2ex]
$(cs)^*(\bar{s}\bar{c})^*$  & 5  & [1;~2;~3]   & $4472$ & \\
$(cs)^*(\bar{s}\bar{c})^*$  & 6  & [1;~2;~4]   & $4437$ & $4406$ \\[2ex]
$K_1$  & 7  & [1;~2;~5]   & $4569$ & \\
  & 8  & [1;~2;~6]   & $4405$ & $4405$ \\[2ex]
$K_3$  & 9  & [1;~2;~9]   & $4454$ & \\
  & 10  & [1;~2;~10]   & $4467$ & $4418$ \\[2ex]
\multicolumn{4}{c}{Complete coupled-channels:} & $4108$
\end{tabular}
\end{ruledtabular}
\end{table}

\begin{figure}[!t]
\includegraphics[width=0.49\textwidth, trim={2.3cm 2.0cm 2.0cm 1.0cm}]{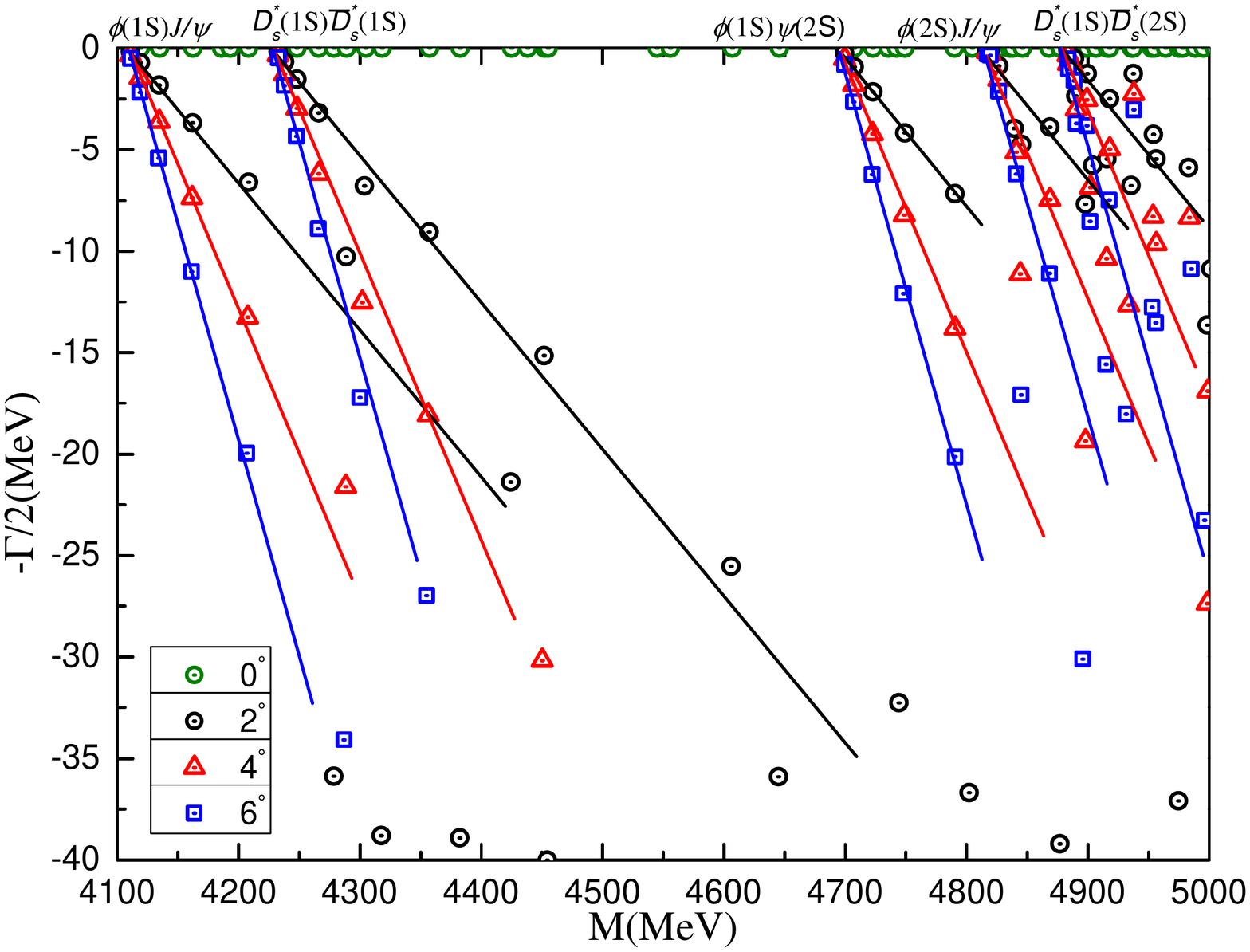}
\caption{\label{PP9} The complete coupled-channels calculation of the $c\bar{c}s\bar{s}$ tetraquark system with $I(J^P)=0(2^+)$ quantum numbers. We use the complex-scaling method of the chiral quark model varying $\theta$ from $0^\circ$ to $6^\circ$.}
\end{figure}

{\bf The $\bm{I(J^P)=0(2^+)}$ sector:} Table~\ref{GresultCC9} shows all the channels that can contribute to the mass of the highest spin state of $c\bar{c}s\bar{s}$ tetraquark. No bound state is found in the single- and coupled-channels calculations, with the lowest mass indicating the value of $J/\psi \phi$ theoretical threshold, $4108$ MeV. The results of a complete coupled-channels calculation using a complex-range method are shown in Fig.~\ref{PP9}. Therein, one can find that $J/\psi \phi$ and $D^*_s \bar{D}^*_s$ scattering states are well presented in $4.1-5.0$ GeV, and no stable resonance pole is acquired.


\begin{table*}[!t]
\caption{\label{GresultCCT} Summary of resonance structures found in the $c\bar{c}q\bar{q}$ $(q=u,\,d,\,s)$ tetraquark systems. The first column shows the isospin, total spin and parity of each singularity. If available, the second column lists well known experimental states, which may be identified in our theoretical framework. The third column refers to the dominant configuration components, particularly, $H$: hidden color, $Di$: diquark-antidiquark, $K$: K-type. Theoretical resonances are presented with the following notation: $E=M+i\Gamma$ in the last column (unit: MeV).}
\begin{ruledtabular}
\begin{tabular}{lccc}
~ $I(J^P)$ & Experimental state & Dominant Component   & Theoretical resonance~~ \\
\hline
\multicolumn{4}{c}{$c\bar{c}q\bar{q}$ $(q=u,\,d)$ tetraquarks}\\
~~$0(0^+)$ &  & $J/\psi \omega(18\%)+H(28\%)+K(46\%)$   & $4660+i5.6$~~  \\[2ex]
~~$0(1^+)$  & $X(3872)$ & $D\bar{D}^*(45\%)+Di(15\%)+K(37\%)$   & $3916+i0.8$~~ \\
            & $X(3940)$  & $H(30\%)+Di(14\%)+K(47\%)$   & $3950+i18.6$~~ \\
            & $X(4630)$  & $J/\psi \omega(19\%)+H(22\%)+Di(23\%)+K(36\%)$   & $4567+i2.0$~~ \\
            & $X(4630)$  & $J/\psi \omega(14\%)+H(31\%)+K(48\%)$   & $4572+i1.6$~~ \\
            & $X(4685)$  & $D\bar{D}^*(31\%)+H(18\%)+Di(16\%)+K(35\%)$   & $4690+i6.2$~~ \\[2ex]
~~$0(2^+)$  & $X(4630)$  & $Di(25\%)+K(72\%)$   & $4570+i0.6$~~ \\
                  &  $X(4630)$ & $H(13\%)+Di(38\%)+K(49\%)$   & $4667+i2.2$~~ \\[2ex]
~~$1(0^+)$  & $X(4020)$, $X(4050)$, $X(4055)$, $X(4100)$  & $J/\psi \rho(12\%)+D\bar{D}(25\%)+Di(28\%)+K(29\%)$   & $4036+i1.4$~~\\
                   &  & $J/\psi \rho(25\%)+D\bar{D}(14\%)+Di(30\%)+K(24\%)$   & $4515+i10.8$~~\\
                   &  & $J/\psi \rho(12\%)+D\bar{D}(27\%)+Di(30\%)+K(23\%)$   & $4548+i2.6$~~ \\
                   &  & $D^* \bar{D}^*(28\%)+H(18\%)+Di(24\%)+K(30\%)$   & $4784+i4.0$~~\\[2ex]
~~$1(1^+)$  & $Z_c(3900)$  & $D\bar{D}^*(44\%)+Di(20\%)+K(34\%)$   & $3917+i1.1$~~\\
                   & $X(4100)$, $X(4160)$, $Z_c(4200)$  & $D\bar{D}^*(25\%)+D^* \bar{D}^*(12\%)+Di(30\%)+K(21\%)$   & $4142+i6.2$~~ \\
                   &  & $J/\psi \rho(29\%)+H(12\%)+Di(16\%)+K(43\%)$   & $4560+i1.6$~~\\
                   &  & $D\bar{D}^*(27\%)+D^* \bar{D}^*(14\%)+Di(43\%)+K(10\%)$   & $4668+i1.2$~~\\
                   &  & $D\bar{D}^*(25\%)+D^* \bar{D}^*(13\%)++Di(32\%)+K(20\%)$   & $4704+i5.2$~~\\
                   &  & $D\bar{D}^*(28\%)+H(18\%)+Di(19\%)+K(35\%)$   & $4710+i3.0$~~\\[2ex]
~~$1(2^+)$  &  & $Di(31\%)+K(66\%)$   & $4689+i2.4$~~ \\
\hline
\multicolumn{4}{c}{$c\bar{c}s\bar{s}$ tetraquarks}\\
~~$0(0^+)$  & $X(3960)$  & $D_s \bar{D}_s(19\%)+D^*_s \bar{D}^*_s(18\%)+Di(42\%)+K(14\%)$ & $3995+i6.4$~~  \\
                    & $X(4700)$  & $D^*_s \bar{D}^*_s(47\%)+Di(46\%)$   & $4720+i8.4$~~ \\
                    &  & $D^*_s \bar{D}^*_s(30\%)+H(12\%)+Di(34\%)+K(24\%)$   & $4838+i12.4$~~ \\
                    &  & $D^*_s \bar{D}^*_s(18\%)+H(27\%)+Di(15\%)+K(40\%)$   & $4891+i8.0$~~ \\
                    &  & $D^*_s \bar{D}^*_s(44\%)+Di(41\%)+K(11\%)$   & $4988+i19.6$~~\\[2ex]
~~$0(1^+)$  & $X(4350)$  & $D_s \bar{D}^*_s(31\%)+Di(26\%)+K(35\%)$   & $4340+i15.2$~~ \\
                    &  & $D_s \bar{D}^*_s(21\%)+D^*_s \bar{D}^*_s(11\%)+Di(39\%)+K(23\%)$   & $4808+i6.2$~~ \\
                    &  & $D_s \bar{D}^*_s(32\%)+D^*_s \bar{D}^*_s(16\%)+Di(48\%)$   & $4937+i6.0$~~ \\
                    &  & $D_s \bar{D}^*_s(30\%)+D^*_s \bar{D}^*_s(9\%)+Di(41\%)+K(12\%)$   & $4962+i2.4$~~
\end{tabular}
\end{ruledtabular}
\end{table*}

\section{Summary}
\label{sec:summary}

The charmonium-like tetraquarks $c\bar{c}q\bar{q}$ $(q=u,\,d)$ and $c\bar{c}s\bar{s}$ with spin-parity $J^P=0^+$, $1^+$ and $2^+$, and isospin $I=0$ or $1$, are comprehensively investigated by means of complex-scaling range of a chiral quark model, which includes one-gluon exchange, linear-screened confining and Goldstone-boson exchanges between light quarks, along with a high efficient numerical approach, Gaussian expansion method. Meanwhile, all possible tetraquark arrangements allowed by quantum numbers are considered, particularly, there are singlet- and hidden-color meson-meson configurations, diquark-antidiquark arrangements with their allowed color triplet-antitriplet and sextet-antisextet wave functions, and K-type configurations.

Several narrow resonances are obtained in a fully coupled-channel computation, and they are summarized in Table~\ref{GresultCCT}. Therein, we collect quantum numbers, tentative assignments to experimental charmonium-like signals, dominant configuration components and pole position. Several conclusions are drawn below.

First of all, some well known exotic resonances may be identified as molecules in our study. Particularly, the $X(3872)$ is a $D \bar{D}^*$ molecule in the $I(J^P)=0(1^+)$ channel; besides, there is also considerable exotic components, $15\%$ diquark-antidiquark and $37\%$ K-type configurations. The $Z_c(3900)$ state has a similar nature, whose dominant component is $D \bar{D}^*$ in $I(J^P)=1(1^+)$. The recently reported $X(3960)$ in $I(J^P)=0(0^+)$ channel is identified as a coupling among $D_s \bar{D}_s$ $(19\%)$, $D^*_s \bar{D}^*_s$ $(18\%)$ and exotic channels $(56\%)$. The $X(4350)$ in $I(J^P)=0(1^+)$ is mainly a coupling between $D_s \bar{D}^*_s$ $(31\%)$ and exotic channels $(61\%)$, which include diquark-antidiquark $(26\%)$ and K-type $(35\%)$ configurations. The $X(4685)$, also in $I(J^P)=0(1^+)$, is a coupling between $D \bar{D}^*$ $(31\%)$ and exotic channels $(69\%)$, which contain hidden-color, diquark-antidiquark and K-type configurations. The $X(4700)$ state with $I(J^P)=0(0^+)$ quantum numbers mainly consists of $D^*_s \bar{D}^*_s$ $(47\%)$ and diquark-antidiquark $(46\%)$ channels.

Secondly, four compact $c\bar{c}q\bar{q}$ tetraquarks are obtained. In particular, the $X(3940)$ state can be identified as a tetraquark state with $I(J^P)=0(1^+)$, and its nature is a coupling among hidden-color $(30\%)$, diquark-antidiquark $(14\%)$ and K-type $(47\%)$ configurations. Furthermore, two narrow resonances at $4.57$ GeV and $4.67$ GeV are obtained in the $I(J^P)=0(2^+)$ sector. Their dominant components $(>97\%)$ are exotic structures, \emph{i.e.} hidden-color, diquark-antidiquark and K-type channels, and they may be compatible with the $X(4630)$. The fourth state has a mass of $4.69$ GeV and quantum numbers $I(J^P)=1(2^+)$. It is almost a combination of diquark-antidiquark $(31\%)$ and K-type $(66\%)$ structures.

Thirdly, two resonances, whose complex energies are $4036+i1.4$ MeV and $4142+i6.2$ MeV, are obtained in the isovector channel, and the spin-parity is $0^+$ and $1^+$, respectively. The lower resonance has $12\%$ $J/\psi \rho$ and $25\%$ $D\bar{D}$ component, and it may be compatible with the $X(4020)$, $X(4050)$, $X(4055)$ and $X(4100)$ experimental candidates. Meanwhile, the $X(4100)$, $X(4160)$ and $Z_c(4200)$ signals may be identified as the other resonance, whose dominant di-meson channels are $D\bar{D}^*$ $(25\%)$ and $D^* \bar{D}^*$ $(12\%)$.

Furthermore, several narrow resonances, which exceed present experimental data, are obtained within an energy region from $4.5$ GeV to $5.0$ GeV. Their pole positions, quantum numbers and dominant components are listed in Table~\ref{GresultCCT}, which will be useful in further experimental investigations of charmonium-like states.


\begin{acknowledgments}
Work partially financed by Zhejiang Provincial Natural Science Foundation under Grant No. LQ22A050004; National Natural Science Foundation of China under Grant Nos. 11535005 and 11775118; Ministerio Espa\~nol de Ciencia e Innovaci\'on under grant No. PID2019-107844GB-C22; the Junta de Andaluc\'ia under contract Nos. Operativo FEDER Andaluc\'ia 2014-2020 UHU-1264517, P18-FR-5057 and also PAIDI FQM-370.
\end{acknowledgments}


\bibliography{ccqq}

\end{document}